\documentclass[%
 reprint,
 amsmath,amssymb,
 aps,
 floatfix,
 ]{revtex4-2}

\usepackage{graphicx}      
\usepackage{dcolumn}       
\usepackage{bm}            
\usepackage[utf8]{inputenc}
\usepackage{textgreek}
\usepackage{array}
\usepackage{tabularray}
\usepackage{hyperref}      
\usepackage{float}
\usepackage{booktabs}
\usepackage{subcaption}
\usepackage{epstopdf}  
\usepackage{enumitem}
\usepackage{tikz}
\usepackage{caption} 
\usetikzlibrary{shapes.geometric, arrows.meta, positioning, calc}
\usepackage{footmisc}
\usepackage{cleveref}

\begin{document}

\preprint{APS/123-QED}

\title{Conditional Wasserstein GAN for Simulating Neutrino Event Summaries using Incident Energy of Electron Neutrinos}

\author{S. Dipthi}
\email{dipthi.s2022@vitstudent.ac.in}
\affiliation{Department of Physics, School of Advanced Sciences, VIT Chennai, India}

\author{Kalyani Desikan}
\email{kalyani.desikan@vit.ac.in} 
\affiliation{Department of Mathematics, School of Advanced Sciences, VIT Chennai, India
 }

\date{\today}

\begin{abstract} 
Event simulation for electron neutrino interactions plays a foundational role in precision measurements in particle physics experiments, yet the computational demand of traditional Monte Carlo methods remains a significant challenge, especially for complete, high-dimensional event reconstruction. In this study, we present a generative model based on the Conditional Wasserstein Generative Adversarial Network (CW-GAN) framework. This architecture is conditioned on the input neutrino energy. It utilizes a Wasserstein loss function, stabilized by a gradient penalty, to learn the complex mapping from a latent space to structured kinematic data. Our model is tailored to replicate the full multidimensional kinematics of electron neutrino interactions as described by the GENIE event generator. Our focus is specifically on the Inverse Beta Decay (IBD-CC), Neutral Current (NC), and $\nu_e - e^-$ elastic scattering processes (NuEElastic), spanning an energy window of $10$--$31$~MeV. Our approach abandons variable reduction schemes and instead generates the entire summary ntuple, enabling holistic event-by-event modeling. Training is performed separately for each of the three interaction types, with rigorous convergence monitoring over $100$--$300$ epochs per channel. We perform a rigorous quantitative validation against held-out GENIE test datasets. The generated samples demonstrate fidelity, reproducing the 1D marginal distributions for all kinematic variables with  statistical compatibility, and successfully capturing the complex non-linear correlations between them. This work offers a scalable and efficient alternative to traditional MC event generation, providing full-spectrum kinematic simulation for key electron neutrino interaction channels while drastically reducing computational overhead.
\end{abstract}

\keywords{Neutrino Event Simulation, GENIE, Monte Carlo Generators, General Adversarial Networks}

\maketitle

\section{\label{sec:intro}Introduction}

Monte Carlo (MC) event generators are essential computational tools that simulate the fundamentally probabilistic nature of particle physics interactions, which are governed by quantum mechanics. Instead of calculating a single, deterministic outcome, these generators act as sophisticated ``dice-rolling'' engines, using random numbers to determine the likelihood of an interaction, the specific physics channel that occurs (e.g., elastic scattering vs. charged-current), and the resulting kinematics of all final-state particles. For neutrino-nucleus scattering, this simulation becomes even more complex as it must also model Final State Interactions (FSI), a random cascade where particles created in the initial interaction re-interact with other nucleons before exiting the nucleus, resulting in a complex, multi-particle final state. 
\\
\newline
MC simulations are indispensable in neutrino physics primarily because the incident neutrino itself is unobservable; experimentalists can only measure the kinematic effect of an interaction (the final-state particles) and not the cause, such as the original neutrino's energy. Simulations bridge this gap by serving as a high-fidelity ``answer key,'' providing the essential link for event reconstruction, which connects the visible detector signals back to the unobservable incident neutrino properties. Furthermore, these simulations are the only reliable method for quantifying and correcting for critical experimental limitations, such as detector acceptance and efficiency (i.e., ``blind spots''), and for accurately modeling and subtracting competing signals from non-neutrino background sources like cosmic rays, thereby isolating the true neutrino signal.

The landscape of neutrino MC generators is dominated by several key frameworks, each tailored  for specific energy ranges and experimental needs. The most prominent include GENIE \cite{Andreopoulos2009GENIE}, a comprehensive framework used by DUNE; NEUT\cite{Hayato:2021clg}, the primary generator for the Japanese T2K and Hyper-Kamiokande experiments; and NuWro \cite{GOLAN2012499}, often used by theorists to test specific nuclear models. Other important tools include GiBUU \cite{lalakulich2011neutrinonucleusreactionsgibuu}, a specialized transport model providing high-fidelity simulation of nuclear FSI, and MARLEY \cite{GARDINER2021108123}, which focuses specifically on low-energy supernova neutrino interactions in liquid argon. This study focuses on GENIE because it provides a comprehensive final summary ntuple for each interaction, detailing all final-state particles and kinematic variables. Furthermore, it is a versatile framework capable of generating events for the diverse array of nuclear interaction channels (like Elastic Scattering (NuEElastic), Charged Current (IBD-CC), and Neutral Current (NC)) relevant to our study.

It is important to note that Monte Carlo generators like GENIE are computationally intensive due to the complexity of simulating a multitude of particle interactions and the stochastic nature of these processes. Each event simulation requires random sampling of numerous physical variables and detailed tracking of particle propagation through nuclear matter, which demands significant processing power and time \cite{PhysRevLett.120.042003}. Consequently, generating statistically meaningful predictions often involves running millions of such simulations, making the computational load substantial and necessitating optimization techniques and high-performance computing resources to achieve timely results \cite{Andreopoulos2009GENIE}.
\\
\newline
To address the significant computational bottleneck of traditional Monte Carlo methods, the field has increasingly turned to deep generative models as a means of creating computationally-efficient ``surrogate models'' \cite{PhysRevD.106.096020}. Among these, Generative Adversarial Networks (GANs), first introduced by Goodfellow et al. \cite{10.5555/2969033.2969125,goodfellow2017nips2016tutorialgenerative}, are particularly well-suited for this task.

Generative Adversarial Networks (GANs) represent a powerful class of unsupervised machine learning models designed to learn and replicate complex data distributions. The framework often operates as a ``minimax" game between two competing neural networks: a Generator ($G$), which acts as a ``forger'' learning to map a random latent noise vector $z$ to synthetic data samples $(\tilde{x})$, and a Discriminator ($D$), which acts as a ``detective'' trained to distinguish these synthetic samples from real-world training data $(x)$. 

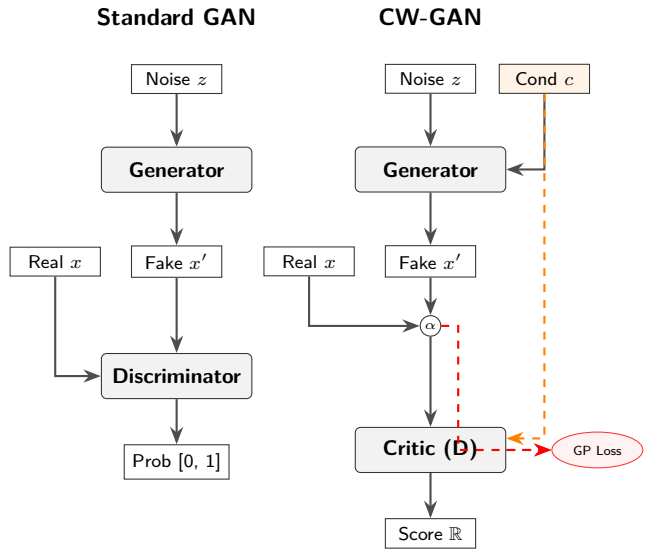
\begin{figure}[h]
\begin{tikzpicture}[
    scale=0.75,
    node distance=0.7cm,
    var/.style={rectangle, draw=black!80, fill=white, minimum width=1.2cm, font=\scriptsize\sffamily},
    model/.style={rectangle, rounded corners=2pt, draw=black!80, fill=gray!10, minimum width=2cm, minimum height=0.6cm, font=\footnotesize\bfseries\sffamily},
    loss/.style={ellipse, draw=red!80, fill=red!5, font=\tiny\sffamily},
    op/.style={circle, draw=black!80, inner sep=1pt, font=\tiny},
    arrow/.style={-Stealth, thick, draw=black!70},
    lbl/.style={font=\small\bfseries\sffamily}
]

\node[lbl] (v_title) {Standard GAN};

\node (v_z) [var, below=0.4cm of v_title] {Noise $z$};
\node (v_g) [model, below=of v_z] {Generator};
\node (v_x) [var, below=of v_g] {Fake $x'$};

\node (v_real) [var, left=0.4cm of v_x] {Real $x$};
\node (v_d) [model, below=1cm of v_x] {Discriminator};
\node (v_out) [var, below=0.6cm of v_d] {Prob [0, 1]};

\draw [arrow] (v_z) -- (v_g);
\draw [arrow] (v_g) -- (v_x);
\draw [arrow] (v_x) -- (v_d);
\draw [arrow] (v_real) |- (v_d);
\draw [arrow] (v_d) -- (v_out);

\begin{scope}[xshift=4.5cm]
    \node[lbl] (c_title) {CW-GAN};

    \node (c_z) [var, below=0.4cm of c_title] {Noise $z$};
    \node (c_c) [var, fill=orange!10, right=0.3cm of c_z] {Cond $c$};
    
    \node (c_g) [model, below=of c_z] {Generator};
    \node (c_x) [var, below=of c_g] {Fake $x'$};
    
    \node (c_interp) [op, below=0.5cm of c_x] {$\alpha$};
    \node (c_real) [var, left=0.4cm of c_x] {Real $x$};
    
    \node (c_crit) [model, below=1.2cm of c_interp] {Critic (D)};
    \node (c_out) [var, below=0.6cm of c_crit] {Score $\mathbb{R}$};
    
    \node (c_gp) [loss, right=0.6cm of c_crit] {GP Loss};

    \draw [arrow] (c_z) -- (c_g);
    \draw [arrow] (c_c) |- (c_g.east);
    \draw [arrow] (c_g) -- (c_x);
    
    \draw [arrow] (c_x) -- (c_interp);
    \draw [arrow] (c_real) |- (c_interp);
    \draw [arrow] (c_interp) -- (c_crit);
    
    \draw [arrow, dashed, orange] (c_c) |- ($(c_crit.north east)!0.5!(c_crit.east)$);
    
    \draw [arrow] (c_crit) -- (c_out);
    \draw [arrow, red, dashed] (c_crit) -- (c_gp);
    \draw [arrow, red, dashed] (c_interp.east) -- ++(0.3,0) |- (c_gp.west);
\end{scope}

\end{tikzpicture}

\caption{Schematic representation of the GAN architectures.}
\end{figure}

The standard GAN objective function is defined as \cite{10.5555/2969033.2969125}:
\begin{equation}
\begin{split}
    \min_G \max_D V(D, G) &= \mathbb{E}_{x \sim p_{\text{data}}(x)}[\log D(x)] \\
    &+ \mathbb{E}_{z \sim p_{z}(z)}[\log(1 - D(G(z)))]
\end{split}
\end{equation}

These networks are trained in opposition, where the Generator is optimized to produce samples that ``fool'' the Discriminator, and the Discriminator is optimized to improve its detection accuracy. In a successful training equilibrium, the Generator becomes so proficient that its output distribution $P_{\text{gen}}$ is indistinguishable from the real data distribution $P_{\text{real}}$.

Despite their proven success in computer vision, the application of standard GANs to High Energy Physics (HEP) \cite{10.21468/SciPostPhys.14.4.079}, particularly neutrino event generation, has been fraught with challenges. Early applications highlighted the challenge of reproducing physically viable results, primarily due to the distinct complexity of particle physics data \cite{Otten2021,fu2023generativemodelssimulationkamlandzen}. Unlike pixel-based images, tabular physics data is characterized by high-dimensionality, sparse data regions, and complex, non-Gaussian, multi-modal distributions (such as the multi-peaked $Q^2$ variable we identified). Standard GAN architectures are prone to training instability, often suffering from ``mode collapse''- where the generator only learns to reproduce a few common event types - or complete training divergence (\textit{NaN or Not a Number} losses occur when the loss function produces an undefined floating-point value due to numerical instability, such as exploding gradients, $\log(0)$, or $0/0$ operations. This causes weight updates to propagate NaN values throughout the network, resulting in complete and irreversible training collapse.) failing to capture the strict physical correlations essential for a valid simulation.\cite{ashrapov2020tabulargansunevendistribution}

While recent work by \citet{bonilla2025generativeadversarialneuralnetworks} has demonstrated the promise of standard Conditional GANs (CGANs) for reproducing final lepton (muon) kinematics from NuWro simulations, our research pursues a distinct and more comprehensive framework. We have developed a single, robust\textit{ Conditional Wasserstein GAN with Gradient Penalty (CW-GAN)} architecture, chosen specifically for its enhanced training stability, which we demonstrate is capable of modeling complete, event-level summaries rather than only final lepton kinematics. We validate this architecture's robustness by training it on three separate, datasets generated using GENIE: 

a) Neutrino-Electron Elastic Scattering (NuEElastic)

b) Neutral Current (NC) 

c) Charged Current - Inverse Beta Decay (IBD-CC), 

where a) and b) were generated using normalized solar neutrino flux and c) was generated using a DSNB flux \cite{PhysRevD.103.043003,santos2025diffusesupernovaneutrinobackground}.

The first, electron neutrino-electron elastic scattering (NuEElastic):
\begin{equation}
    \nu_e + e^- \to \nu_e + e^-
\end{equation}
represents one of the most precisely understood processes in neutrino physics. Its purely leptonic nature, with theoretical cross-sections that have no hadronic or nuclear uncertainties at leading order, makes it an ideal benchmark for validating generative approaches \cite{PhysRevD.101.033006}.

We then extend this validation to the semi-leptonic Charged Current  Inverse Beta Decay (IBD-CC) process:
\begin{equation}
    \bar{\nu}_e + p \to e^+ + n
\end{equation}
This interaction, involving an antineutrino interacting with a proton target, serves as the primary detection channel for reactor antineutrinos and the Diffuse Supernova Neutrino Background (DSNB), introducing the complexities of hadronic physics \cite{PhysRevD.103.043003}.
\footnotetext{NaN (Not a Number) losses indicate numerical instability, such as exploding gradients or division by zero, leading to an undefined loss value and training failure.}

Finally, we consider a Neutral Current (NC) interaction:
\begin{equation}
    \nu_x + N \to \nu_x + N
\end{equation}
where the neutrino scatters off a target nucleon without changing flavor. This channel is crucial for measuring total neutrino flux. This multi-channel approach allows us to test the CW-GAN's robustness across purely leptonic, semi-leptonic CC, and NC processes, each presenting distinct kinematic signatures and physical complexities \cite{10.1093/acprof:oso/9780198508717.001.0001}.
\\
\newline
The transition from conventional GANs to Conditional Wasserstein GANs represents a significant methodological advancement motivated by both theoretical and practical considerations. Traditional GANs, while successful, often suffer from training instabilities, mode collapse, and convergence difficulties due to the Jensen-Shannon divergence used in their loss functions \cite{arjovsky2017wassersteingan}. Wasserstein GANs address these fundamental limitations by replacing the Jensen-Shannon divergence with the Wasserstein distance (Earth Mover's distance), which provides a more meaningful and stable measure of distributional differences. This theoretical improvement translates to smoother gradients, more stable training dynamics, and better convergence properties. The gradient penalty mechanism, which distinguishes Wasserstein GANs with Gradient Penalty (WGAN) from the original WGAN implementation, further enhances training stability. Rather than relying on weight clipping, which can lead to vanishing gradients or prolonged training times, the gradient penalty approach enforces the crucial Lipschitz constraint \cite{zhou2019lipschitzgenerativeadversarialnets} by penalizing gradients with large norm values.
\\
\newline
The WGAN objective function is given by \cite{HEJAZI20233681}, where $L$ is the total WGAN loss used to train the critic (discriminator). 
$\mathbb{P}_r$ denotes the distribution of real training samples, 
$\mathbb{P}_g$ denotes the distribution of generated (fake) samples produced by the generator, 
and $\hat{x}$ represents interpolated samples drawn uniformly along straight lines between 
real and generated data points, with $\mathbb{P}_{\hat{x}}$ being their distribution. 
The scalar $\lambda$ is the gradient penalty coefficient controlling the strength of the 
regularization term.

\begin{equation}
\begin{split}
    L &= \underbrace{\mathbb{E}_{\tilde{x} \sim \mathbb{P}_g} [D(\tilde{x})] - \mathbb{E}_{x \sim \mathbb{P}_r} [D(x)]}_{\text{Critic Loss}} \\
      &+ \underbrace{\lambda \mathbb{E}_{\hat{x} \sim \mathbb{P}_{\hat{x}}} [(\|\nabla_{\hat{x}} D(\hat{x})\|_2 - 1)^2]}_{\text{Gradient Penalty}}
\end{split}
\end{equation}

This regularization technique ensures that the discriminator maintains the 
1-Lipschitz property required for optimal transport theory - a mathematical 
framework that measures the distance between probability distributions - while 
avoiding the pathological behaviors associated with weight clipping, a naive 
stabilization approach that hard-limits network weights to a fixed range but 
leads to vanishing gradients and reduced model capacity. \cite{gulrajani2017improvedtrainingwassersteingans}.

The conditional aspect of our approach enables physics-informed generation by incorporating explicit dependencies on experimental parameters, specifically the incident neutrino energy ($E_\nu$) as the conditioning vector. Unlike unconditional GANs that generate random samples from the learned distribution, Conditional GANs allow for controlled generation based on specified input conditions. 

In the context of neutrino physics, this conditioning mechanism is essential. It enables the network to learn the distinct, energy-dependent correlations that govern the kinematics for each interaction channel - from the recoil electron spectrum in elastic scattering to the positron and neutron kinematics in IBD-CC. This ensures that the generated events are consistent with the underlying physics across the relevant energy range.
\\
\newline
Section II introduces the CW-GAN methodology and describes the technical details of our approach, including the conditioning framework, network architectures, and training procedures. Section III presents comprehensive validation and testing, including 1D histograms, joint correlations, and quantitative metrics of the CW-GAN against GENIE. Section IV details results for all three interaction channels under flux conditions relevant to each process (solar $\nu_e$ for NuEElastic/NC, DSNB for IBD-CC). Section V summarizes the findings and discusses the implications for future ML-based neutrino event generation.

\section{METHODOLOGY}

Our goal is to train a Conditional Wasserstein GAN with gradient penalty (CW-GAN) that can accurately learn the complex, high-dimensional joint probability distribution $P(\mathbf{x} | E_\nu)$ of our training datasets, where $\mathbf{x}$ is the complete set of final-state kinematic variables and $E_\nu$ is the incident neutrino energy. The model must not only reproduce the 1D distributions of individual variables but also capture their intricate, physically-driven correlations. The ultimate aim is to produce a stable generator that can, upon request, generate vast new event summaries that are statistically indistinguishable from the original GENIE data, achieving this orders of magnitude faster than traditional MC simulation. We implement our (CW-GAN) model and perform numerical analysis using PyTorch.
\\
\newline
To accomplish this, our methodology is structured into the following five key stages:

\begin{enumerate}[label=\Alph*.]
    \item Dataset and Variables
    \item Data Preprocessing
    \item Model Architecture Definition: Conditional Wasserstein GAN with Gradient Penalty (CW-GAN).
    \item Loss and Training 
    \item Generation of Data
\end{enumerate}

\subsection{Dataset and Variables}
\begin{figure}[htbp]
    \centering
    \includegraphics[width=0.9\linewidth]{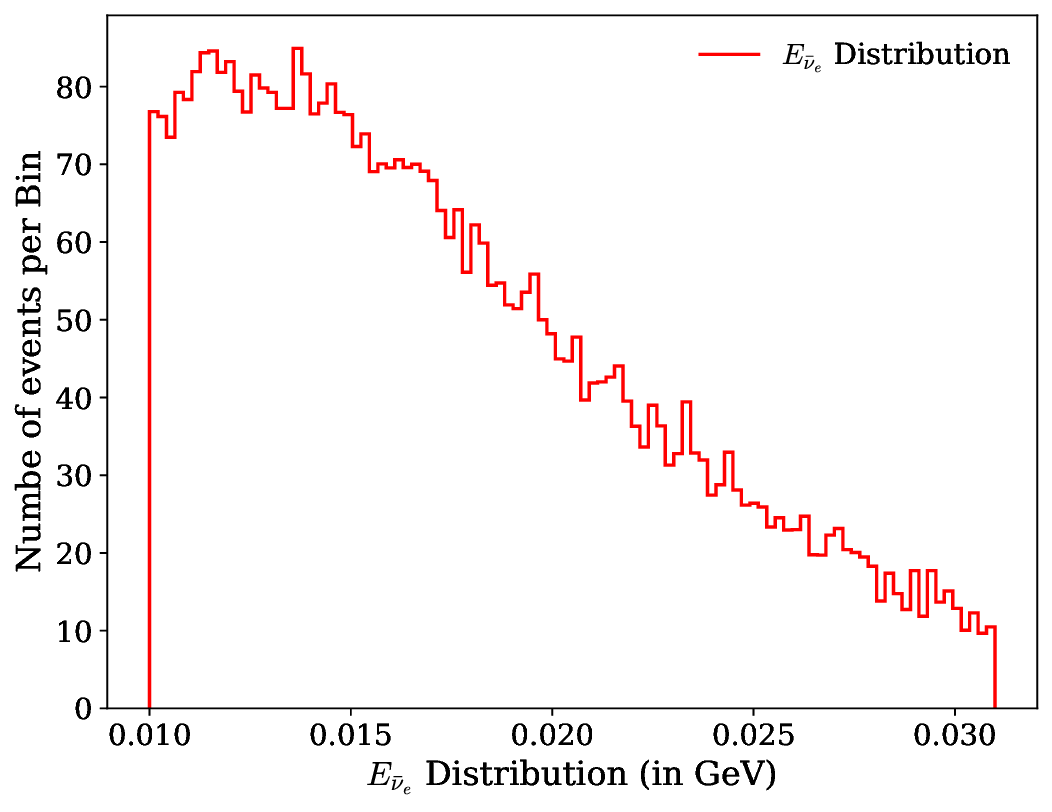}
    
    \vspace{0.2cm} 
    
    \includegraphics[width=0.9\linewidth]{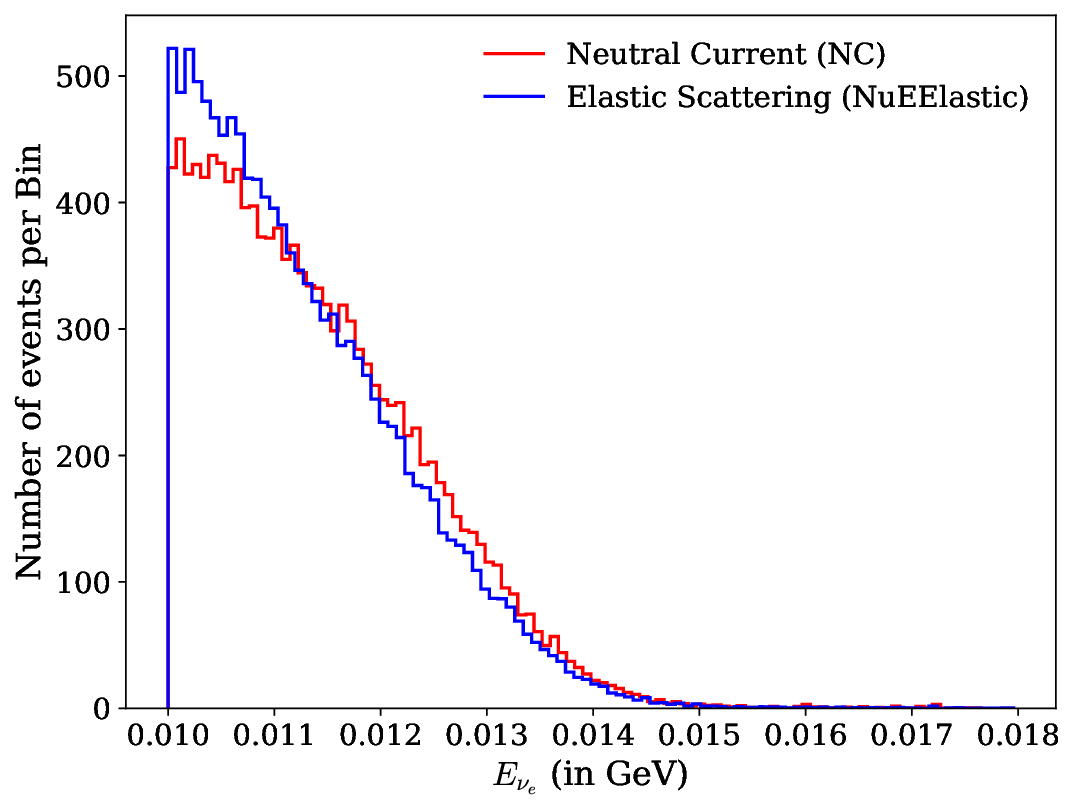}
    
    \caption{a) Energy distribution for IBD-CC. b) Energy distribution for NC and NuEElastic scattering.}
    \label{fig:energy_distributions}
\end{figure}

The training, testing, and validation datasets for all three interaction channels were generated using GENIE v3.04 Monte Carlo framework \cite{Andreopoulos2009GENIE}. The energy distributions of the events used to train the model are shown in Fig \ref{fig:energy_distributions}. GENIE provides a comprehensive simulation of neutrino-nucleus interactions, and its output is stored in a summary ntuple (.gst) format. This ntuple provides a complete ``event record,'' containing a total of 81 distinct attributes that describe every aspect of the simulated interaction. 

A complete definition of all 81 attributes can be found in the appendix and is also available in the official GENIE manual \cite{Andreopoulos2009GENIE}. For our generative model, we isolate the final-state kinematic variables and high-level observables as the target for the GAN, while using $E_\nu$ as the conditioning input.

To prepare this comprehensive ntuple data for our machine learning framework, we followed a two-step data conversion pipeline. The primary GENIE event records, initially in .root format, were first processed using the \texttt{gntpc} utility to generate the flat summary .gst.root ntuples. These were subsequently converted into a whitespace-delimited plaintext (.dat) format, which serves as the direct input for our Python-based preprocessing and training scripts. This entire procedure was performed for all three distinct physics datasets under investigation: Inverse Beta Decay (IBD-CC) with a DSNB flux, Neutral Current (NC) scattering with a solar flux, and Neutrino-Electron Elastic Scattering (NuEElastic) with a solar flux. 
\\
\newline
To ensure robust and unbiased validation of our model's generalization capabilities, we generated a dataset of 100,000 datapoints per channel (300,000 total), randomly split 70:30 into 70,000 training events and 30,000 test events. Training datasets were used exclusively for fitting transformers and optimizing the GAN, while test datasets were held out for all final validation and analysis, including histogram comparisons and quantitative metrics.

\subsection{Data Preprocessing}
To prepare the data for the generative model, we first performed a ``column-wise'' separation of the full 81-variable ntuple into three distinct categories. First, we identified all Constant columns (e.g., $Z, A, \text{tgt}$), which contain no variance across the dataset and were therefore excluded from the training process. Second, we isolated the single Conditioning variable ($E_\nu$), which was removed from the list of variables to be generated and instead provided as a conditional input to both the Generator and Discriminator. Finally, the remaining set of Target attributes comprising all high-level observables (like $Q^2, W, y_s$) and final-state kinematic 4-momenta (like $E_l, px_l, px_f$, etc.) were isolated to serve as the high-dimensional output distribution that the GAN must learn to generate. During the final post-processing stage, these excluded \textit{constant columns} are re-inserted into the file, along with the conditioning $E_\nu$ values and the CW-GAN generated Target variables, to reconstruct a complete event record that fully matches the original GENIE format.
\\
\newline
The raw GENIE data is not suitable for direct input into a GAN. The kinematic variables have highly non-Gaussian, heavily skewed distributions (some multi-modal) and exist on vastly different numerical scales (e.g., $E_\nu$ in GeV vs. $Q^2$ in $\text{GeV}^2$). Standard normalization techniques like min-max scaling are insufficient, as they only rescale the range but fail to address the underlying skew, leading to unstable GAN training. To mitigate this, we employed the PowerTransformer technique from scikit-learn \cite{pedregosa2018scikitlearnmachinelearningpython}, which applies a non-linear transformation to make each variable’s distribution more Gaussian-like. Specifically, we utilize the Yeo-Johnson transformation \cite{10.1093/biomet/87.4.954}, which is selected over the Box-Cox transformation \cite{9174711} for its ability to handle non-positive data (e.g., momenta). The Yeo-Johnson transformation for a variable $x_i$ is defined as:

\begin{equation}
\psi(\lambda_j, x_{i,j}) = 
\begin{cases} 
\frac{(x_{i,j}+1)^\lambda_j - 1}{\lambda_j} & \text{if } \lambda \neq 0, x_i \geq 0 \\
\ln(x_{i,j}+1) & \text{if } \lambda = 0, x_i \geq 0 \\
-\frac{(-x_{i,j}+1)^{2-\lambda_j} - 1}{2-\lambda_j} & \text{if } \lambda_j \neq 2, x_i < 0 \\
-\ln(-x_{i,j}+1) & \text{if } \lambda_j = 2, x_i < 0 
\end{cases}
\end{equation}

During the fit process, the transformer finds the optimal hyperparameter $\lambda_j$ for each variable by maximizing the log-likelihood of the transformed data under a Gaussian distribution. The resulting data $x_{i,j}^{(\lambda_j)}$ is then standardized to have zero mean and unit variance, providing a stable, well-behaved target for the GAN.

The incident neutrino energy, $E_\nu$, was treated differently from all other variables as it serves as the basis for our conditional model. It was not included in the PowerTransformer group; instead, it was isolated to be the single conditioning variable for the CW-GAN. To make it an ideal input for the neural networks, $E_\nu$ was normalized 
to the range  $[-1, 1]$. This transformation preserves the shape of the original energy distribution while ensuring all input conditional values are on a consistent scale. The exact transformation applied to $E_\nu$ is given by:

\begin{equation}
    E'_\nu = 2 \cdot \frac{E_\nu - E_{\min}}{E_{\max} - E_{\min}} - 1
\end{equation}

where $E_{\min}$ and $E_{\max}$ are the minimum and maximum neutrino energies in the training sample respectively, and $E'_\nu$ is the resulting scaled value provided as input to both the Generator and the Critic (Discriminator).

\subsubsection{Special Handling of Variables due to Numerical Instability}
A critical finding during our experiments was that the PowerTransformer (Yeo-Johnson) transformation, while effective for most attributes, failed when applied to attributes with specific numerical properties, leading to unstable training or physically inaccurate results. We identified one such case.
For the \textit{NuEElastic dataset}, a significant numerical instability was found in the $Q^2$ variable, whose values were of an extremely small magnitude ($\sim 10^{-5}$). This resulted in the GAN learning the overall pattern (shape) of the distribution but failing to converge to the correct values. We corrected this with a more aggressive pre-scaling by a factor of $10^5$, moving the data into a stable numerical range that the PowerTransformer could model effectively.

This pre-scaling was applied before the PowerTransformer during preprocessing. This step was then deterministically reversed during post-processing, dividing the generated $Q^2$ by $10^5$, to return the final generated values to their correct physical scale.

\subsection{Model Architecture - CW-GAN}
Our generative model is a Conditional Wasserstein GAN with Gradient Penalty (CW-GAN), an architecture chosen for its demonstrated superior training stability over standard GAN frameworks when dealing with complex distributions. The model consists of two main neural networks, a Generator and a Discriminator (Critic), both constructed as Multi-Layer Perceptrons (MLPs) incorporating residual connections and LayerNorm for stable gradient flow.
\\ \\
\textbf{Generator (G):} The generator is purpose-built to map a 32-dimensional latent variable $z$ and a one-dimensional conditioning variable $c$ (such as an energy parameter) into a 
full-sized vector representing the target scientific observables. The latent vector 
$z$ is sampled from a standard normal distribution $z \sim \mathcal{N}(0,1)$, 
capturing random variation, while the conditioning variable $c$ infuses the desired 
scientific or contextual information. Both inputs are independently embedded into 
128-dimensional feature space via a \textit{Latent FC} layer (for $z$) and a 
\textit{Cond FC} layer (for $c$), concatenated ($\textcircled{\scriptsize c}$), and passed through a series of 3 residual generative blocks (\textit{GenBlock~1} $\rightarrow$ 
\textit{GenBlock~2} $\rightarrow$ \textit{GenBlock~3}) that refine and integrate 
the latent and conditional information at each stage. 
\\
\newline
Each block receives the conditioning signal $c$ as a secondary input (dashed arrows) and applies a dense layer with bias, layer normalization, dropout, leaky rectified non-linear activation function, and skip connections, supporting robust learning and stable optimization 
through efficient gradient propagation~\cite{li2018visualizinglosslandscapeneural}. Importantly, the 
generator's \textit{Final Linear} layer is a purely linear transformation, with no 
\texttt{tanh} or other bounded activation functions, producing the 
\textit{Generated Output}. This design choice is critical because the 
\texttt{PowerTransformer} preprocessing step produces unbounded, gaussian-like 
target distributions, and a bounded activation function such as \texttt{tanh} would 
artificially constrain the generator's outputs, preventing accurate modeling of 
distribution tails and limiting physical realism.
\\
\begin{figure}
\centering
\begin{tikzpicture}[
    node distance=0.9cm,
    data/.style={rectangle, rounded corners=2pt, draw=black!80, fill=green!10, minimum width=2.5cm, minimum height=0.6cm, font=\small\sffamily, align=center},
    cond/.style={rectangle, rounded corners=2pt, draw=black!80, fill=orange!10, minimum width=1.5cm, minimum height=0.6cm, font=\small\sffamily},
    block/.style={rectangle, draw=black!80, fill=gray!5, thick, minimum width=3.2cm, minimum height=0.8cm, font=\bfseries\small\sffamily},
    op/.style={circle, draw=black!80, inner sep=0pt, minimum size=5mm, font=\small},
    tag/.style={font=\scriptsize\sffamily\itshape, text=black!60},
    arrow/.style={-Stealth, thick, draw=black!70}
]

\node (in_z) [data] at (-1.4,0) {Latent $z$};
\node (in_c) [cond] at (1.4,0) {Cond $c$};

\node (zf) [data, below=0.8cm of in_z] {Latent FC};
\node (cf) [cond, below=0.8cm of in_c] {Cond FC};

\node (cat0) [op] at (0,-2.5) {$\copyright$};

\node (b1) [block, below=0.8cm of cat0] {GenBlock 1};
\node (b2) [block, below=0.8cm of b1] {GenBlock 2};
\node (b3) [block, below=0.8cm of b2] {GenBlock 3};

\node (final) [data, below=0.8cm of b3, minimum width=3.2cm] {Final Linear};
\node (out) [circle, draw, fill=blue!5, below=0.5cm of final] {};
\node at (out.south) [below, tag, yshift=-2pt] {Generated Output};

\draw [arrow] (in_z) -- (zf);
\draw [arrow] (in_c) -- (cf);

\draw [arrow] (zf.south) -- ++(0,-0.3) -| (cat0.north);
\draw [arrow] (cf.south) -- ++(0,-0.3) -| (cat0.north);

\draw [arrow] (cat0) -- (b1);
\draw [arrow] (b1) -- (b2);
\draw [arrow] (b2) -- (b3);
\draw [arrow] (b3) -- (final);
\draw [arrow] (final) -- (out);

\draw [arrow, dashed, black!40] (in_c.east) -- ++(0.8,0) |- (b1.east);
\draw [arrow, dashed, black!40] (in_c.east) -- ++(0.8,0) |- (b2.east);
\draw [arrow, dashed, black!40] (in_c.east) -- ++(0.8,0) |- (b3.east) node[pos=0.5, right, tag] {};

\end{tikzpicture}
\caption{Generator Architecture}
\end{figure}
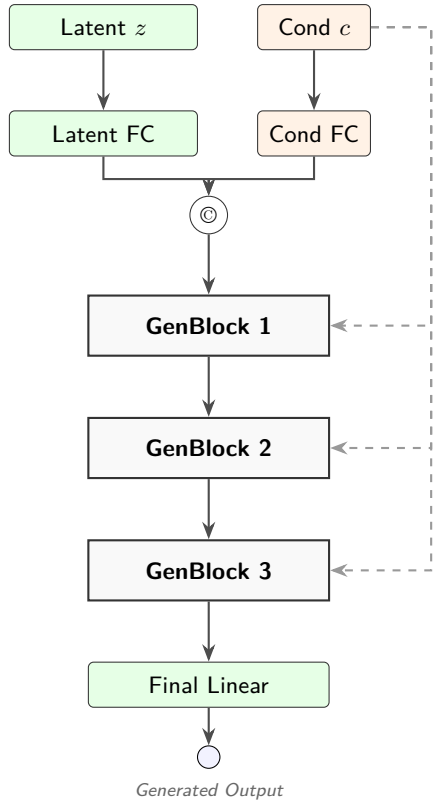

\textbf{Critic (Discriminator (D)):} The critic (discriminator) in CW-GAN terminology adopts a similar multi-layer 
perceptron stack based on discriminative residual blocks. It receives both the data 
sample $x$ (real or generated) and its conditioning variable $c$, processes them 
independently via two parallel embedding branches, and concatenates 
($\textcircled{\scriptsize c}$) their embeddings. The \textit{Data Branch} processes 
$x$ through two successive fully connected layers (each projecting to 128 dimensions) 
with leaky rectified non-linear activations and dropout in between, producing a 
128-dimensional data embedding. The \textit{Cond Branch} independently embeds $c$ 
through a single fully connected layer with leaky rectified non-linear activation, 
also projecting to a 128-dimensional conditioning embedding. The resulting 
256-dimensional concatenated vector is then passed through a sequence of two 
discriminative residual blocks (\textit{DiscBlock~1} $\rightarrow$ 
\textit{DiscBlock~2}), each of which re-injects the raw conditioning signal $c$ 
(dashed arrows) by concatenating ($\textcircled{\scriptsize c}$) it with the block 
input, and then processes the result through a \textit{Main Path} (dense layer with 
bias, layer normalization, dropout, and leaky rectified non-linear activation) and a 
parallel \textit{Skip Path} (a direct linear projection), whose outputs are summed 
($+$) to form the block output, revealing complex relationships between data and 
condition. Unlike a standard GAN discriminator that produces a probability, the critic 
outputs a single, unbounded scalar \textit{Score} via a \textit{Final Linear} layer 
with no activation function. This score is used to compute the Wasserstein distance 
(refer Eq.~\ref{eq:wassertain distance}) between the empirical and model 
distributions, guiding both generator and critic optimization and supporting the 
CW-GAN's core objective.
\begin{figure}
\begin{tikzpicture}[
    node distance=0.8cm,
    data/.style={rectangle, rounded corners=2pt, draw=black!80, fill=blue!10, minimum width=2.5cm, minimum height=0.6cm, font=\small\sffamily, align=center},
    cond/.style={rectangle, rounded corners=2pt, draw=black!80, fill=orange!10, minimum width=1.5cm, minimum height=0.6cm, font=\small\sffamily},
    block/.style={rectangle, draw=black!80, fill=gray!5, thick, minimum width=3cm, minimum height=0.8cm, font=\bfseries\small\sffamily},
    op/.style={circle, draw=black!80, inner sep=0pt, minimum size=5mm, font=\small},
    tag/.style={font=\scriptsize\sffamily\itshape, text=black!60},
    arrow/.style={-Stealth, thick, draw=black!70}
]

\node (in_x) [data] {Input $x$};
\node (in_c) [cond, right=0.8cm of in_x] {Cond $c$};

\node (db) [data, below=0.8cm of in_x] {Data Branch};
\node (cb) [cond, below=0.8cm of in_c] {Cond Branch};

\node (cat0) [op, below=1.2cm of $(db.south)!0.5!(cb.south)$] {$\copyright$};

\node (b1) [block, below=0.8cm of cat0] {DiscBlock 1};
\node (b2) [block, below=1cm of b1] {DiscBlock 2};

\node (final) [data, below=0.8cm of b2, minimum width=3cm] {Final Linear};
\node (out) [circle, draw, fill=red!5, below=0.5cm of final] {};
\node at (out.south) [below, tag] {Score};

\draw [arrow] (in_x) -- (db);
\draw [arrow] (in_c) -- (cb);
\draw [arrow] (db.south) -- ++(0,-0.2) -| (cat0.north);
\draw [arrow] (cb.south) -- ++(0,-0.2) -| (cat0.north);
\draw [arrow] (cat0) -- (b1);
\draw [arrow] (b1) -- (b2);
\draw [arrow] (b2) -- (final);
\draw [arrow] (final) -- (out);

\draw [arrow, dashed, black!40] (in_c.east) -- ++(0.6,0) |- (b1.east);
\draw [arrow, dashed, black!40] (in_c.east) -- ++(0.6,0) |- (b2.east) node[pos=0.5, right, tag]{};

\end{tikzpicture}

\begin{tikzpicture}[
    node distance=1cm,
    data/.style={rectangle, rounded corners=2pt, draw=black!80, fill=blue!10, minimum width=2.2cm, minimum height=0.6cm, font=\small\sffamily, align=center},
    cond/.style={rectangle, rounded corners=2pt, draw=black!80, fill=orange!10, minimum width=1.5cm, minimum height=0.6cm, font=\small\sffamily},
    op/.style={circle, draw=black!80, inner sep=0pt, minimum size=5mm, font=\small},
    tag/.style={font=\scriptsize\sffamily\itshape, text=black!60},
    arrow/.style={-Stealth, thick, draw=black!70}
]

\node (detail_in) [circle, draw, inner sep=1pt] {};
\node at (detail_in.north) [above, tag] {Block Input};

\node (join) [op, below=0.7cm of detail_in] {$\copyright$};

\node (c_in) [cond, right=0.8cm of join] {$c$};

\node (main) [data, below=1cm of join] {Main Path};
\node (skip) [data, right=0.8cm of main, fill=gray!10] {Skip Path};
\node (add) [op, below=1.2cm of main] {$+$};

\draw [arrow] (detail_in) -- (join);
\draw [arrow] (c_in) -- (join);
\draw [arrow] (join) -- (main);
\draw [arrow] (join.south) -- ++(0,-0.4) -| (skip.north);
\draw [arrow] (main) -- (add);
\draw [arrow] (skip.south) -- ++(0,-0.65) -| (add.east);
\draw [arrow] (add) -- ++(0,-0.7) node[below, tag] {Block Output};

\draw [dashed, black!20] ($(main.west |- detail_in.north)+(-0.5,0.4)$) rectangle ($(skip.east |- add.south)+(0.5,-0.4)$);

\end{tikzpicture}

\caption{Discriminator Architecture}
\end{figure}
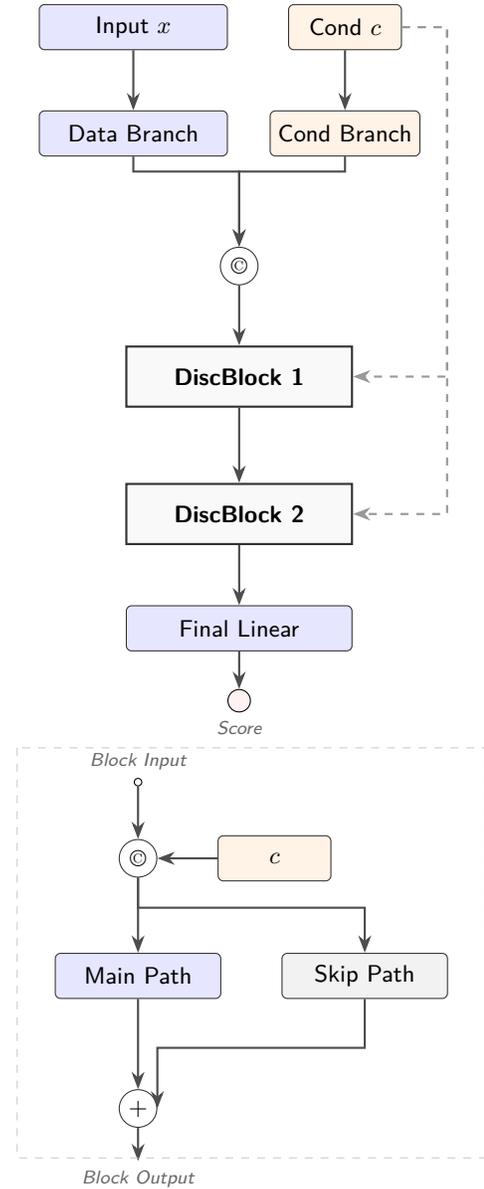

\subsection{Loss and Training}
The network leverages the Wasserstein distance as its adversarial objective, providing smoother gradients and addressing mode collapse typically seen in legacy GAN models. The gradient penalty term ensures the critic adheres to the Lipschitz constraint, fostering stable convergence. Adaptive optimizers and robust weight initialization further reinforce reliable and reproducible adversarial learning.
\\ \\
This model uses the Wasserstein loss to train the generator and critic, measuring the distance between real and generated data distributions via the Earth Mover's (Wasserstein-1) distance. The Wasserstein loss provides smoother gradients and more stable training compared to conventional GANs that use the Jensen–Shannon divergence. In this approach, the critic assigns a real-valued score to each sample instead of a probability, reflecting how closely the generated data matches the real data.
\\ \\
Mathematically, the loss employs statistical expectation, denoted as $\mathbb{E}$. Specifically, $\mathbb{E}_{x \sim \mathbb{P}_r}[D(x)]$ is the expected (mean) value of the critic's output over real samples $x$ drawn from the real data distribution $\mathbb{P}_r$, and $\mathbb{E}_{\tilde{x} \sim \mathbb{P}_g}[D(\tilde{x})]$ is the expected value over generated samples $\tilde{x}$ from the generator's distribution $\mathbb{P}_g$ and $\mathbb{E}_{\hat{x} \sim \mathbb{P}_{\hat{x}}}$ is the expected value over interpolated samples $\hat{x}$. During training, these expectations are estimated as averages over each mini-batch of data. The Wasserstein loss functions are defined as:
\\
\textbf{Critic (Discriminator) Loss:}
\begin{equation}
\begin{split}
    L_D &= \underbrace{\mathbb{E}_{\tilde{x} \sim \mathbb{P}_g} [D(\tilde{x})] - \mathbb{E}_{x \sim \mathbb{P}_r} [D(x)]}_{\text{Wasserstein Estimate}} \\
    &\quad + \underbrace{\lambda \mathbb{E}_{\hat{x} \sim \mathbb{P}_{\hat{x}}} [(\|\nabla_{\hat{x}} D(\hat{x})\|_2 - 1)^2]}_{\text{Gradient Penalty}}
\end{split}
\end{equation}

\textbf{Generator Loss:}
\begin{equation}
    L_G = -\mathbb{E}_{\tilde{x} \sim \mathbb{P}_g} [D(\tilde{x})]
\end{equation}

Here, $D$ is the critic network, which aims to maximize the score gap between real and generated samples, guiding the generator to produce data that minimizes this gap. The use of expectation ensures the loss reflects the average behavior of the model over sampled data, making it both practical for mini-batch optimization and theoretically meaningful for distributional comparison.
\\ \\
For both the generator and the critic, the learning rate is taken as $10^{-4}$ using the \textit{AdamW optimizer}. We set the exponential decay rates for the gradient moving averages at $\beta_1=0.5$ for the first moment and $\beta_2=0.9$ for the second moment to ensure stable convergence during the adversarial training. Dropout rate in all layers is 0.2 to prevent overfitting. For each generator update, the critic is updated five times to better approximate the Wasserstein distance. Gradient penalty regularizes the critic to satisfy the 1-Lipschitz constraint, which stabilizes WGAN training and ensures meaningful Wasserstein distance computation. We use the same neural network architecture uniformly for training across all three datasets; only the preprocessing and postprocessing procedures differ between datasets, while the network structure and parameters remain unchanged.
\subsection{Generation of Data}
After training, synthetic data generation is performed by evaluating the generator in inference (Evaluation State) mode. To ensure direct comparability with the original data, the same conditioning variable values (such as $E_\nu$) from the original dataset are provided to the generator. For each sample, a random latent vector is drawn from a standard normal distribution, and the normalized conditioning values are computed according to the same scaling used during training. These vectors are passed into the generator, yielding output in the normalized feature space.
\\ \\
Since the generator's outputs are still in the normalized domain (due to preprocessing applied during training), each feature's values are denormalized using the corresponding fitted PowerTransformer model. This ensures the synthetic data have the same distributional properties as the original (unscaled) dataset.
The result is written to disk as a formatted data file, preserving both column headers and table structure. This workflow guarantees that the synthetic dataset mirrors the statistical and structural features of the original data while introducing controlled generative variability via the latent space. All postprocessing steps are carefully aligned with those performed during preprocessing, ensuring the validity and consistency of the generated outputs.

\section{Validation and Testing Framework}
\label{sec:III}

The validation of a generative model intended for High-Energy Physics (HEP) simulation presents a unique challenge compared to standard computer vision tasks. Unlike image synthesis, where perceptual quality can be subjective, physics simulations require quantitative, reproducible evaluation grounded in the underlying laws governing particle interactions. Our primary objective is to demonstrate that the Generative Adversarial Network (GAN) functions as a high-fidelity surrogate for the GENIE Monte Carlo (MC) simulator . This necessitates a validation framework that transcends simple statistical comparison. While standard metrics verify that the GAN mimics the \textit{appearance} of the data (marginal distributions), a robust physics-based validation must confirm that the model mimics the \textit{logic} of the data, preserving correlations and adhering to conservation laws.

We employ a hierarchical validation strategy designed to assess the model at four increasing levels of complexity. First, we ensure training stability by monitoring Wasserstein gradients. Second, we evaluate the statistical convergence of the marginal distributions using 1D histograms, Pull distributions, and Earth Mover's Distance (EMD). Third, we test the internal physical consistency of the generated events through interaction-specific kinematic reconstruction tests. Finally, we confirm that the generator maintains high-dimensional consistency by evaluating pairwise variable relationships. All evaluations are performed on a statistically independent, ``held-out'' test sample of GENIE data ($N_{\text{test}}$) strictly excluded from the training process.

\subsection{Training Stability and Wasserstein Loss}
The foundational requirement for a robust generator is stable training dynamics. To mitigate mode collapse, we employ the \textit{Wasserstein distance (Earth Mover's Distance or EMD)} as the objective function. The Wasserstein distance $W(\mathbb{P}_r, \mathbb{P}_g)$ quantifies the minimum ``work'' required to transform the real distribution $\mathbb{P}_r$ into the generated 
distribution $\mathbb{P}_g$ \cite{Kantorovich2006}. Here, $\Pi(\mathbb{P}_r, 
\mathbb{P}_g)$ denotes the set of all joint distributions $\gamma(x, y)$ whose 
marginals are $\mathbb{P}_r$ and $\mathbb{P}_g$ respectively, with each $\gamma$ 
representing a valid transport plan that pairs samples $x \sim \mathbb{P}_r$ with 
samples $y \sim \mathbb{P}_g$. The infimum $\inf_{\gamma \in \Pi(\mathbb{P}_r, 
\mathbb{P}_g)}$ selects the transport plan of minimum cost across all such pairings, 
$\mathbb{E}_{(x,y) \sim \gamma}[\|x - y\|]$ is the expected Euclidean distance 
between paired samples $(x, y)$ under a given transport plan $\gamma$, representing 
the average cost of moving probability mass from $x$ to $y$:
\begin{equation}
    W(\mathbb{P}_r, \mathbb{P}_g) = \inf_{\gamma \in \Pi(\mathbb{P}_r, \mathbb{P}_g)} \mathbb{E}_{(x, y) \sim \gamma} [\|x - y\|].
    \label{eq:wassertain distance}
\end{equation}
We monitor the Critic loss, which approximates $W$, alongside the Generator loss. Successful training is characterized by the stabilization of the Critic loss, indicating that the generator has reached an equilibrium with the discriminator.

\subsection{Statistical Fidelity: Pulls and EMD}
To evaluate the fidelity of the generated event kinematics, we utilize two complementary statistical metrics: \textit{the 1-Wasserstein Distance ($W_1$) and the Mean Absolute Pull (MAP)}. The $W_1$ distance, colloquially known as the Earth Mover’s Distance, quantifies the global dissimilarity between the GAN-generated and GENIE-simulated distributions by calculating the minimum cost required to transform one probability mass function into the other. Unlike traditional divergence measures, $W_1$ provides a smooth, meaningful gradient even when distributions have non-overlapping supports, making it an ideal metric for monitoring training stability and convergence in the high-dimensional phase space of neutrino interactions. 

While $W_1$ assesses global morphology, we quantify local, bin-wise agreement using the Pull statistic, defined for each bin as,

\begin{equation}
\text{Pull}_i = \frac{N_{\text{CW-GAN}}^i - N_{\text{GENIE}}^i}{\sqrt{\sigma_{\text{CW-GAN}}^{2,i} + \sigma_{\text{GENIE}}^{2,i}}}
\label{eq:Pull_definition}
\end{equation}

Where $\text{Pull}_i$ represents the statistical deviation for the $i$-th bin. The terms $N_{\text{CW-GAN}}^i$ and $N_{\text{GENIE}}^i$ denote the normalized counts (or densities) for the CW-GAN generated data and the GENIE reference data within that bin, respectively. The denominator accounts for the total statistical uncertainty, where $\sigma_{\text{CW-GAN}}^{2,i}$ and $\sigma_{\text{GENIE}}^{2,i}$ represent the variances (squared uncertainties) of the CW-GAN and GENIE distributions for the $i$-th bin.

 We aggregate these values into the Mean Absolute Pull (MAP) per epoch to provide a robust, scalar summary of model precision; for a GAN that perfectly reproduces the target distribution, the pull values follow a standard normal distribution, resulting in a theoretical MAP asymptote of $\sqrt{2/\pi} \approx 0.8$. This dual-metric approach allows us to simultaneously verify that the CW-GAN has captured the broad physical features of the interaction while maintaining statistical parity with the underlying Monte Carlo truth.

\subsection{Physics Signature Tests}
\label{sec:III.C Physics Signature Tests}

The ultimate test of a generative model in the physical sciences is its adherence to kinematic constraints. We identify a unique signature test for each interaction type. The decision to utilize different signature tests for each interaction type is rooted in the unique final-state topology of the interactions. A single "one-size-fits-all" test cannot be applied because the observable information varies fundamentally across channels. This also enables us to verify that the CW-GAN has learnt the inherent physical consistencies without them being directly fed into it \cite{de_Oliveira_2017}. 

\subsubsection{Elastic Scattering (NuEElastic): The $E\theta^2$ Limit}
\label{sec:III.C.1}

For neutrino-electron elastic scattering ($\nu + e^- \to \nu + e^-$), the standard energy reconstruction formula is mathematically unstable for validation purposes. The relationship between neutrino energy and lepton kinematics is given by:
\begin{equation}
    E_{\nu} \approx \frac{E_e}{1 - \frac{E_e}{m_e}(1 - \cos\theta)}.
\end{equation}
where $E_e$ is the measured kinetic energy of the recoiling electron, $m_e$ is the 
rest mass of the electron and $\theta$ is the angle between the recoiling electron and 
the direction of the incident neutrino beam. At first glance, one might expect a singularity as $\theta \to 0$, since elastic scattering is highly forward peaked. However, as $\theta \to 0$, $\cos\theta \to 1$, so $(1 - \cos\theta) \to 0$, and the denominator approaches unity, yielding the well behaved limit $E_{\nu} \approx E_e$. The true singularity arises at a finite, 
nonzero angle $\theta^*$, defined by the condition:
\begin{equation}
    1 - \frac{E_e}{m_e}(1 - \cos\theta^*) = 0 \implies E_e(1 - \cos\theta^*) = m_e.
\end{equation}
At this angle, the reconstructed $E_{\nu}$ diverges. This is  a genuine kinematic hard limit imposed by energy-momentum conservation. The condition $E_e(1 - \cos\theta) = m_e$ corresponds precisely to the neutrino transferring its maximum kinematically allowed momentum to the target 
electron. Beyond $\theta^*$, conservation laws are violated and no physical elastic 
scattering events can exist. The denominator vanishing is therefore the mathematical 
signature of this physical boundary.

This behavior poses a challenge for evaluating the CW-GAN. The 
generative model produces angles $\theta$ with small but nonzero deviations from 
their true values. Unlike $\theta^*$, such deviations are physically negligible. 
However, in the neighborhood of $\theta^*$, the denominator of the reconstruction 
formula is near zero, and even infinitesimally small angular errors are amplified into 
macroscopic singularities in the reconstructed $E_{\nu}$, rendering the direct 
energy based evaluation numerically unstable.

To avoid this, we adopt the kinematic variable $E_e\theta^2$, as used by the 
MINERvA and NOvA experiments for elastic scattering signal selection 
\cite{PhysRevD.100.092001}. For small angles, the approximation 
$1 - \cos\theta \approx \theta^2/2$ holds, and substituting this into the kinematic 
boundary condition gives:
\begin{equation}
    E_e \cdot \frac{\theta^2}{2} = m_e \implies E_e\theta^2 = 2m_e.
\end{equation}
This shows that the same kinematic hard limit that caused the singularity in the 
energy reconstruction formula manifests as a clean, finite upper bound in the 
$E_e\theta^2$ representation. The inelasticity parameter $y \in [0, 1]$ of the 
scattering process enforces that all physical elastic scattering events must satisfy:
\begin{equation}
    E_e \theta^2 < 2 m_e \text  {MeV},
    \label{eq:kinematic_cut}
\end{equation}
with no events permitted beyond this wall. The variable $E_e\theta^2$ thus encodes 
the identical physics as the energy reconstruction formula, but in a representation 
that is smooth, bounded, and free of singularities, making it the natural and stable 
choice for evaluating the CW-GAN's performance in the elastic scattering channel.
A valid generative model must produce a distribution that peaks sharply at zero and falls off strictly before this $2m_e$ \textit{kinematic wall}. This confirms the model has learned the precise coupling between the electron's energy and its scattering angle without suffering from the singularity issues of direct reconstruction.

\subsubsection{Inverse Beta Decay (IBD): Energy Reconstruction}
In contrast to Elastic Scattering, the Inverse Beta Decay process ($\bar{\nu}_e + p \to e^+ + n$) is the ``golden channel" for reactor antineutrino detection precisely because it allows for robust energy reconstruction \cite{Ricciardi_2022}. The relatively large mass of the target proton suppresses the forward-peaking behavior seen in electron scattering, avoiding the angular singularity issues described in Sec.~\ref{sec:III.C Physics Signature Tests}. \cite{Ge_2022}\cite{Munteanu_2020}

For IBD-CC, the primary figure of merit is the resolution with which the incident neutrino energy can be recovered from the visible positron. We therefore validate the GAN using the standard reconstruction formula \cite{PhysRevD.60.053003}:
\begin{equation}
    E_\nu^{\text{rec}} \approx E_e + \frac{m_n^2 - m_p^2 - m_e^2}{2 m_p}.
\end{equation}

where the incident antineutrino energy $E_\nu^{\text{rec}}$ is reconstructed from the measured positron energy $E_e$ alone, and $m_n$,  $m_p$, and $m_e$ are the rest masses of the neutron, proton, and positron respectively, and the second term represents a small, constant mass threshold correction arising from the mass difference between the initial and final state 
baryons.

We calculate the fractional residual $(E_\nu^{\text{rec}} - E_\nu^{\text{true}})/E_\nu^{\text{true}}$. A distribution centered at zero with a width consistent with the intrinsic physics width can confirm that the GAN has correctly learned the three-body kinematics and energy partitioning required for reactor flux measurements. The near-perfect overlap of the CW-GAN and GENIE reconstructed energy distributions with the ideal response line directly confirms that the residuals are centered at zero, without requiring explicit histogram analysis of the fractional residuals.

\subsubsection{Neutral Current (NC): $Q^2$ Consistency}
Neutral Current elastic scattering ($\nu + p \to \nu + p$) presents a unique challenge: the outgoing neutrino is experimentally invisible. This creates an ``under-constrained'' system where the incident neutrino energy $E_\nu$ cannot be reconstructed on an event-by-event basis. Consequently, the standard linearity checks used for IBD are impossible.

Instead, we validate the self consistency of the generated events using the \textit{momentum transfer Squared} ($Q^2$). In NC interactions, $Q^2$ can be derived exclusively from the kinetic energy ($T_p$) of the recoil nucleon, independent of the neutrino energy, as defined in the MiniBooNE and MINERvA NC analyses \cite{PhysRevD.82.092005,higuera2014neutral}:
\begin{equation}
    Q^2_{\text{rec}} = 2 M_p T_p.
\end{equation}
where $Q^{2}_{\text{rec}}$ is the reconstructed squared four-momentum transfer, $M_p$ is the mass of the proton, and $T_p$ is the kinetic energy of the recoiling proton. The generative model simultaneously produces the high-level kinematic observable $Q^2$ and the low-level proton momentum vectors $\vec{p}_p$ as part of the joint output manifold. By performing a posterior calculation of $Q^2_{\text{rec}}$ from the generated $\vec{p}_p$, we observe that the reconstructed values align with the directly generated $Q^2$ labels. This convergence demonstrates that the CW-GAN has successfully internalized the latent physical constraints and the underlying conservation laws governing the momentum transfer, effectively mapping the non-linear relationship between high-level descriptors and low-level particle kinematics.

\subsection{Joint Correlations}
While 1D histograms ensure that individual variables are modeled correctly, they are insufficient for verifying high-dimensional correlations. To validate that the CW-GAN captures the multidimensional structure of the phase space, we analyze specific 2D projections chosen to probe the underlying kinematics of each interaction type \cite{10.21468/SciPostPhys.7.6.075}.

\textbf{Elastic Scattering (NuEElastic) (6 Pairs):}
For NuEElastic, the correlations are strictly governed by the electron-neutrino scattering kinematics. We evaluate:
\begin{itemize}
    \item \textbf{Kinematic Boundaries:} $(\nu, E_l)$, $(E_l, Q^2)$, and $(\nu, Q^2)$. These pairs verify the energy transfer logic, ensuring the recoil electron energy $E_l$ never exceeds the incident neutrino energy $\nu$ and respects the momentum transfer ($Q^2$) limits.
    \item \textbf{Transverse Isotropy:} $(p_{x}^l, p_{y}^l)$. This confirms the rotational symmetry of the outgoing electron in the transverse plane.
    \item \textbf{Longitudinal Boost:} $(p_{z}^\nu, p_{z}^l)$. This verifies the forward-scattering nature of the interaction.
    \item \textbf{Inelasticity:} $(Q^2, y)$. This validates the relationship between momentum transfer and the fraction of energy transferred to the lepton.
\end{itemize}

\textbf{Inverse Beta Decay (7 Pairs):}
For IBD, validation requires checking the energy sharing between the positron and the neutron. We evaluate:
\begin{itemize}
    \item \textbf{Energy Conservation:} $(\nu, E_l)$ and $(E_l, E_f)$. These pairs probe how the incident energy is partitioned between the lepton ($E_l$) and the final-state nucleon ($E_f$).
    \item \textbf{Interaction Regime:} $(E_l, Q^2)$ and $(Q^2, W)$. These verify that the interaction remains within the quasi-elastic regime ($W \approx M_p$) and respects the dependence of lepton energy on momentum transfer.
    \item \textbf{Spatial Symmetry:} $(p_{x}^l, p_{y}^l)$ and $(p_{x}^f, p_{y}^f)$. We check both lepton and hadron transverse planes to ensure no artificial angular bias is introduced.
    \item \textbf{Beam Correlation:} $(p_{z}^\nu, p_{z}^l)$. Validates the longitudinal momentum transfer from the beam to the lepton.
\end{itemize}

\textbf{Neutral Current (7 Pairs):}
Since the outgoing neutrino is invisible in NC events, correlations must be anchored to the recoil nucleon. We evaluate:
\begin{itemize}
    \item \textbf{Recoil Kinematics:} $(\nu, E_f)$, $(E_f, Q^2)$, and $(\nu, Q^2)$. These are the critical checks for NC; since $Q^2$ is experimentally derived from the nucleon energy $E_f$, the $(E_f, Q^2)$ plot must show a tight, deterministic correlation.
    \item \textbf{Hadronic Physics:} $(Q^2, W)$ and $(y, Q^2)$. These constrain the invariant mass and inelasticity to physically allowed regions for elastic scattering.
    \item \textbf{Nucleon Geometry:} $(p_{x}^f, p_{y}^f)$ and $(p_{z}^\nu, p_{z}^f)$. These Validate that the recoil nucleon retains the correct angular distribution relative to the incident beam.
\end{itemize}

\section{Results}

The performance of the CW-GAN is evaluated by comparing the generated event kinematics against a hold-out test dataset derived from GENIE simulations. The evaluation focuses on three criteria: (1) stability of the adversarial training, (2) precise reproduction of 1D marginal distributions verified by statistical pulls, and (3) the preservation of complex physical correlations (2D joint distributions) inherent to neutrino interaction physics.

	\subsection{ Training Stability (Loss Metrics)}
\begin{figure}[h!]
    \centering
    \begin{subfigure}{\linewidth}
        \centering
        \includegraphics[width=\linewidth]{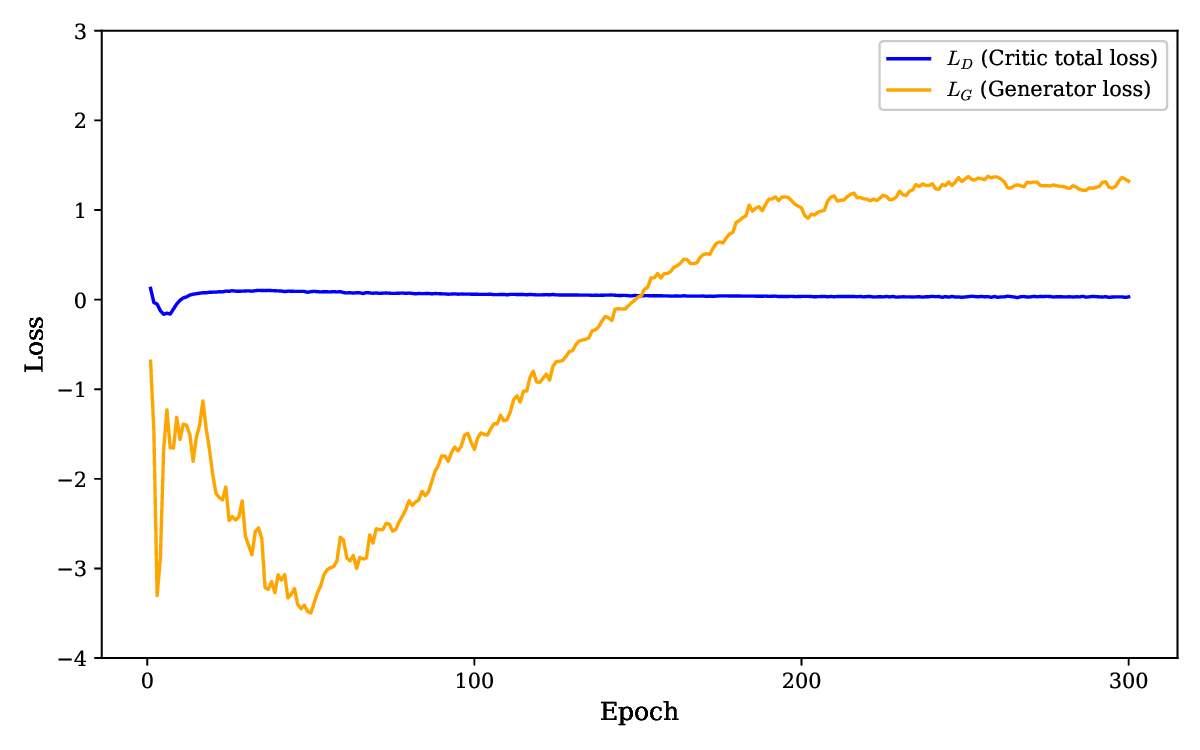}
        \caption{NuEElastic: Losses}
    \end{subfigure}
    
    \vspace{0.2cm} 
    
    \begin{subfigure}{\linewidth}
        \centering
        \includegraphics[width=\linewidth]{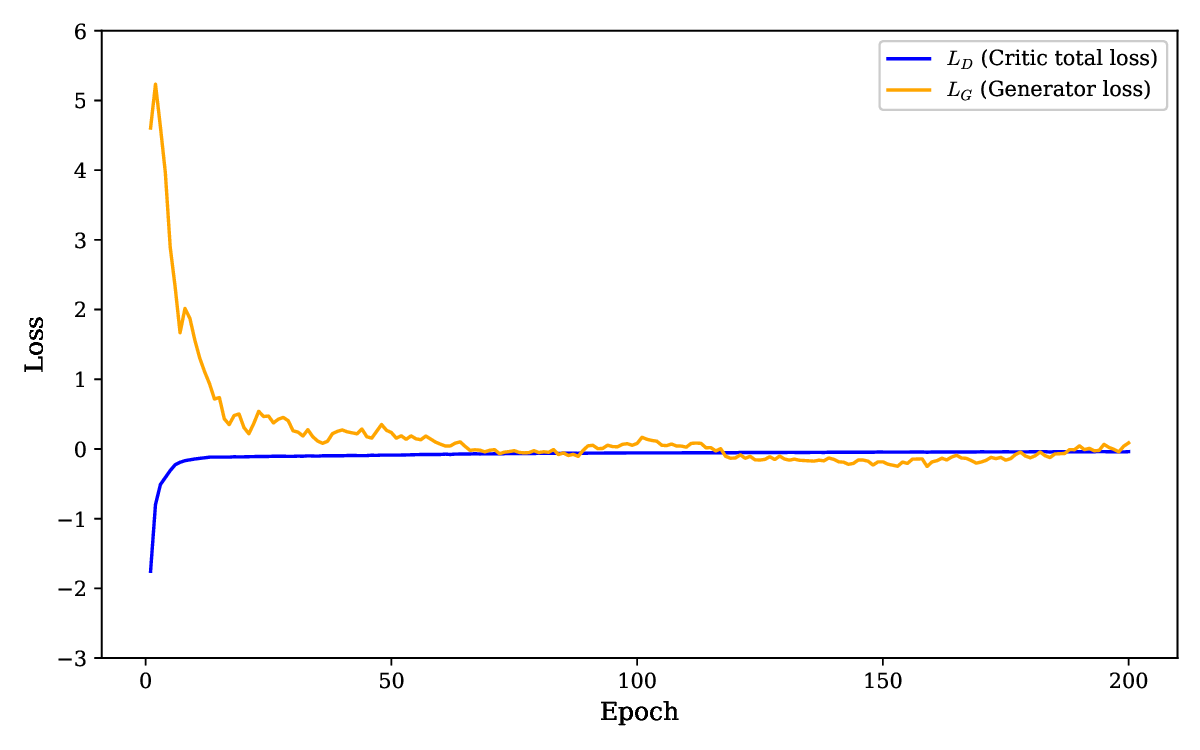}
        \caption{IBD-CC: Losses}
    \end{subfigure}
    
    \vspace{0.2cm} 
    
    \begin{subfigure}{\linewidth}
        \centering
        \includegraphics[width=\linewidth]{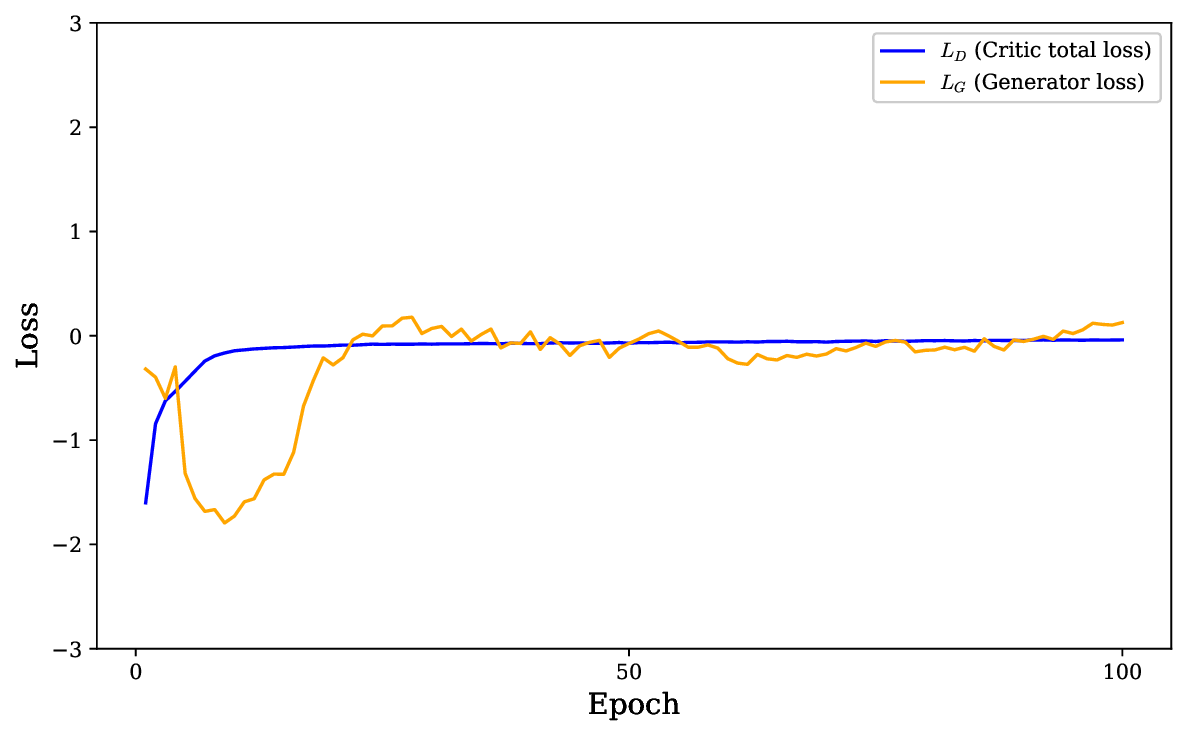}
        \caption{NC: Losses}
    \end{subfigure}
    
    \caption{Training Stability. Evolution of Critic and Generator losses over epochs for all three interaction types.}
    \label{fig:loss_metrics}
\end{figure}

The stability of the adversarial training is the primary indicator of model health in CW-GAN architectures. Figure \ref{fig:loss_metrics} displays the evolution of the Critic and Generator losses for NuEElastic, IBD-CC, and NC channels. The evolution of the adversarial game is monitored through the loss functions of the Critic ($L_D$) and the Generator ($L_G$). In the Wasserstein framework, the Critic does not output a probability of authenticity but rather a scalar value representing the ``criticism'' or distance metric of the sample. The objective is to maximize the divergence between the scores assigned to real and fake data, thereby approximating the Wasserstein distance, while the generator attempts to minimize this distance. 

In all three cases, we observe the characteristic convergence pattern of a WGAN optimized with a gradient penalty. The Critic loss (blue) drops rapidly in the initial epochs as it learns to distinguish random noise from physical events. Subsequently, the Generator loss (orange) stabilizes, indicating that the model has found an equilibrium where the generated distributions are statistically close to the truth. We observe no signs of mode collapse or diverging gradients, suggesting the hyperparameters and architecture are well-tuned for the 10-31 MeV energy window. A critical finding from the loss analysis is the absence of mode collapse. In standard GANs, mode collapse is often signaled by a sudden, permanent drop in generator loss or a cyclic behavior where the generator hops between different modes of the distribution \cite{che2017moderegularizedgenerativeadversarial}. The loss curves in Fig \ref{fig:loss_metrics} show an asymptotic approach to equilibrium, characteristic of a healthy WGAN training session. This suggests that the generator is maintaining diversity in its output., a claim that is further substantiated by the statistical coverage analysis in the subsequent sections.

\begin{figure*}[t!]
    \centering
    \begin{subfigure}{0.32\textwidth}
        \centering
        \includegraphics[width=\linewidth]{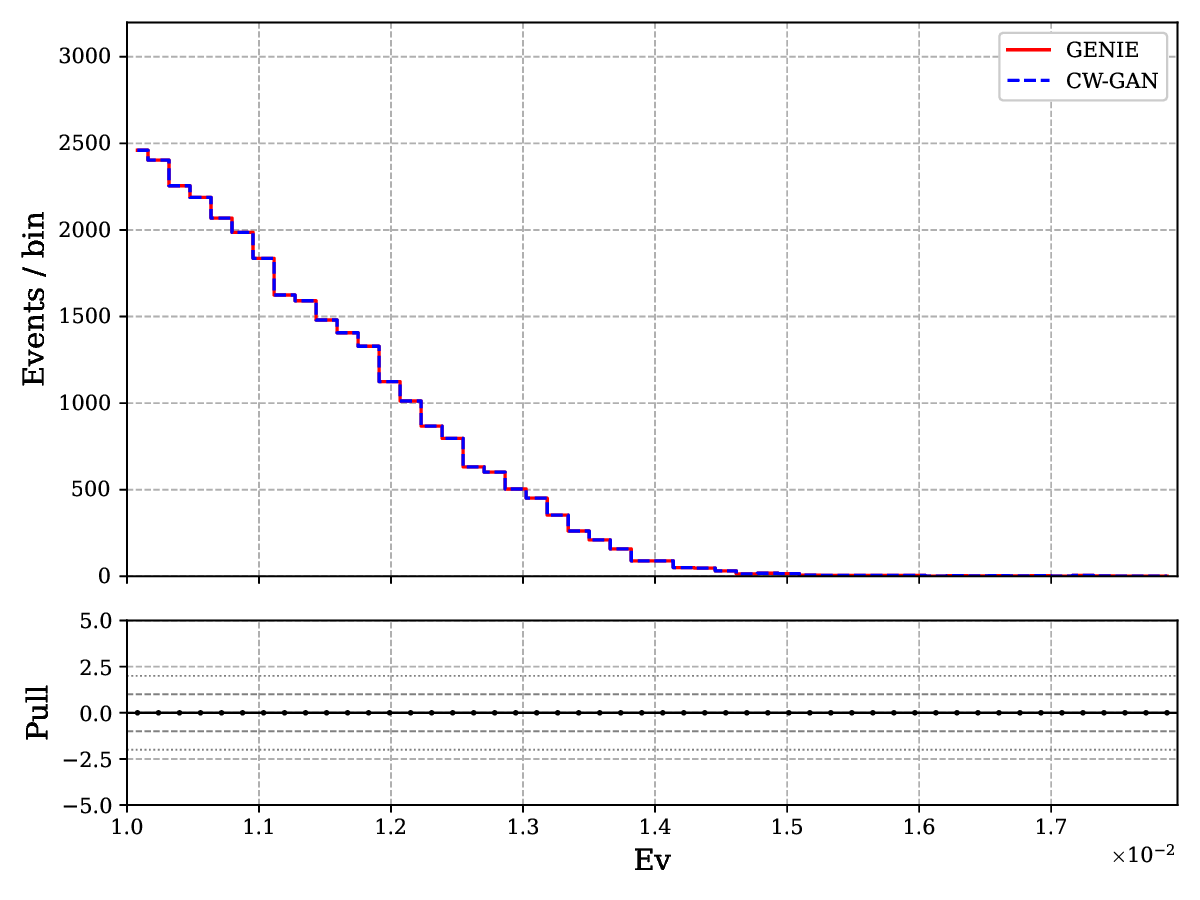}
        \caption{NuEElastic: $E_\nu$}
        \label{fig:es1d_Ev}
    \end{subfigure}
    \hfill
    \begin{subfigure}{0.32\textwidth}
        \centering
        \includegraphics[width=\linewidth]{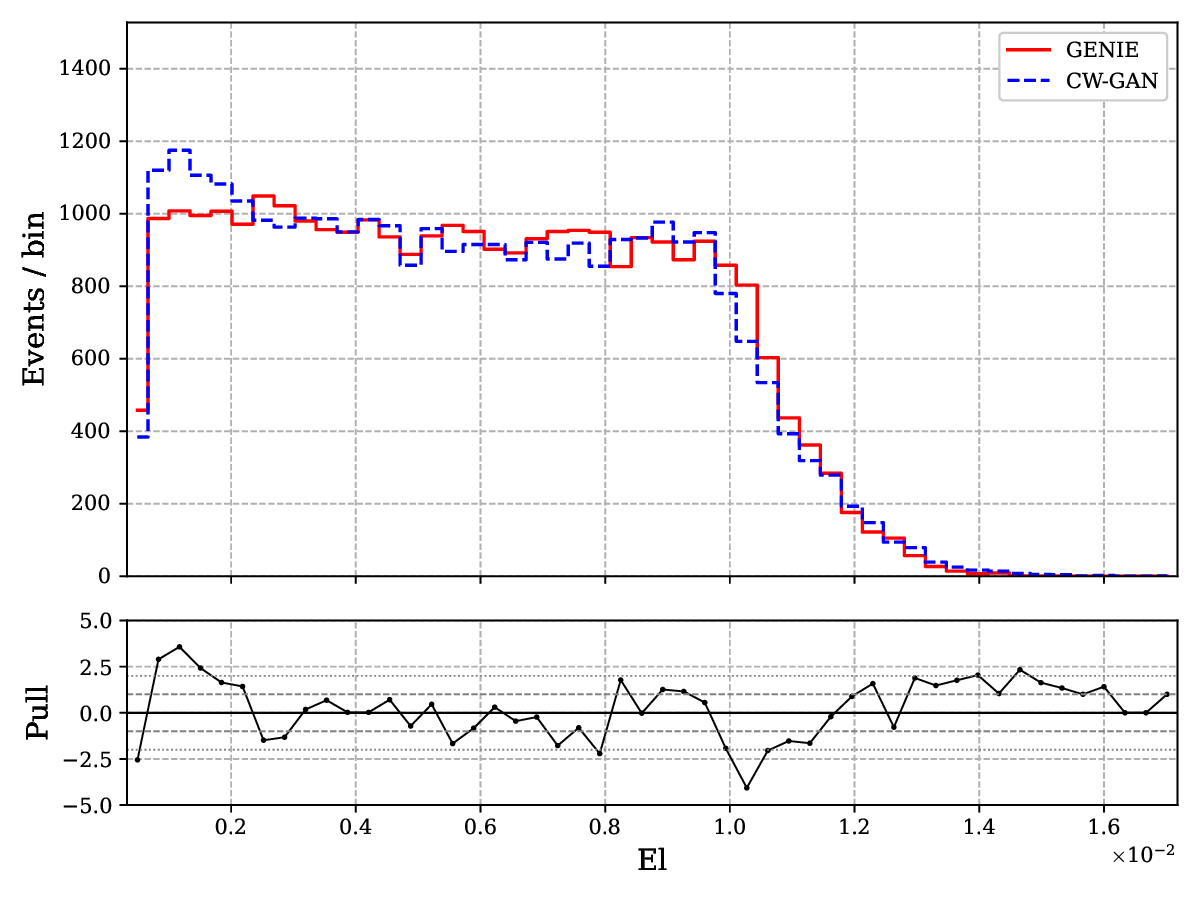}
        \caption{NuEElastic: El}
        \label{fig:es1d_El}
    \end{subfigure}
    \hfill
    \begin{subfigure}{0.32\textwidth}
        \centering
        \includegraphics[width=\linewidth]{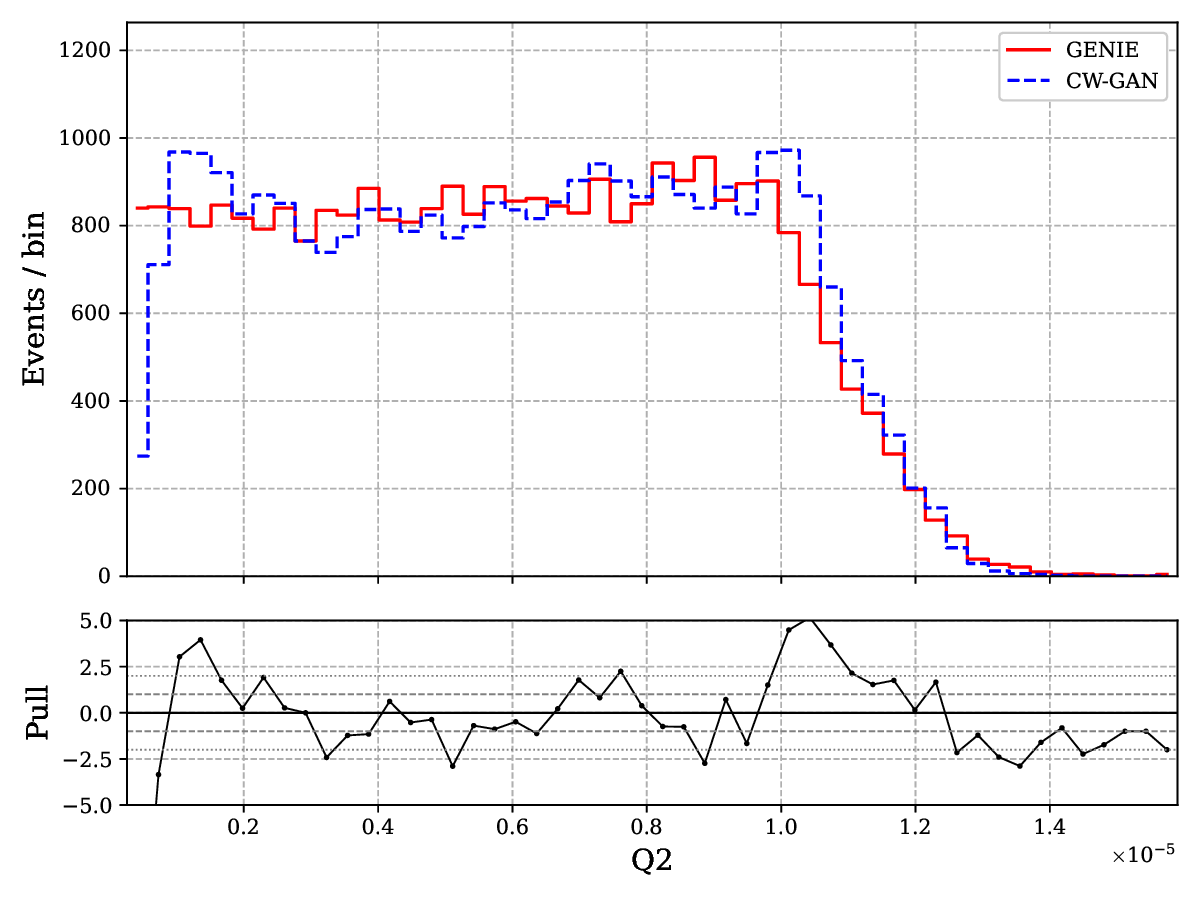}
        \caption{NuEElastic: Q2}
        \label{fig:es1d_Q2}
    \end{subfigure}
    \vspace{0.3cm}
    \begin{subfigure}{0.32\textwidth}
        \centering
        \includegraphics[width=\linewidth]{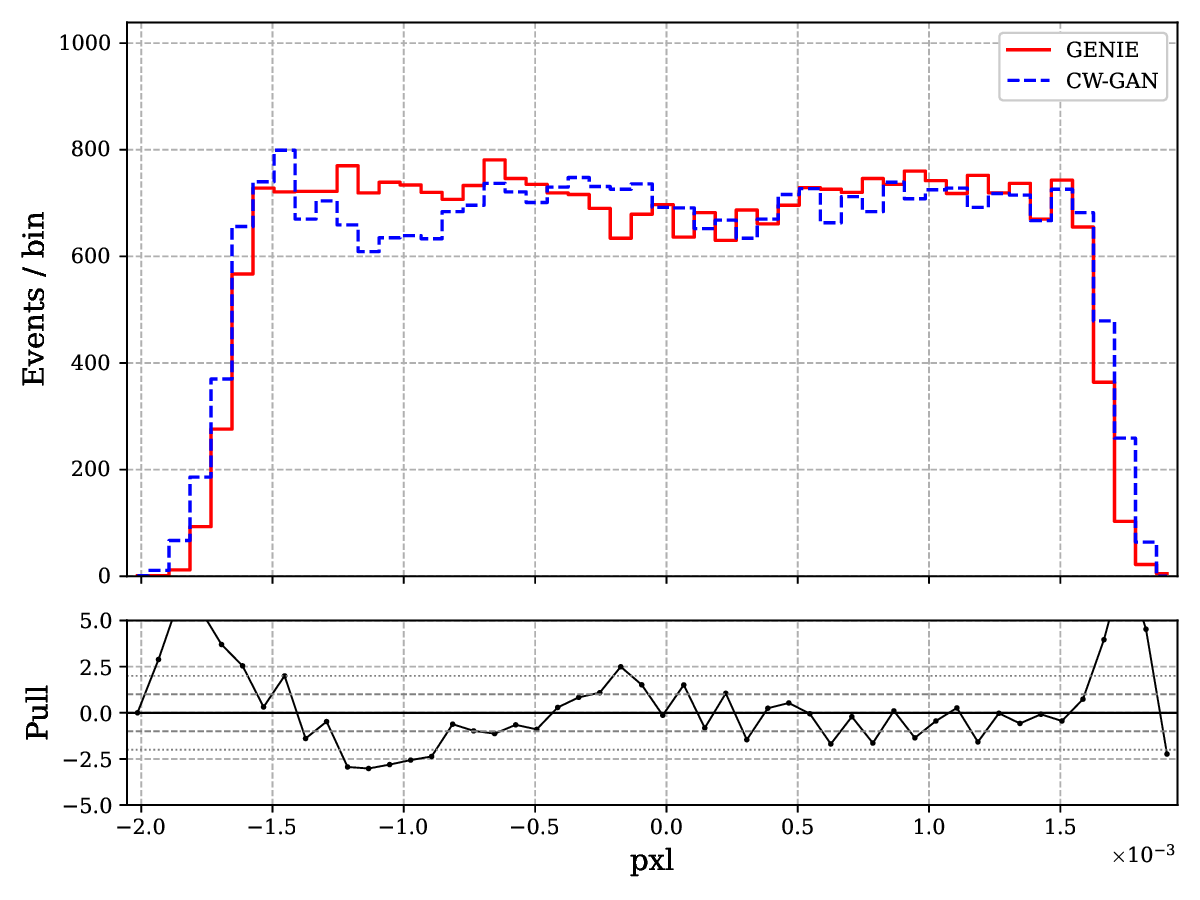}
        \caption{NuEElastic: pxl}
        \label{fig:es1d_pxl}
    \end{subfigure}
    \hfill
    \begin{subfigure}{0.32\textwidth}
        \centering
        \includegraphics[width=\linewidth]{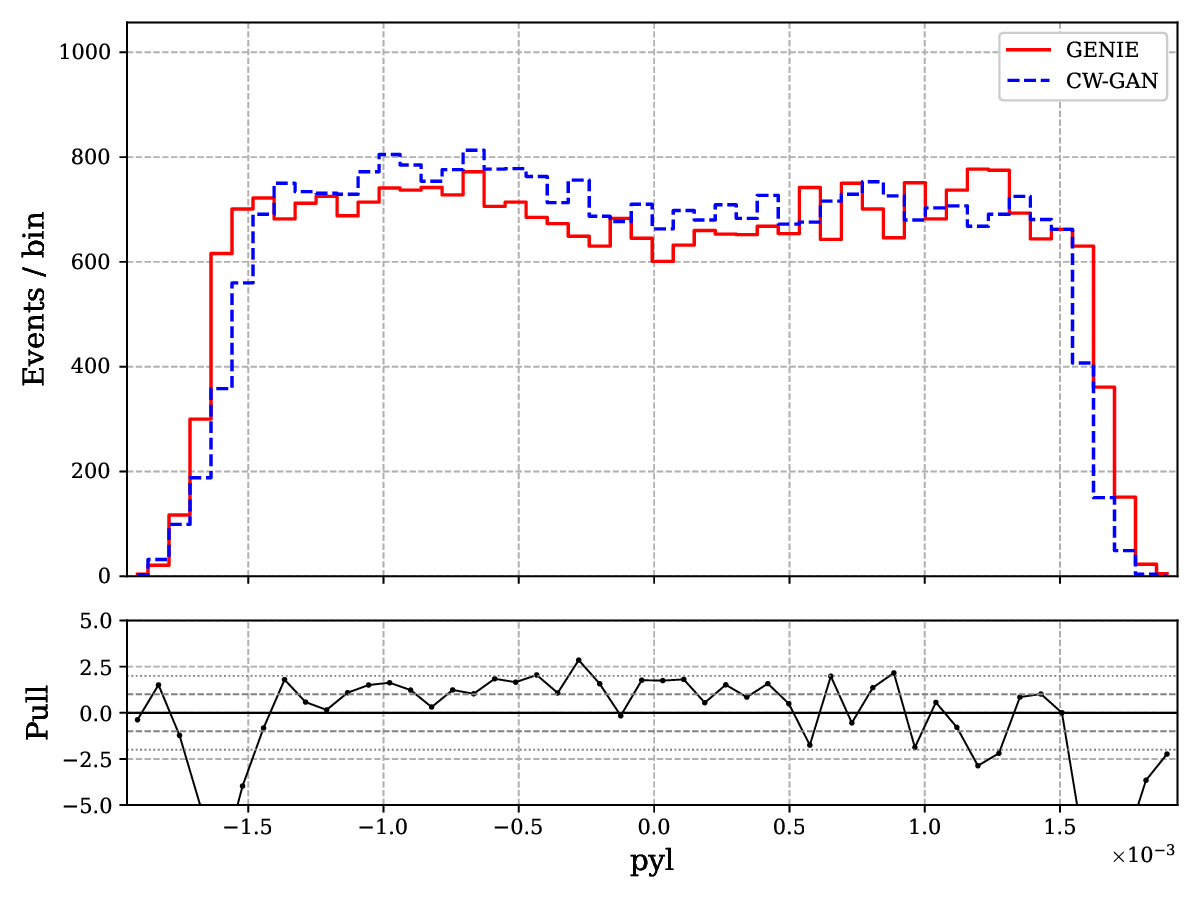}
        \caption{NuEElastic: pyl}
        \label{fig:es1d_pyl}
    \end{subfigure}
    \hfill
    \begin{subfigure}{0.32\textwidth}
        \centering
        \includegraphics[width=\linewidth]{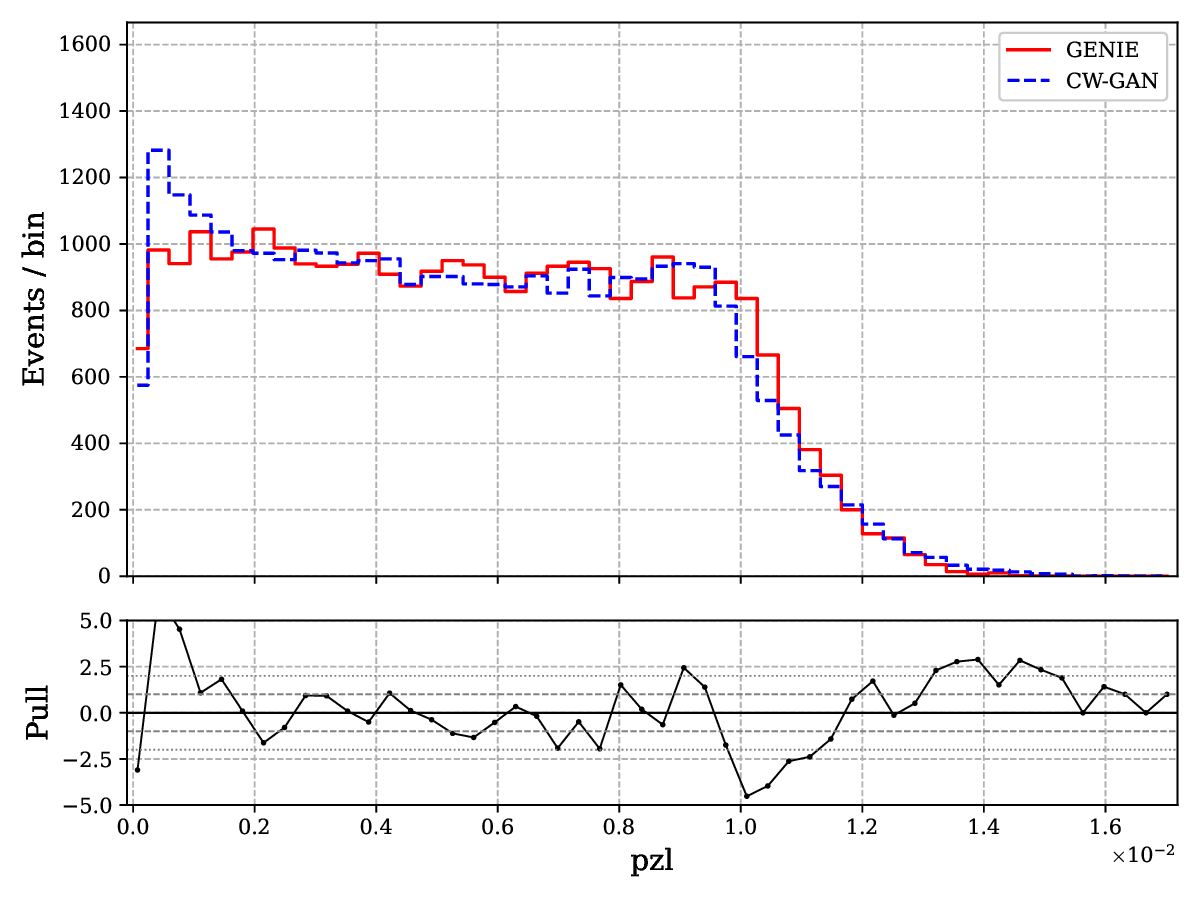}
        \caption{NuEElastic: pzl}
        \label{fig:es1d_pzl}
    \end{subfigure}
    \vspace{0.3cm}
    \begin{subfigure}{0.32\textwidth}
        \centering
        \includegraphics[width=\linewidth]{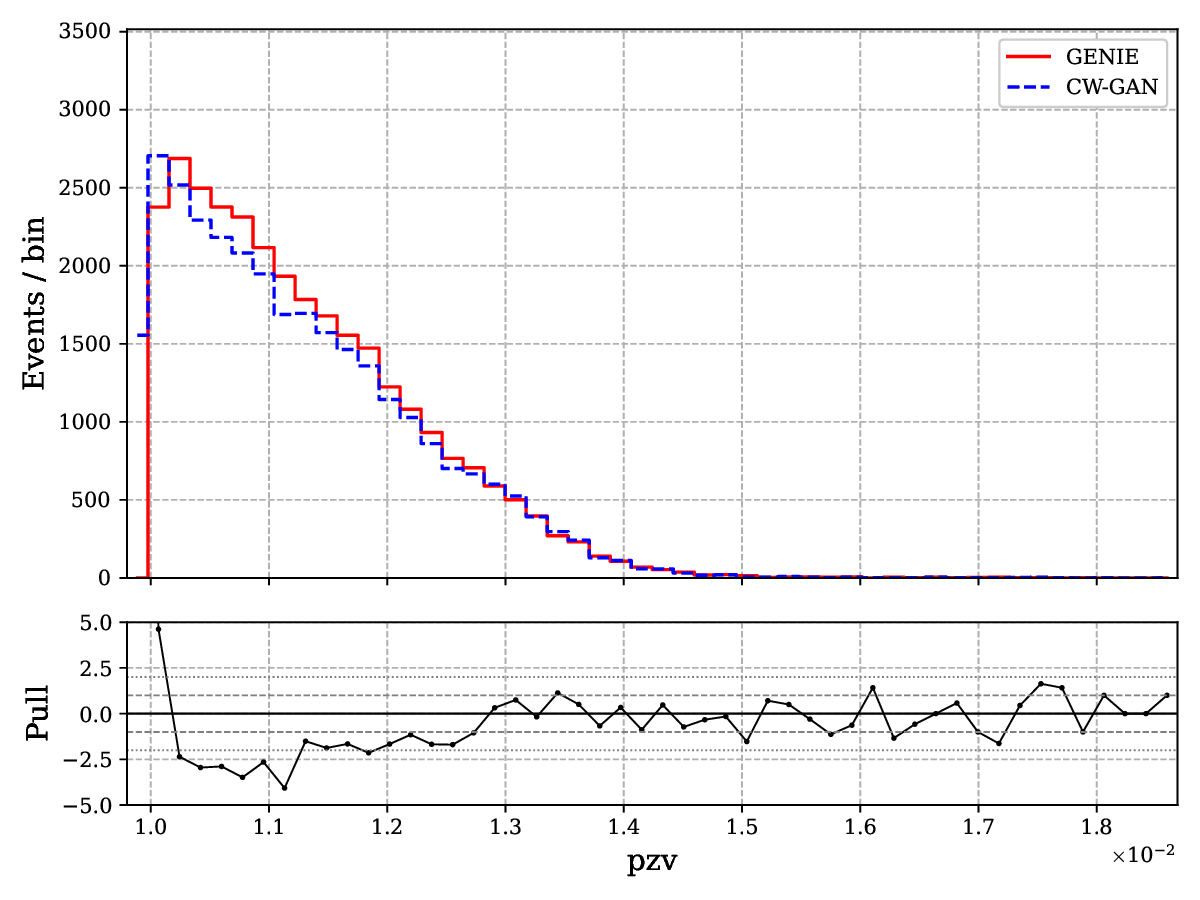}
        \caption{NuEElastic: pzv}
        \label{fig:es1d_pzv}
    \end{subfigure}
    \hspace{0.5cm}
    \begin{subfigure}{0.32\textwidth}
        \centering
        \includegraphics[width=\linewidth]{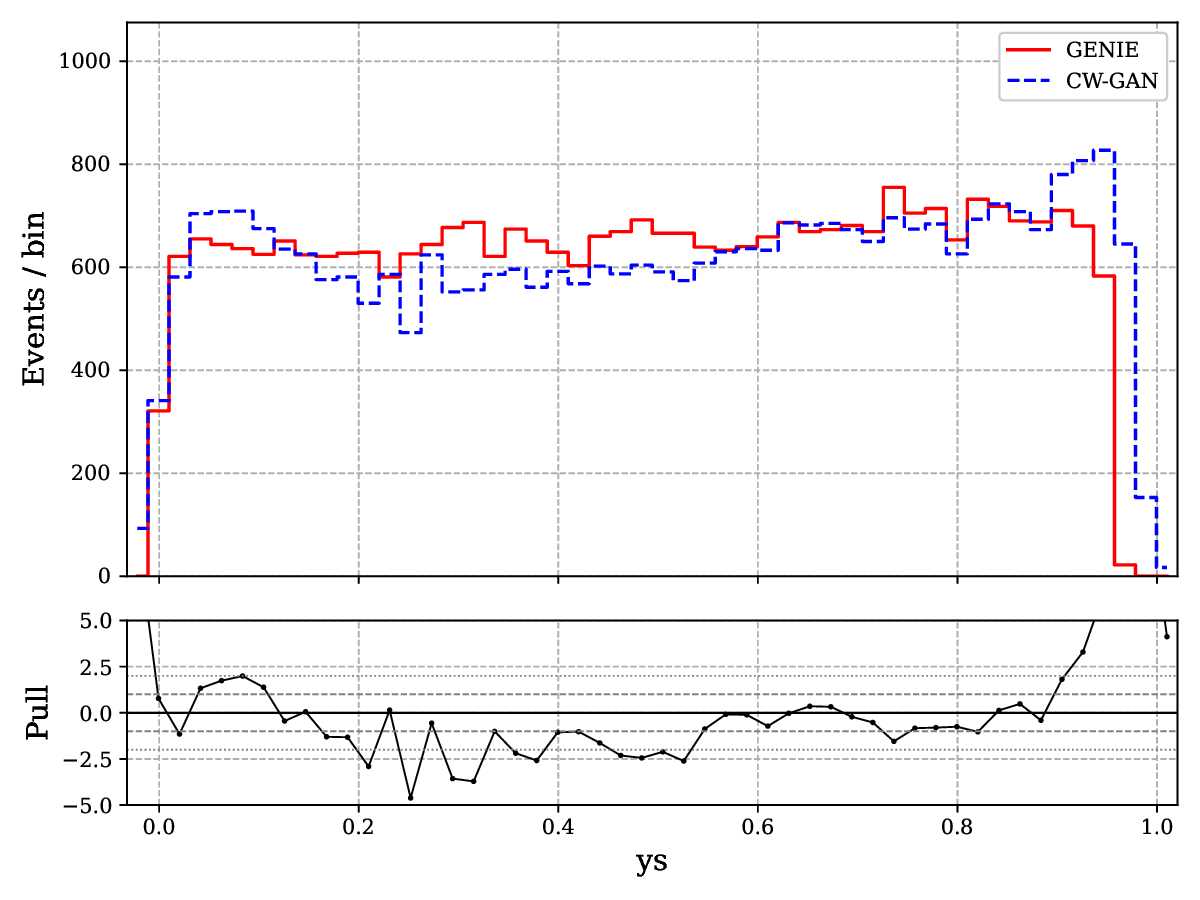}
        \caption{NuEElastic: ys}
        \label{fig:es1d_ys}
    \end{subfigure}

    \caption{(Elastic Scattering \textit{(NuEElastic)} Variables: 1D Marginal Distributions for the 8 kinematic variables. The red line represents the GENIE truth data, while the blue dashed line represents the CW-GAN generated events. The pull values (bottom panels) represent statistical agreement which is maintained close to $1\sigma$}
    \label{fig:marginals_es}
\end{figure*}

\subsection{ Statistical Fidelity (1D Histograms and Pulls)}

The first tier of physical validation is the reproduction of the one-dimensional marginal distributions for every kinematic variable. This verifies that the generator has learned the correct frequency of occurrence for specific values of physical observables, such as energy spectra and angular distributions. We evaluate this using the GENIE held-out test set as the "Ground Truth."

Before analyzing the distributions, it is crucial to address the numerical challenges identified during the preprocessing phase. The raw kinematic variables from GENIE span vast orders of magnitude from neutrino energies in the GeV range ($10^0$) to momentum transfers ($Q^2$) in the $10^{-5}$ GeV$^2$ range for elastic scattering. As detailed in the \ref{sec:III}.A , direct training on these raw values leads to numerical instability.

The momentum transfer $Q^2$ in neutrino-electron scattering is extremely small. A scaling factor of $10^5$ was applied to shift these values into a numerical range ($\mathcal{O}(1)$) accessible to the PowerTransformer and the neural network weights.

The results presented are for the post-processed data, where these scaling factors have been reversed, returning the variables to their physical units (GeV, GeV$^2$, etc.).

We assess the marginal distributions for the complete feature set: 8 variables for NuEElastic, 21 for IBD-CC, and 18 for NC. Figures \ref{fig:marginals_es}, \ref{fig:marginals_ibd_part1}, \ref{fig:nc_1d_1} present the comparison between GENIE (Truth) and CW-GAN (Generated) events.

The results demonstrate exceptional agreement. The CW-GAN generated distributions (red/dashed) overlay the GENIE truth data (blue/solid) almost perfectly across orders of magnitude in event density. Crucially, the model captures the high-energy tails of the momentum distributions ($p_x, p_y, p_z$) and the sharp spectral features of the energy variables ($E_l, E_\nu$).

To quantify this agreement, we calculate the pull values for each bin $i$, defined in Eq. \ref{eq:Pull_definition}. The sub-panels in the Figs. \ref{fig:marginals_es}, \ref{fig:marginals_ibd_part1} and \ref{fig:nc_1d_1} display these pull distributions. 

\begin{figure}[h]
    \centering
    \begin{subfigure}{\linewidth} 
        \centering
        \includegraphics[width=\linewidth]{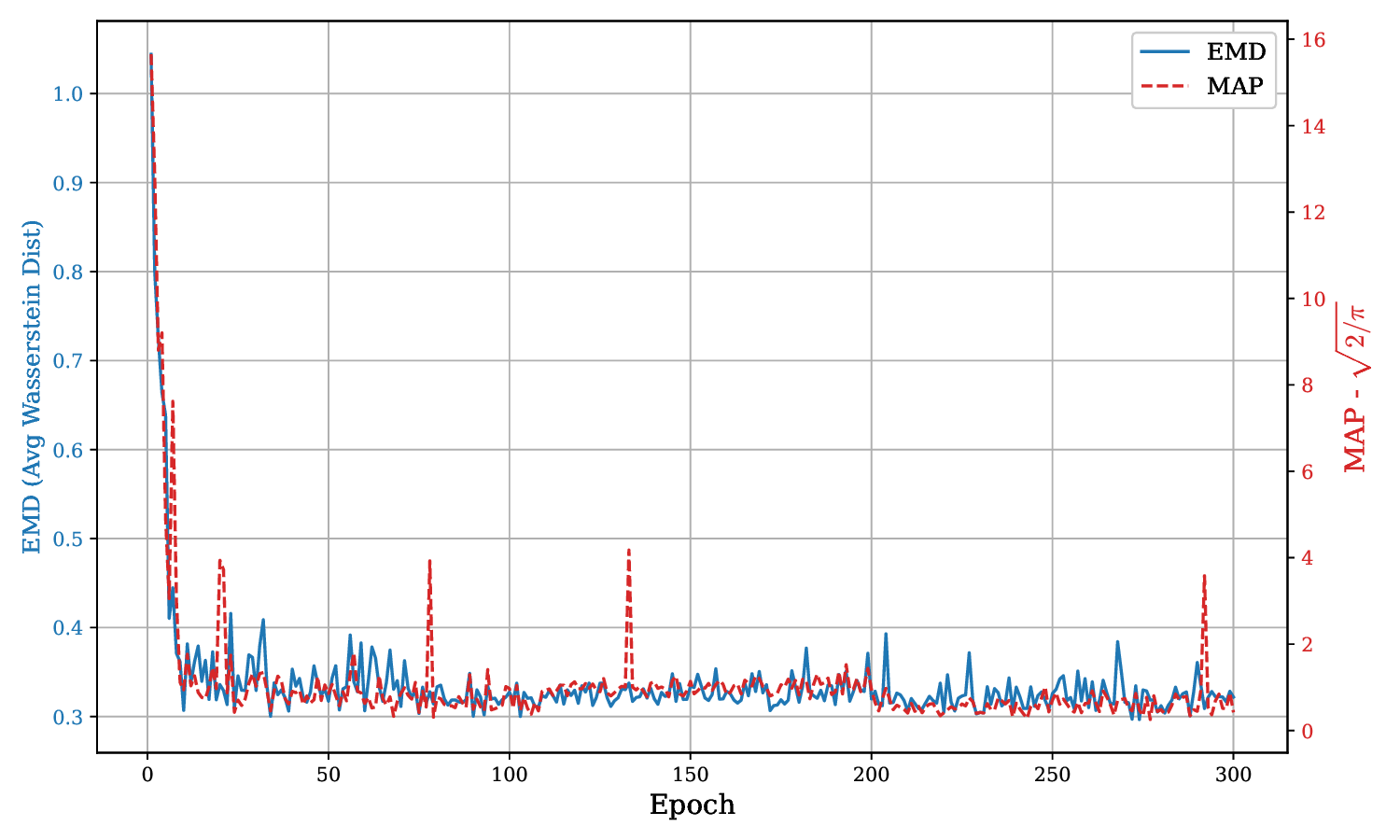}
    \end{subfigure}

    
    \centering

    \begin{subfigure}{\linewidth} 
        \centering
        \includegraphics[width=\linewidth]{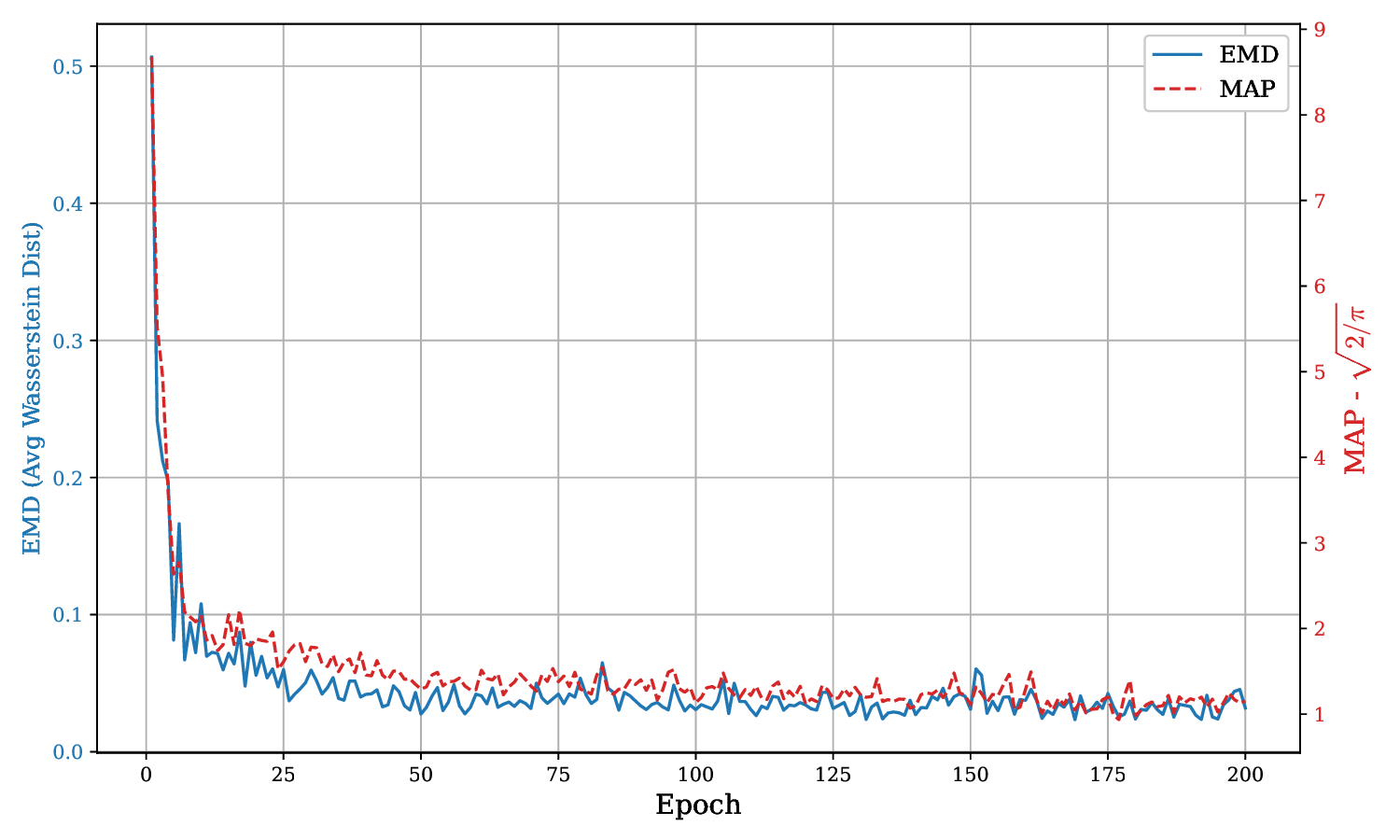}
    \end{subfigure}

    \begin{subfigure}{\linewidth} 
        \centering
        \includegraphics[width=\linewidth]{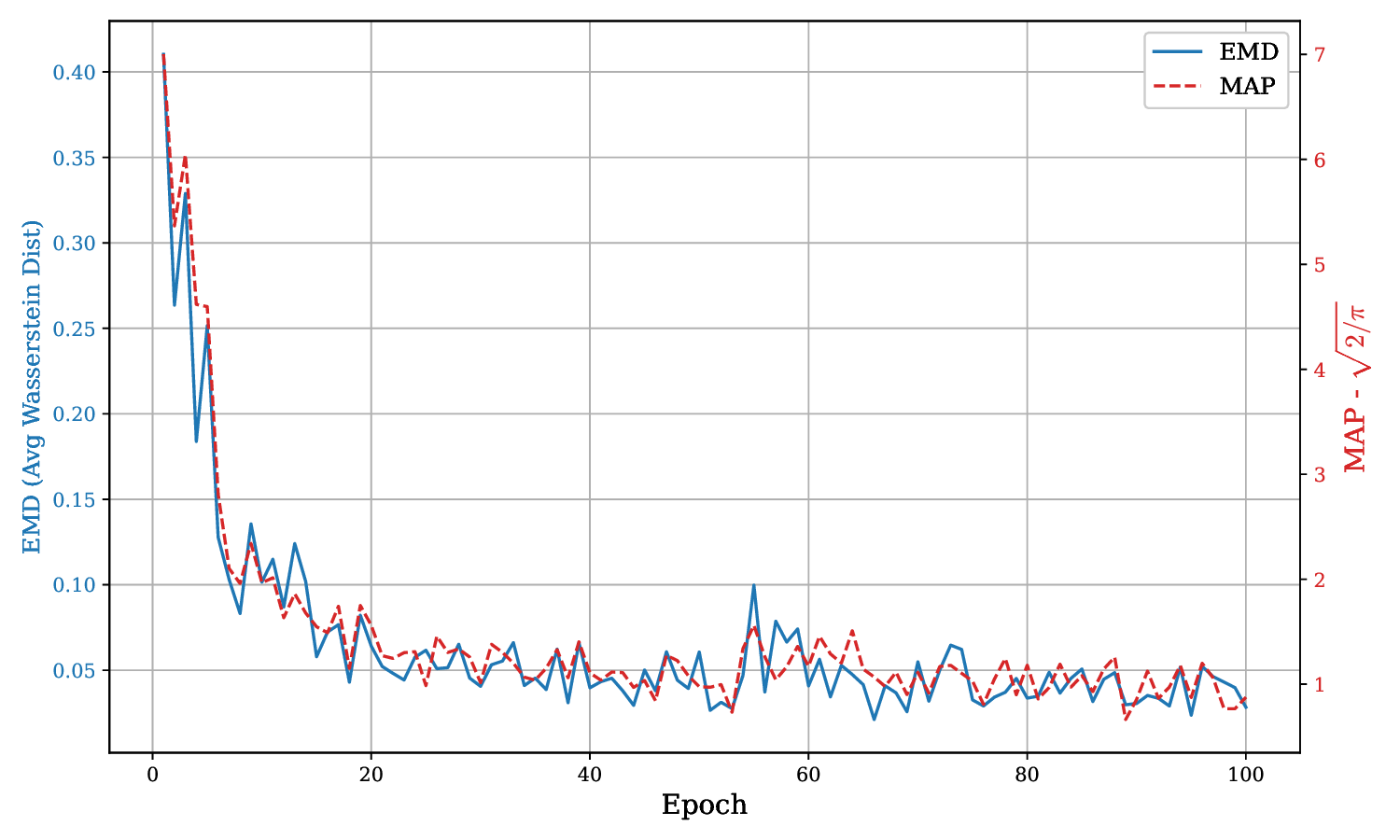}
    \end{subfigure}
    
    \caption{EMD and MAP score evolution over training epochs for NuEElastic, IBD-CC, and NC datasets (top to bottom)}
    \label{fig:global_stats}
\end{figure}

The generated incident neutrino energy distribution matches the input solar spectrum, confirming that the conditioning mechanism ($z|E_\nu$) effectively governs the physics. For Lepton Energy ($E_l$), the model accurately reproduces the weak interaction cross-section, including the difficult-to-learn high-energy tail. The Momentum Transfer ($Q^2$) results validate our pre-scaling strategy, capturing the zero-peak structure and logarithmic tail without discretization artifacts. Finally, the symmetric Gaussian distributions of transverse momenta ($p_{xl}, p_{yl}$) confirm that the CW-GAN autonomously learned the physical rotational invariance without explicit constraints.
\\ \\
In Fig \ref{fig:marginals_ibd_part1} we validate the conditioning mechanism by observing that the incident neutrino energy generated ($E_\nu$) and its longitudinal momentum ($p_{zv}$) perfectly match the Horiuchi DSNB input spectrum.
\\ \\
Figures \ref{fig:nc_1d_1}(a) and \ref{fig:nc_1d_2}(b) validate the Neutral Current ($\nu + p \to \nu + p$) channel, where the outgoing neutrino remains invisible. The model precisely reproduces the exponential decay of the \textbf{Recoil Proton} ($E_f$) kinetic energy. Capturing this slope is vital for experimental sensitivity studies, particularly for estimating backgrounds in Dark Matter searches where detector thresholds are critical. Additionally, the correct modeling of the \textbf{Invisible Neutrino} transverse momenta ($p_{xf}, p_{yf}$) confirms that the generator internally enforces momentum conservation ($\vec{p}_{T,\nu'} = -\vec{p}_{T,p}$) despite the particle's unobservability.
\\ \\ 
While visual inspection of histograms is necessary, it is qualitative. To rigorously quantify the convergence of the generated distributions toward the true physical distributions, we employ the Earth Mover's Distance (EMD) and the Mean Absolute Pull (MAP).
\\ \\
The Earth Mover's Distance measures the minimum ``work'' required to transform the generated distribution into the target distribution. The plot in Fig \ref{fig:global_stats} (plot corresponding to left axis) shows a characteristic exponential decay in EMD values, stabilizing after approximately 40 epochs. The very low final values indicate that the generated distributions closely match the true physical data.

As shown in Fig.~\ref{fig:global_stats} (right axis), the MAP scores show a steep drop in the early epochs, followed by noisy oscillations around a low mean value which is a characteristic of the adversarial minimax dynamics between the generator and critic. As shown in Table~\ref{tab:convergence}, the CW-GAN achieves low EMD values across all three datasets: 0.34, 0.046, and 0.064 for NuEElastic, IBD, and NC respectively, confirming that the generated distributions closely follow the true physical data across the full kinematic range. The MAP values, hovering between 1.0 - 2.0 across all datasets, are slightly above the ideal value of 
unity, which is expected given the statistical fluctuations inherent in the Monte Carlo training data itself. Since the model is trained on finite sampled data rather than the true underlying distribution, a MAP marginally above 1 reflects the model faithfully learning the statistical noise present in the training set, rather than indicating a systematic bias.

\begin{table}[h]
\centering
\normalsize 
\caption{Convergence Metrics }
\label{tab:convergence}

\begin{tabular*}{\columnwidth}{@{\extracolsep{\fill}} lccc @{}}
\toprule
Dataset & EMD & MAP & Epoch \\
\midrule
NuEElastic & 0.336489 & 1.897165 & 300 \\
IBD        & 0.045825 & 2.256705 & 200 \\
NC         & 0.064262 & 2.237644 & 100\\
\bottomrule
\end{tabular*}
\end{table}

\begingroup
\let\clearpage\relax
\let\newpage\relax

\begin{figure*}[p]
    \centering
    \begin{subfigure}{0.32\textwidth} \includegraphics[width=\linewidth]{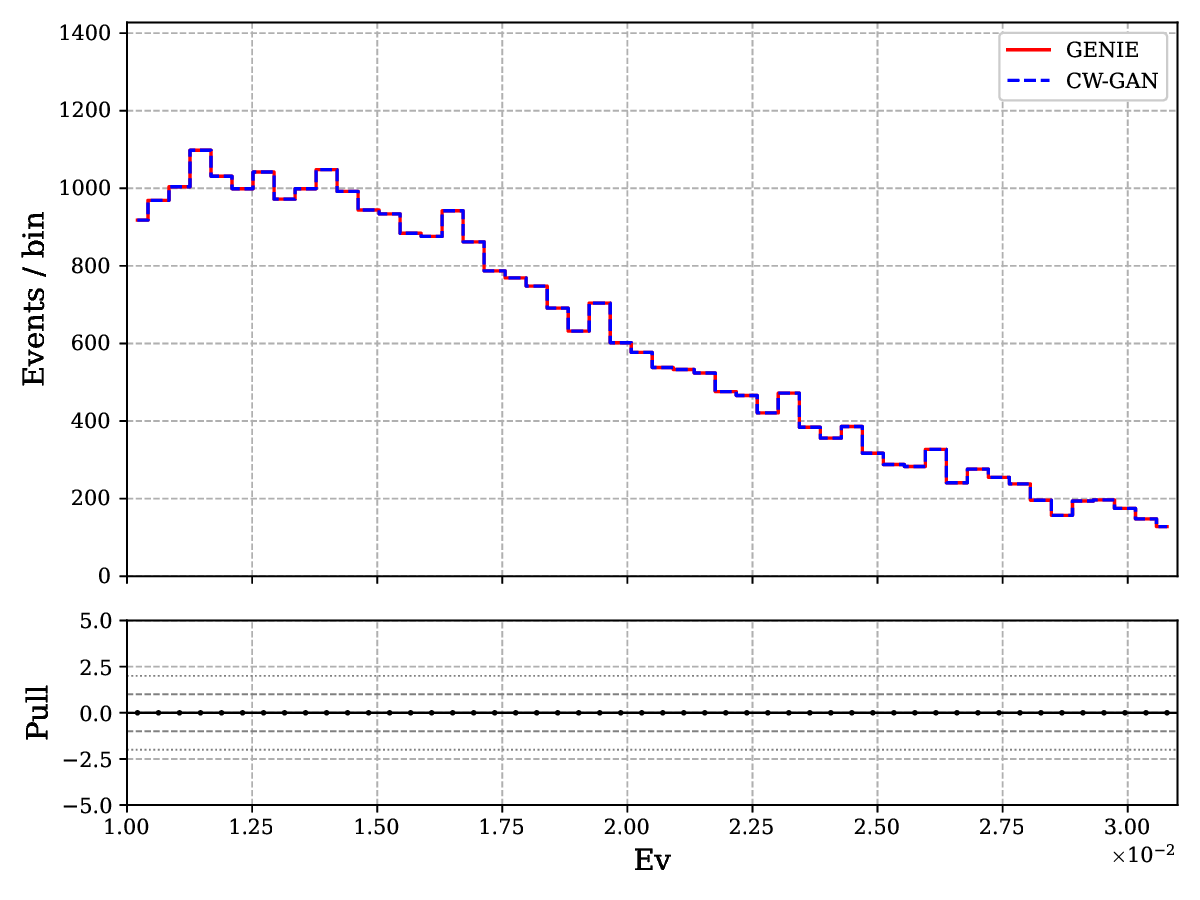} \caption{IBD-CC: Ev} \label{fig:ibd1d_Ev} \end{subfigure}
    \begin{subfigure}{0.32\textwidth} \includegraphics[width=\linewidth]{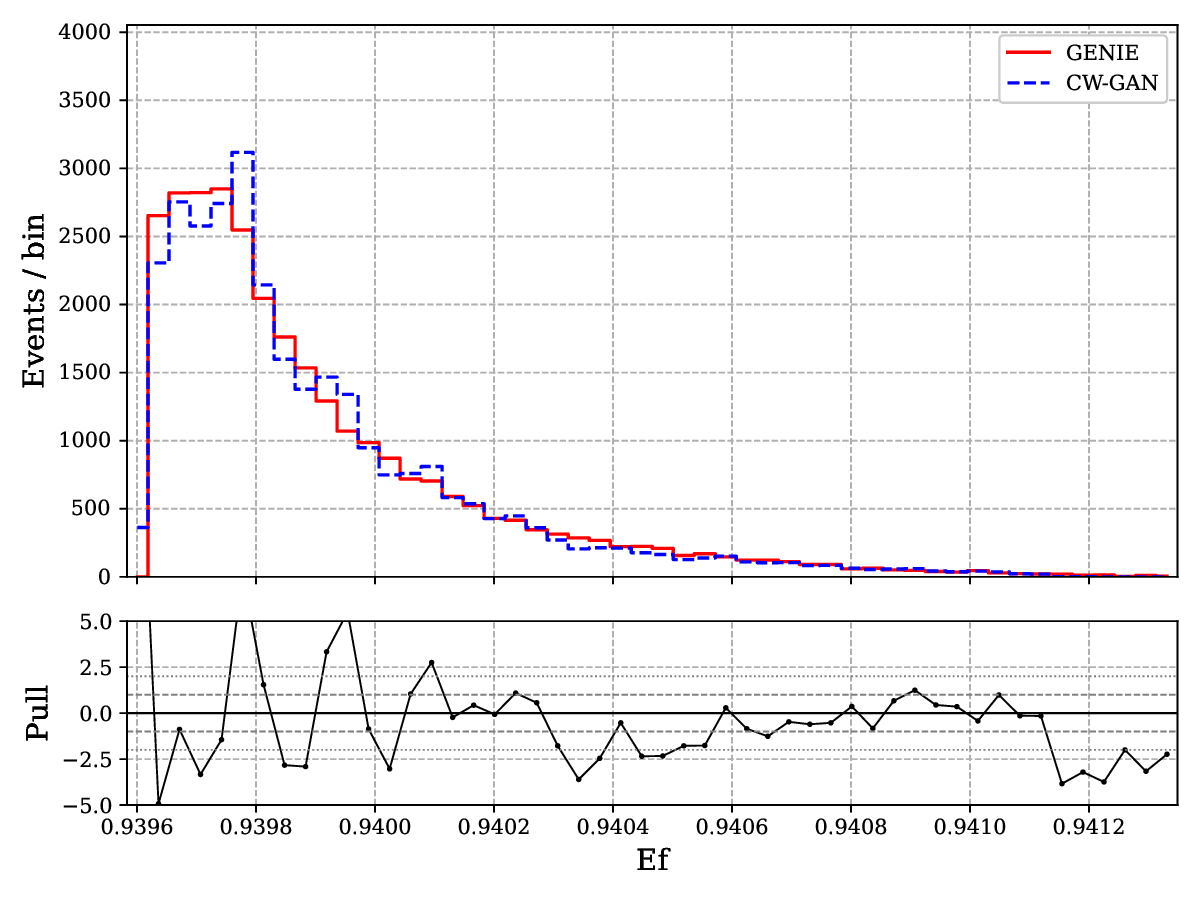} \caption{IBD-CC: Ef} \label{fig:ibd1d_Ef} \end{subfigure} \hfill
    \begin{subfigure}{0.32\textwidth} \includegraphics[width=\linewidth]{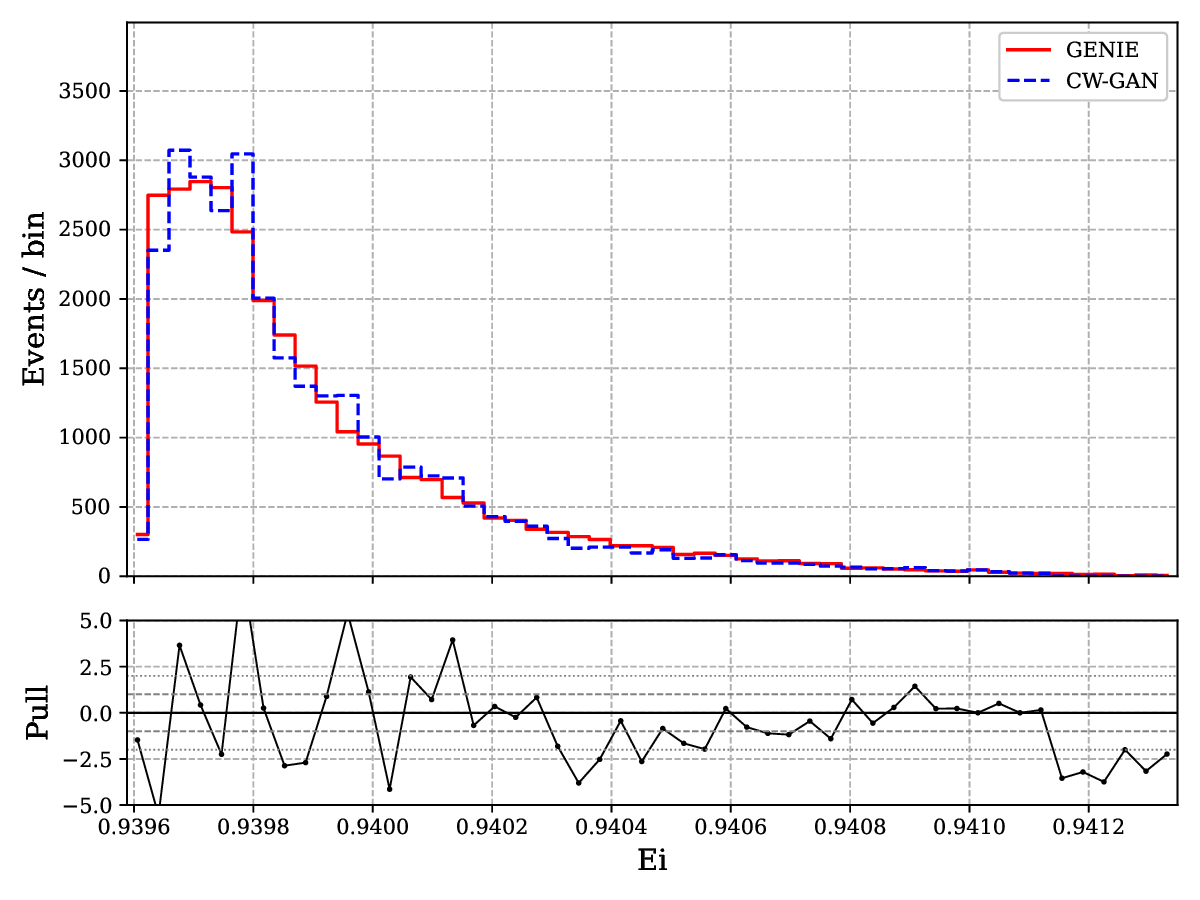} \caption{IBD-CC: Ei} \label{fig:ibd1d_Ei} \end{subfigure} \hfill

    \vspace{0.2cm}
    \begin{subfigure}{0.32\textwidth} \includegraphics[width=\linewidth]{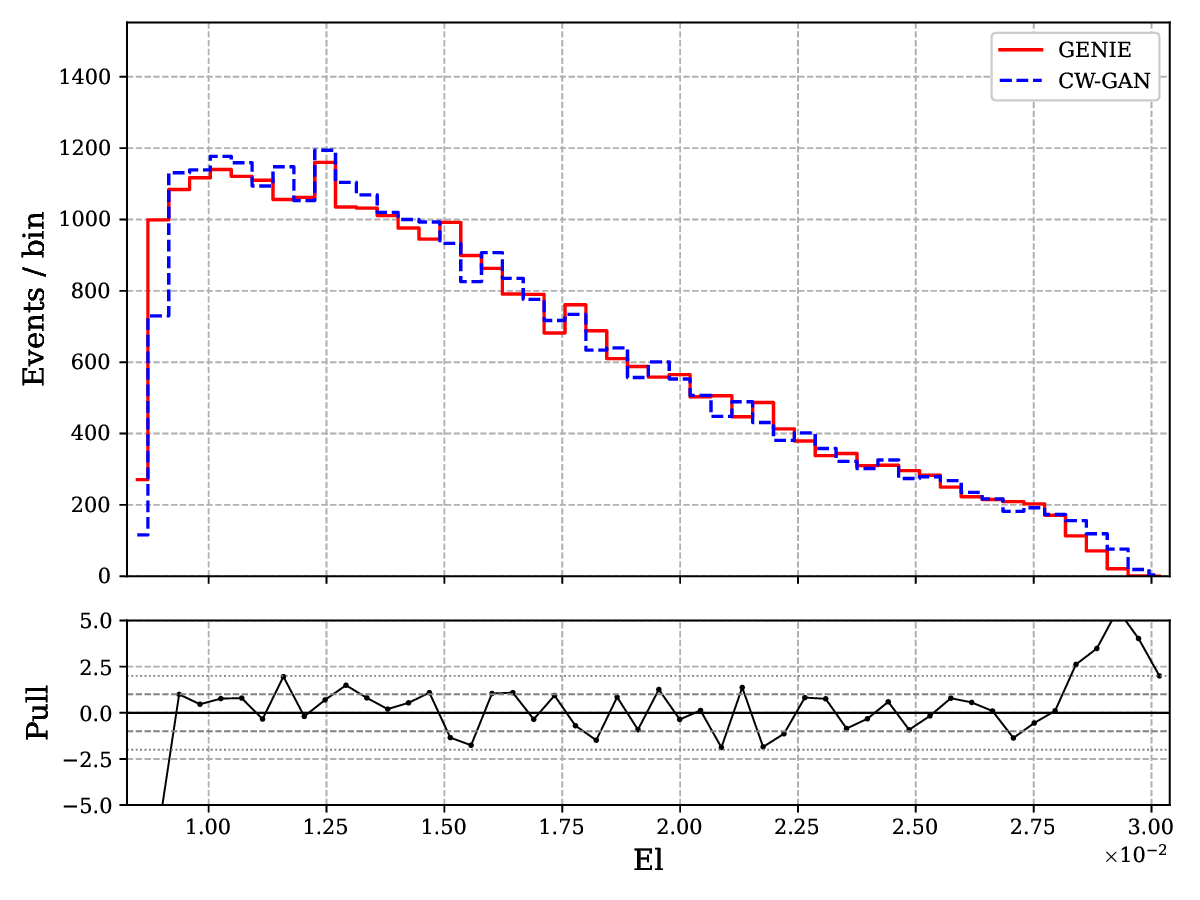} \caption{IBD-CC: El} \label{fig:ibd1d_El} \end{subfigure}
    \begin{subfigure}{0.32\textwidth} \includegraphics[width=\linewidth]{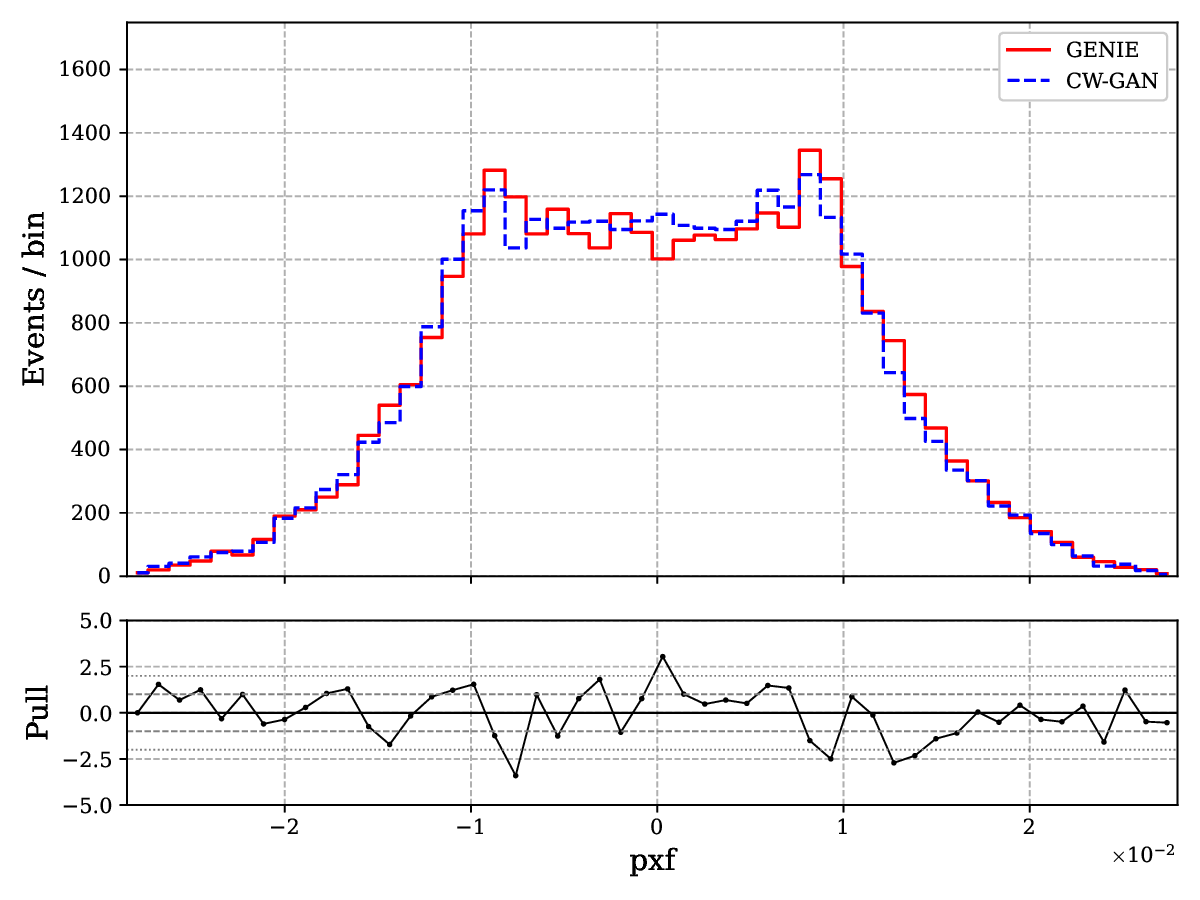} \caption{IBD-CC: pxf} \label{fig:ibd1d_pxf} \end{subfigure} \hfill
    \begin{subfigure}{0.32\textwidth} \includegraphics[width=\linewidth]{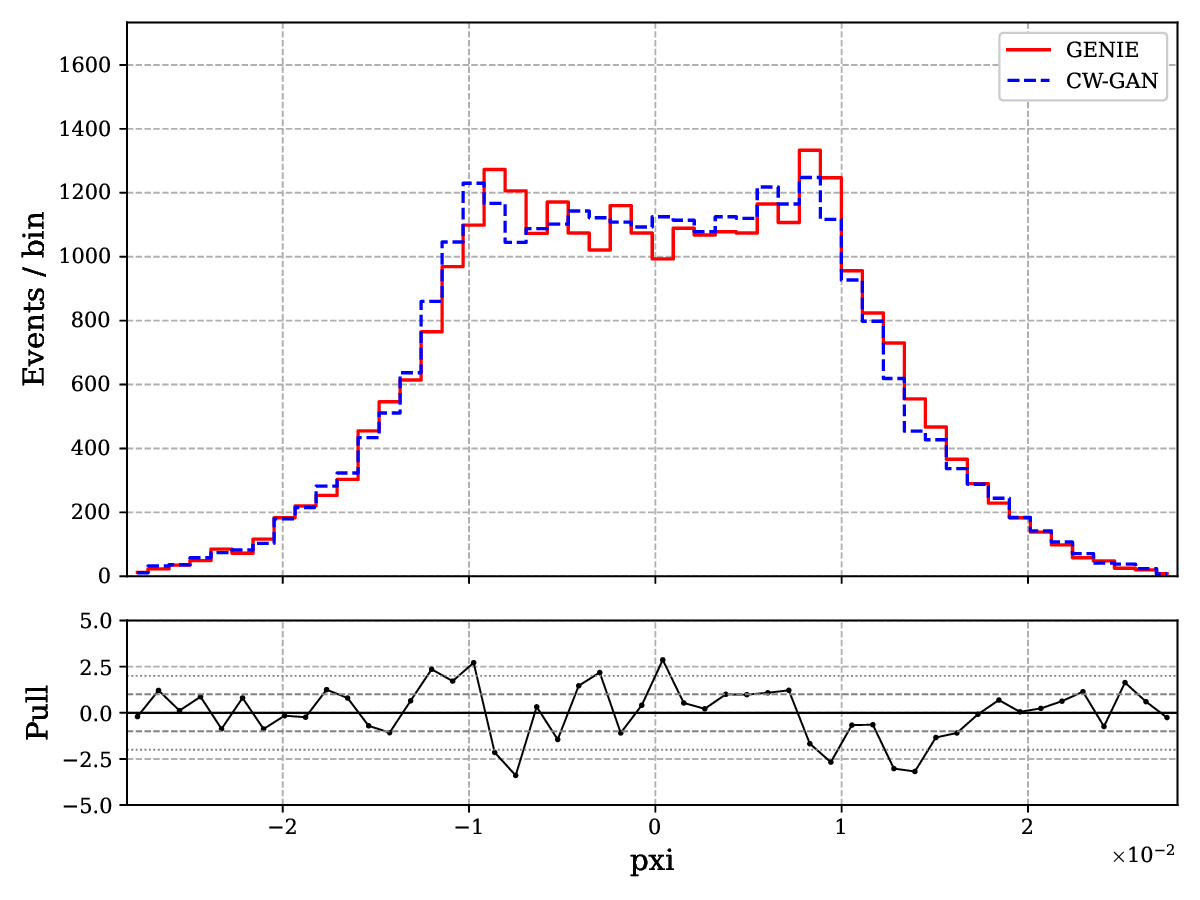} \caption{IBD-CC: pxi} \label{fig:ibd1d_pxi} \end{subfigure} \hfill

    \vspace{0.2cm}
    \begin{subfigure}{0.32\textwidth} \includegraphics[width=\linewidth]{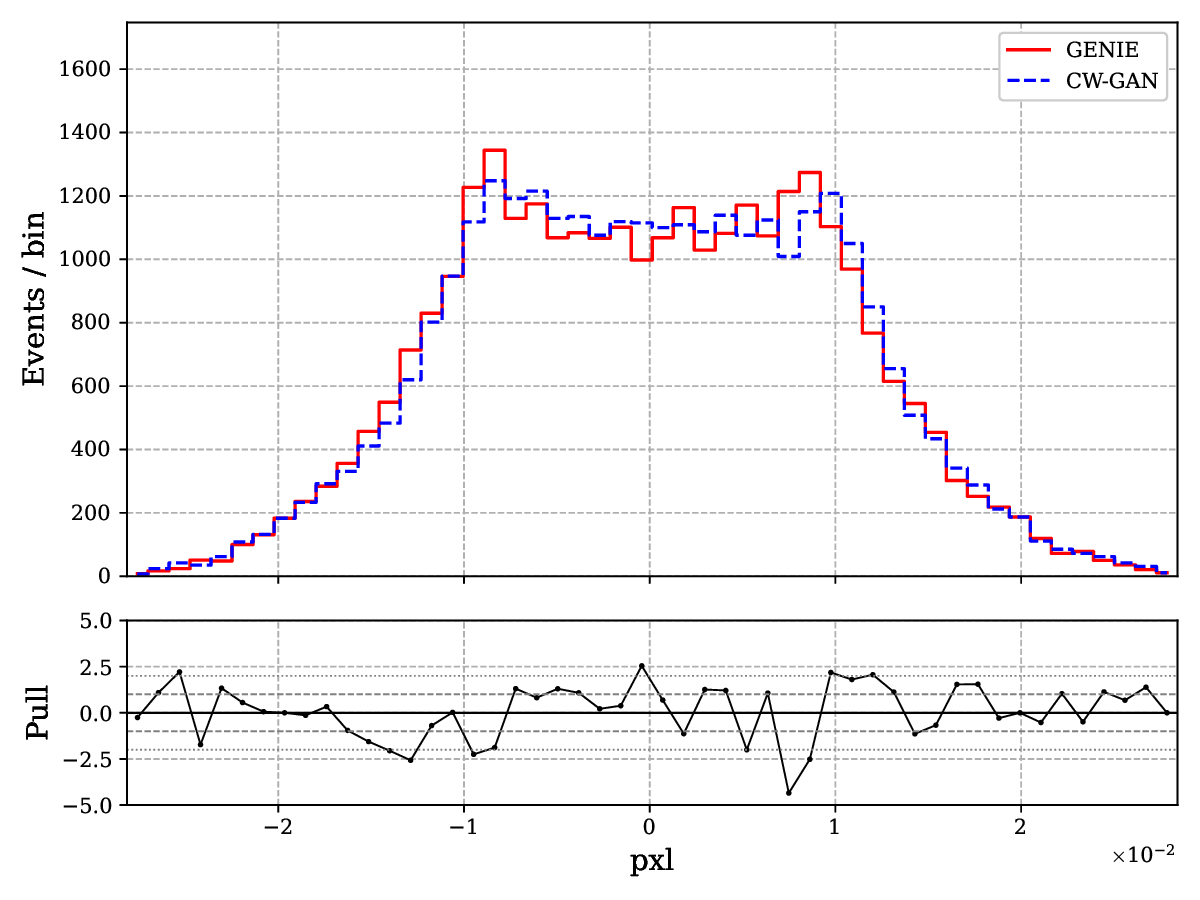} \caption{IBD-CC: pxl} \label{fig:ibd1d_pxl} \end{subfigure}
    \begin{subfigure}{0.32\textwidth} \includegraphics[width=\linewidth]{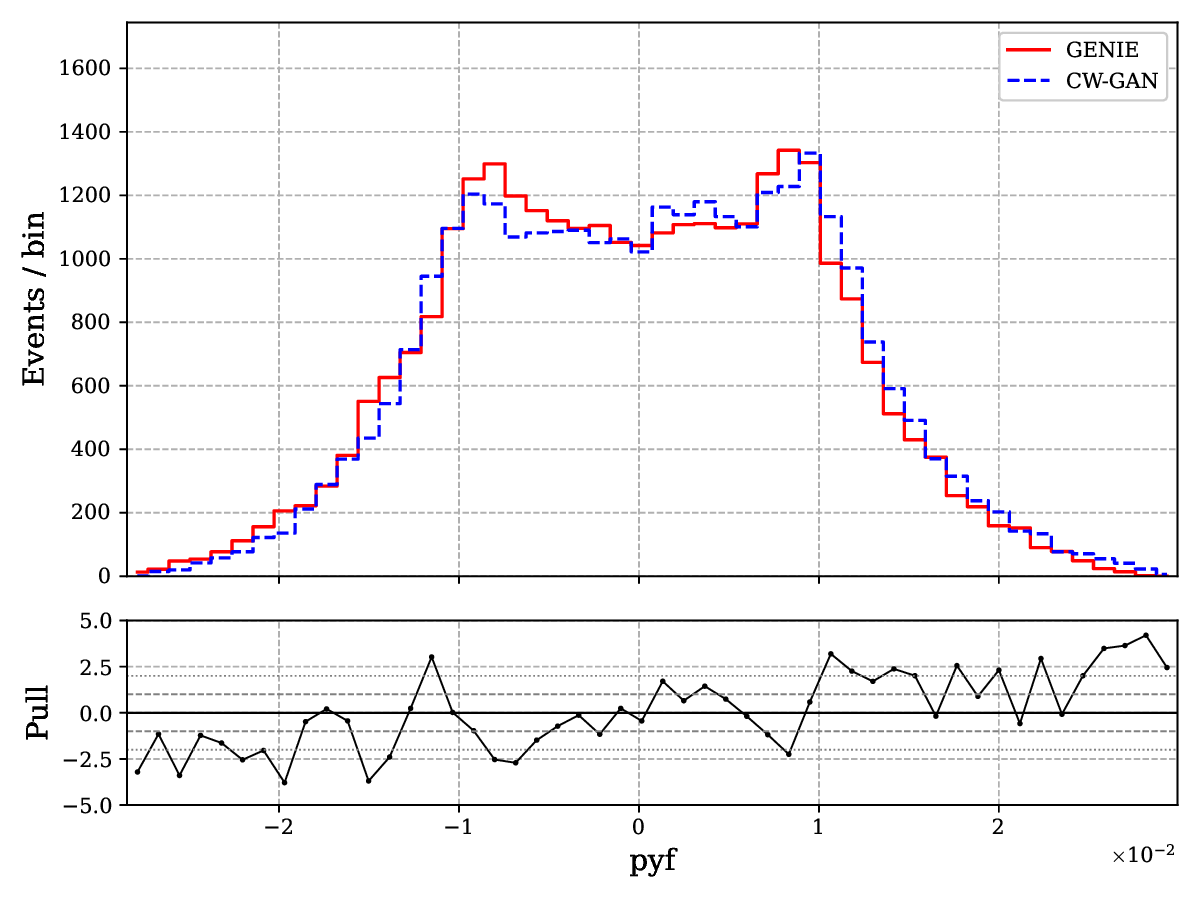} \caption{IBD-CC: pyf} \label{fig:ibd1d_pyf} \end{subfigure} \hfill
    \begin{subfigure}{0.32\textwidth} \includegraphics[width=\linewidth]{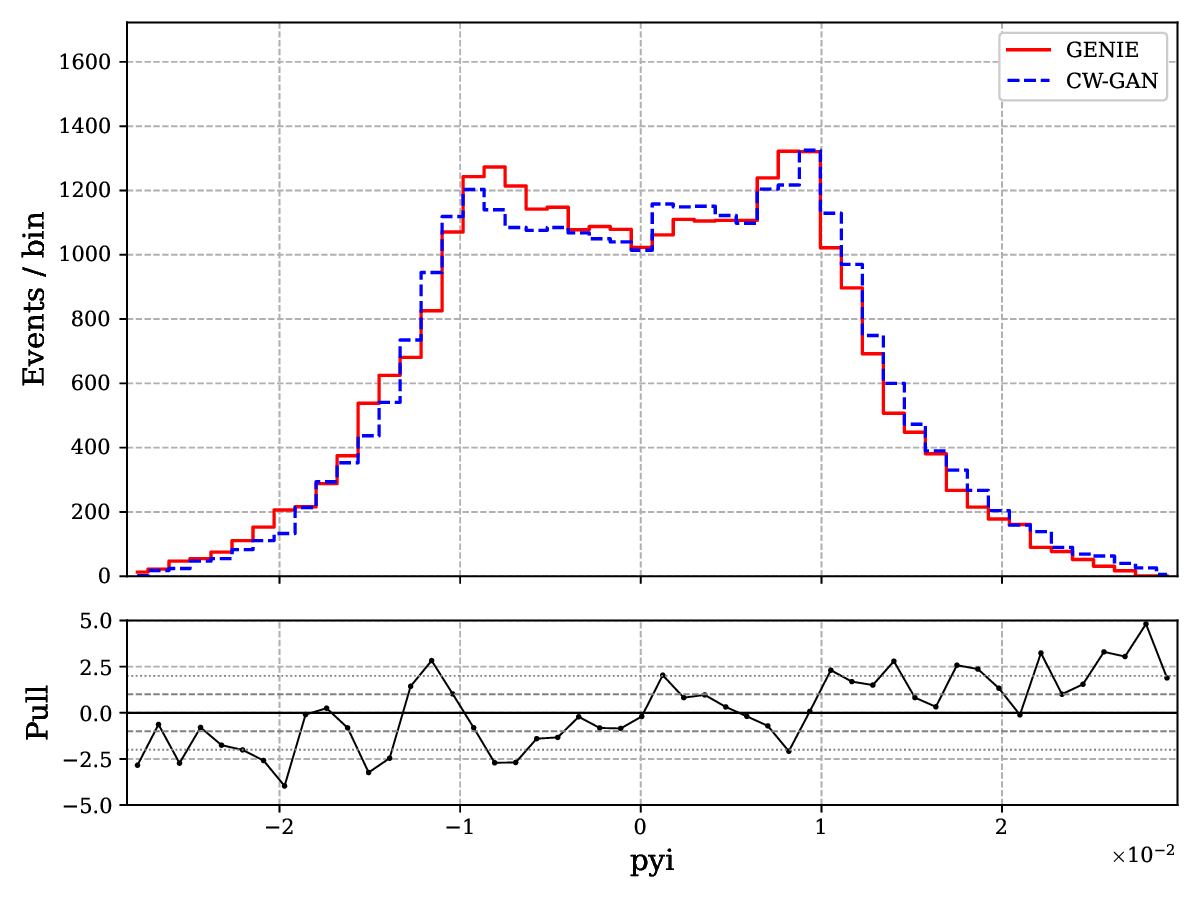} \caption{IBD-CC: pyi} \label{fig:ibd1d_pyi} \end{subfigure} \hfill

    \vspace{0.2cm}
    \begin{subfigure}{0.32\textwidth} \includegraphics[width=\linewidth]{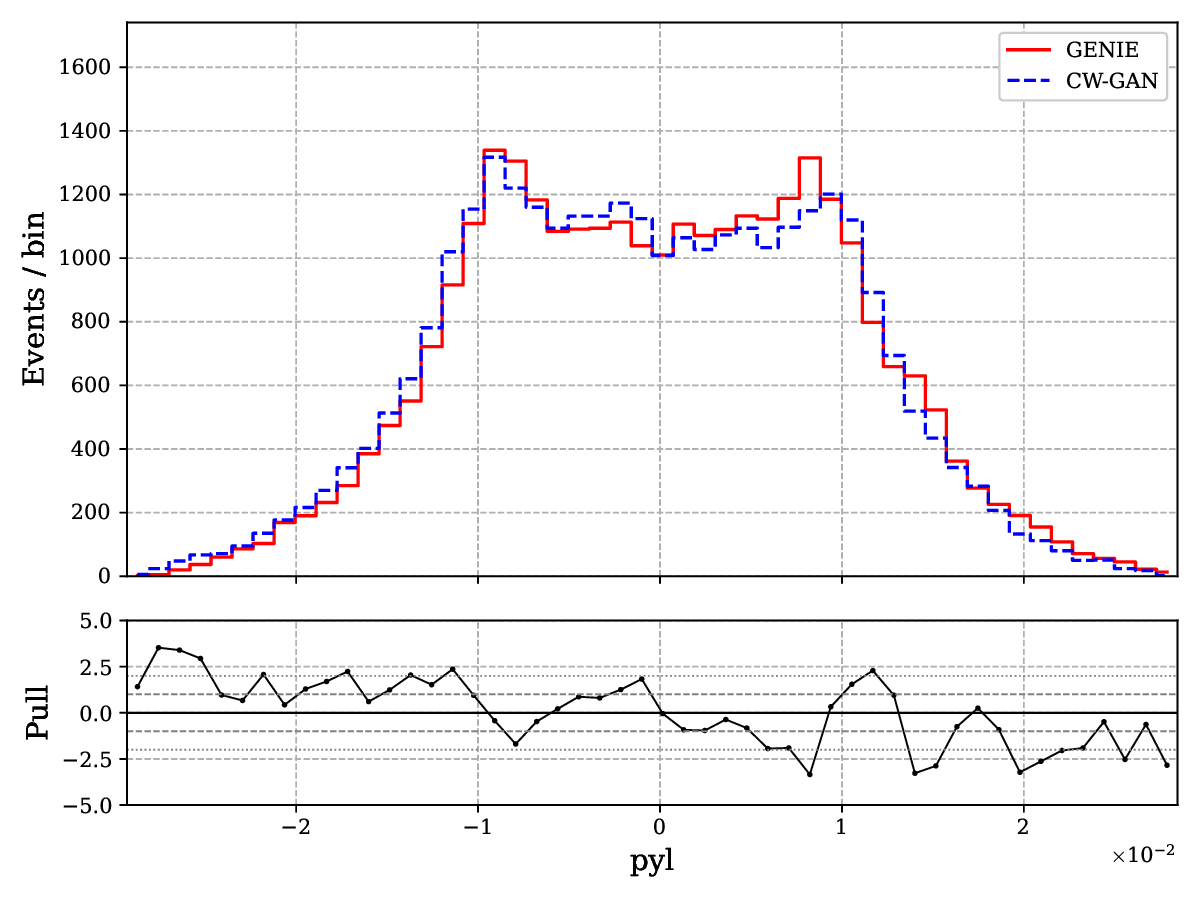} \caption{IBD-CC: pyl} \label{fig:ibd1d_pyl} \end{subfigure}
    \begin{subfigure}{0.32\textwidth} \includegraphics[width=\linewidth]{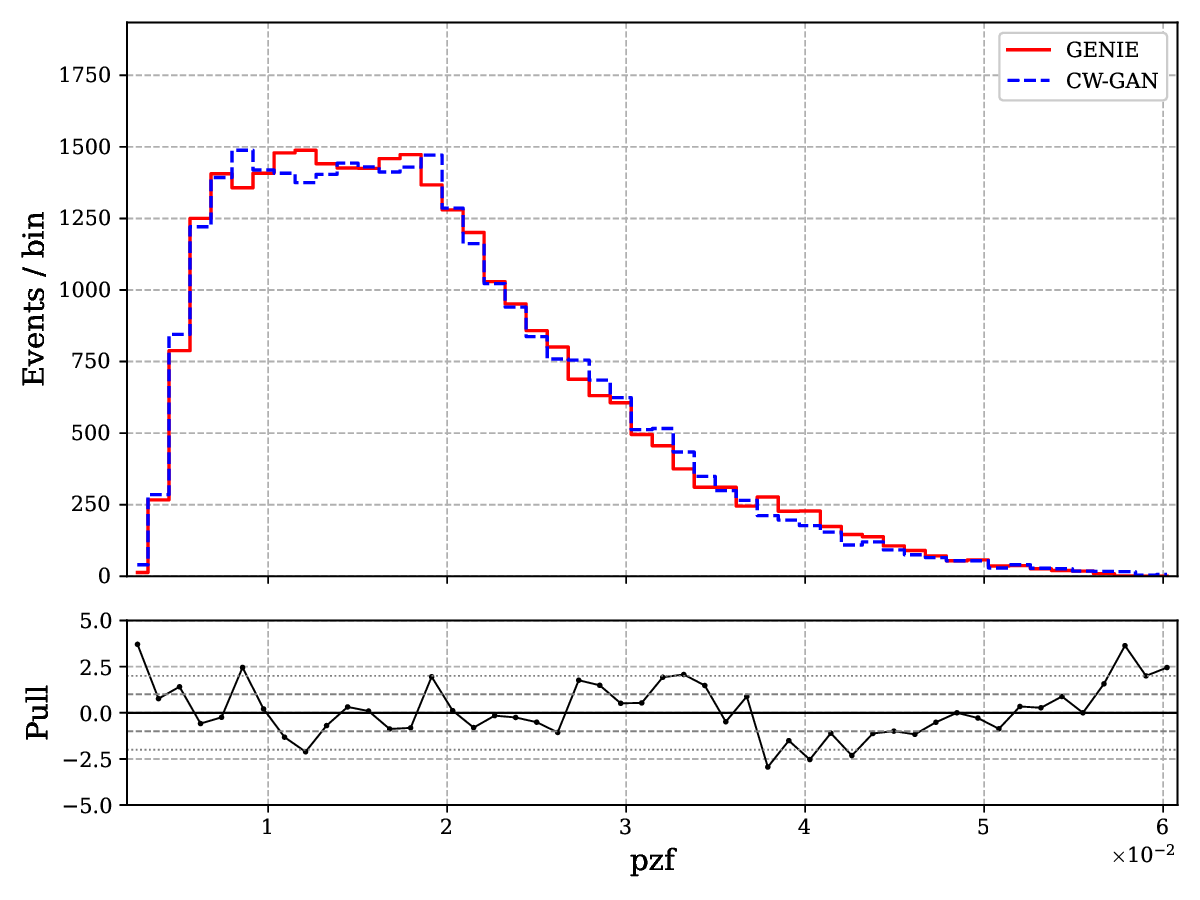} \caption{IBD-CC: pzf} \label{fig:ibd1d_pzf} \end{subfigure} \hfill
    \begin{subfigure}{0.32\textwidth} \includegraphics[width=\linewidth]{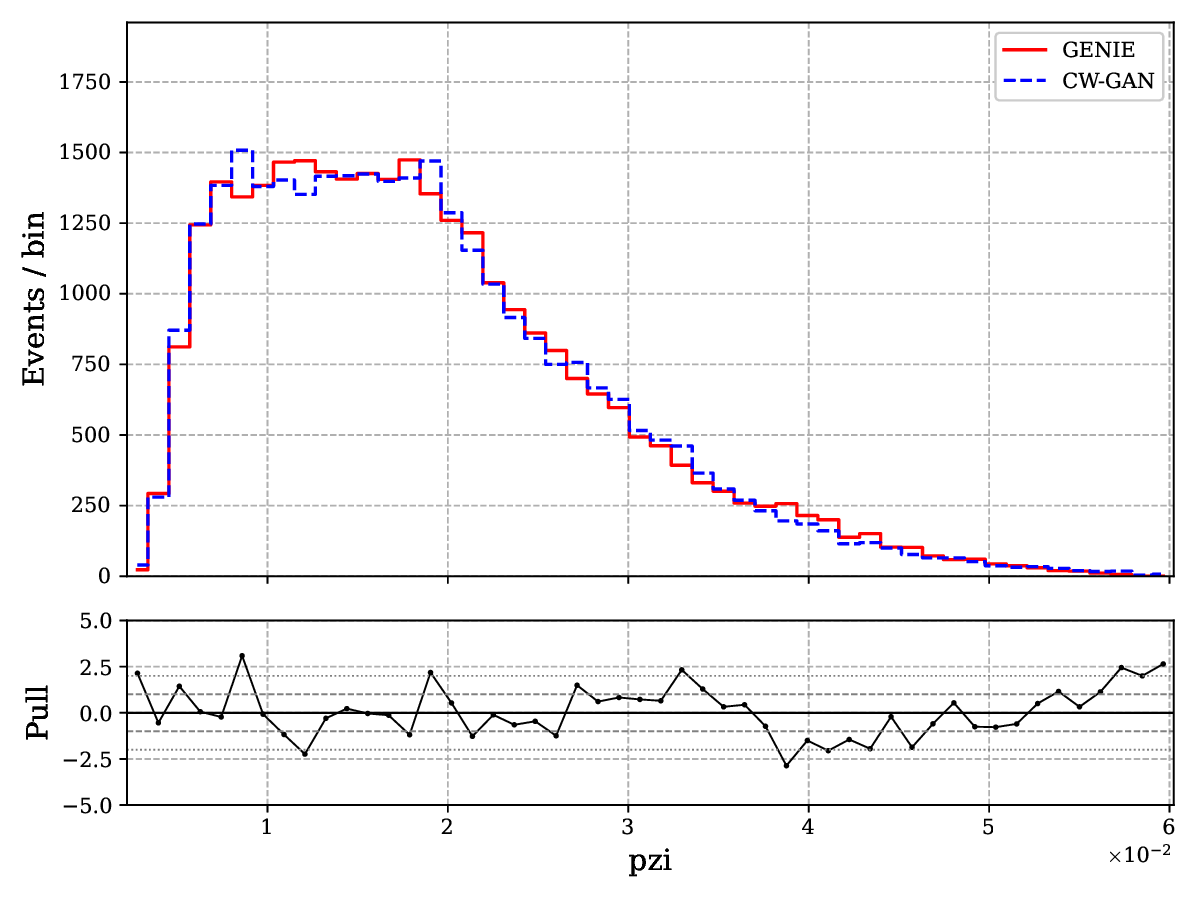} \caption{IBD-CC: pzi} \label{fig:ibd1d_pzi} \end{subfigure} \hfill
    
    \caption{(\subref{fig:ibd1d_Ev}) Inverse Beta Decay: 1D Marginal Distributions for the first 12 kinematic variables. The plots compare generated events against ground truth for energies ($E_\nu, E_f, E_l$) and momentum components, showing precise replication of the multi-body decay kinematics.}
    \label{fig:marginals_ibd_part1}
\end{figure*}

\begin{figure*}[p]
    \ContinuedFloat
    \centering
    \begin{subfigure}{0.32\textwidth} \includegraphics[width=\linewidth]{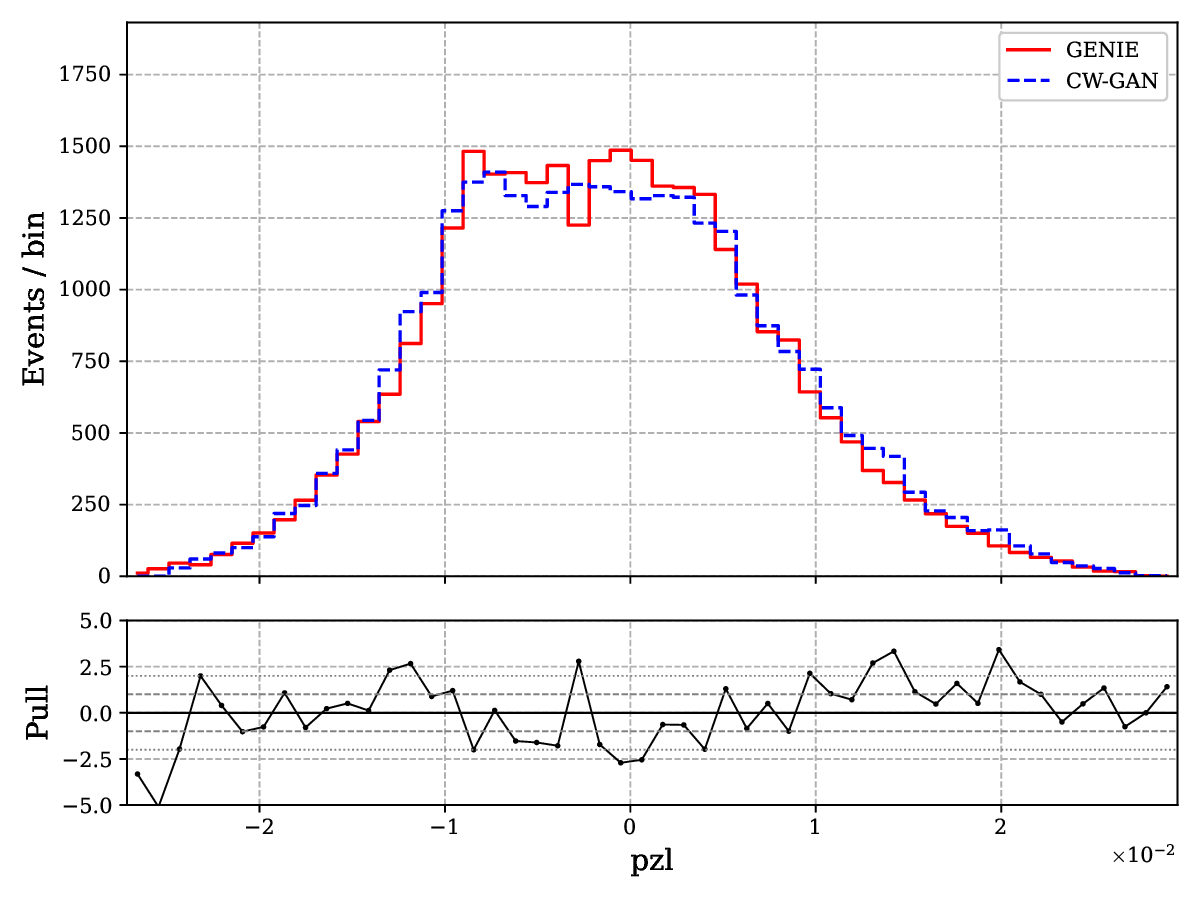} \caption{IBD-CC: pzl} \label{fig:ibd1d_pzl} \end{subfigure}
    \begin{subfigure}{0.32\textwidth} \includegraphics[width=\linewidth]{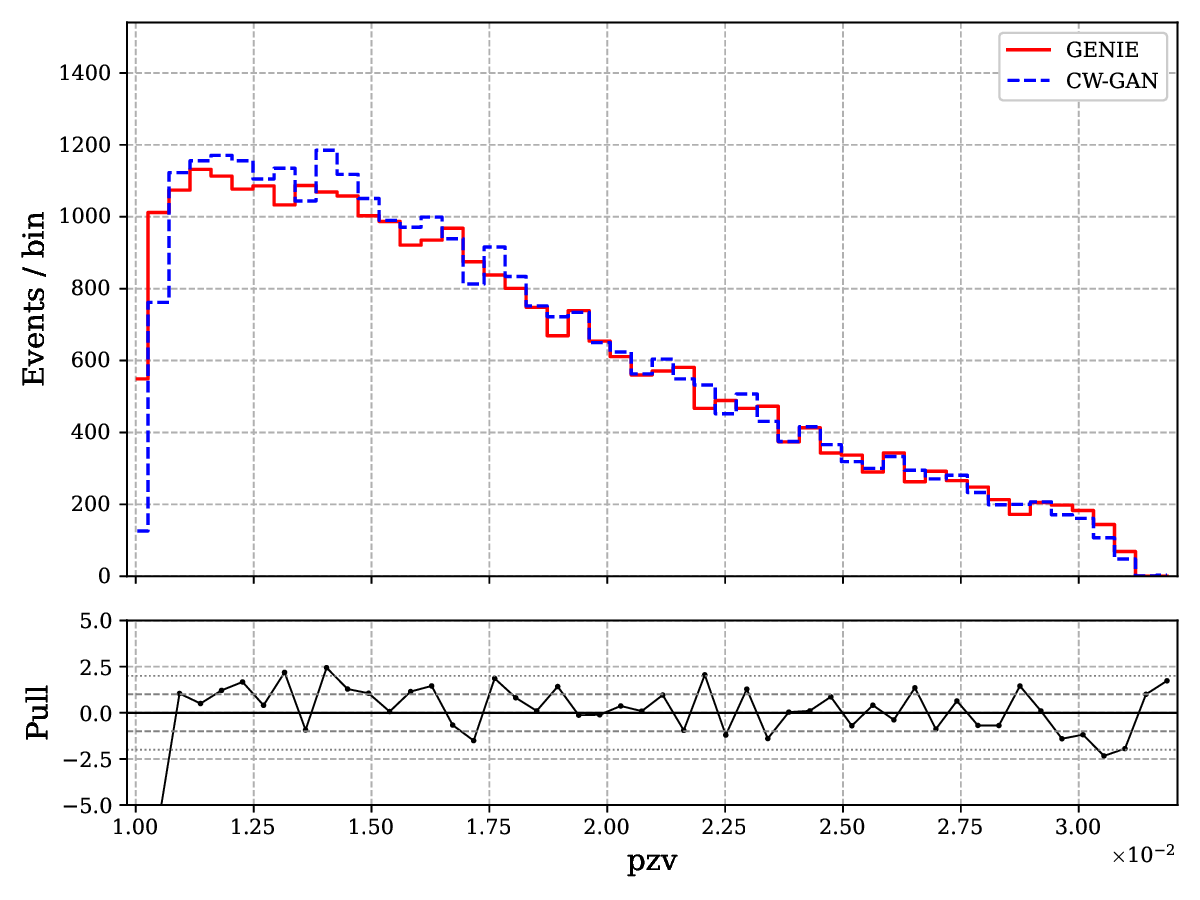} \caption{IBD-CC: pzv} \label{fig:ibd1d_pzv} \end{subfigure} \hfill
    \begin{subfigure}{0.32\textwidth} \includegraphics[width=\linewidth]{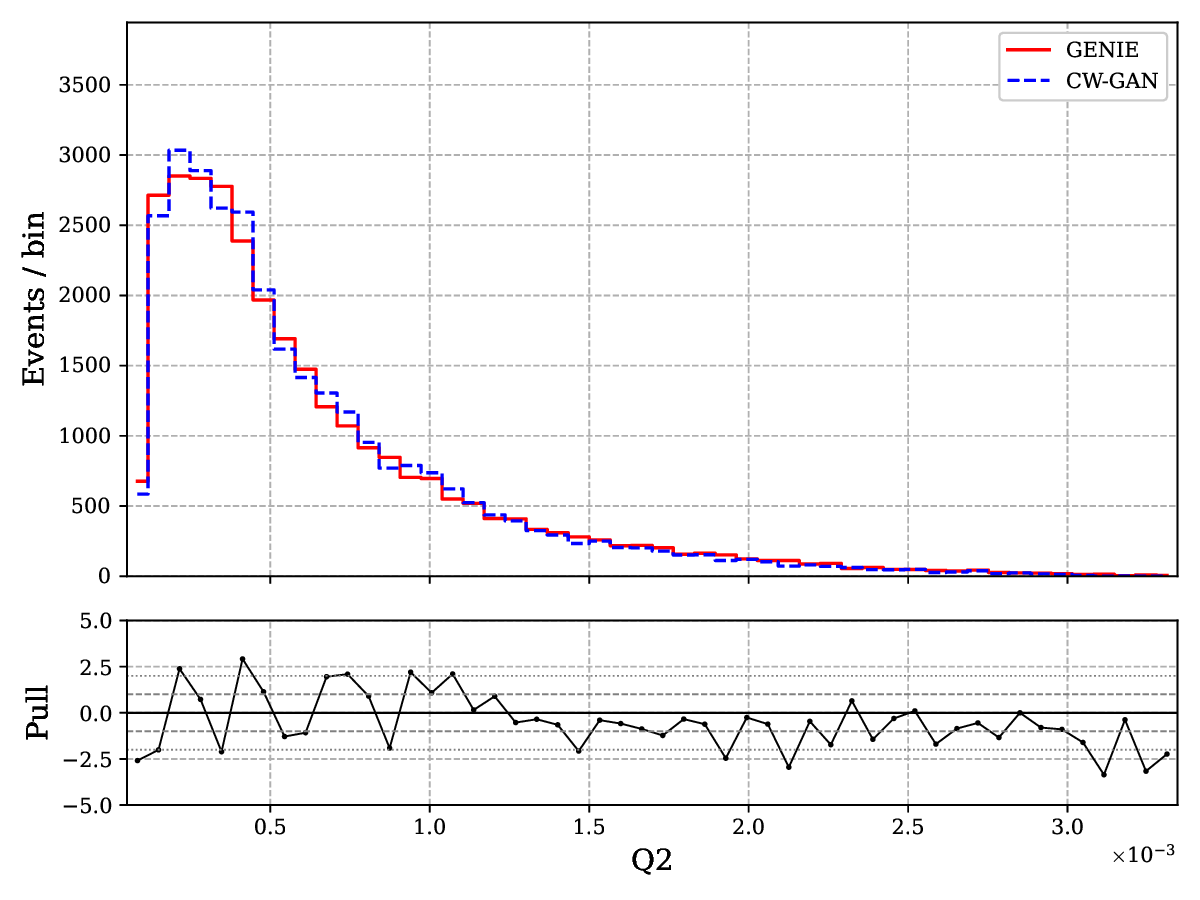} \caption{IBD-CC: Q2} \label{fig:ibd1d_Q2} \end{subfigure} \hfill

    \vspace{0.2cm}
    \begin{subfigure}{0.32\textwidth} \includegraphics[width=\linewidth]{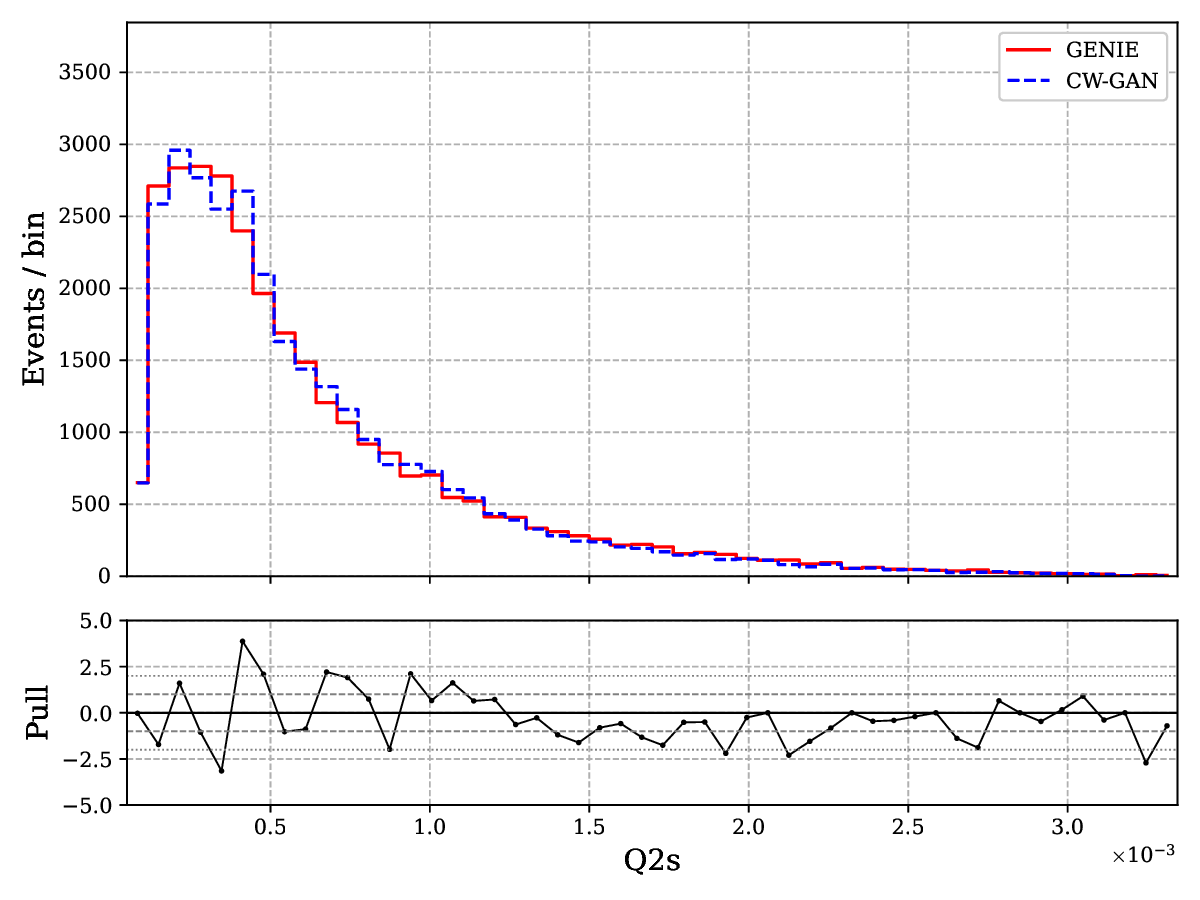} \caption{IBD-CC: Q2s} \label{fig:ibd1d_Q2s} \end{subfigure}
    \begin{subfigure}{0.32\textwidth} \includegraphics[width=\linewidth]{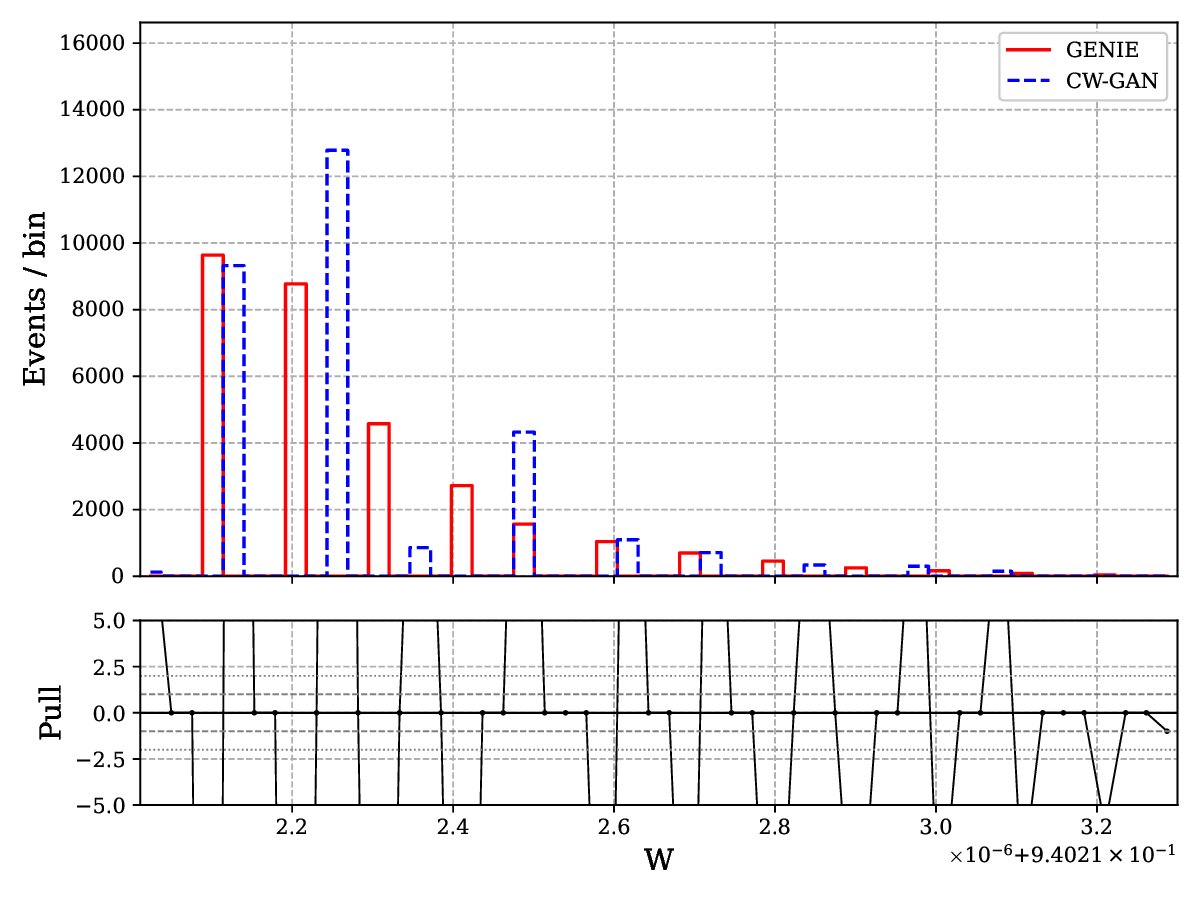} \caption{IBD-CC: W} \label{fig:ibd1d_W} \end{subfigure} \hfill
    \begin{subfigure}{0.32\textwidth} \includegraphics[width=\linewidth]{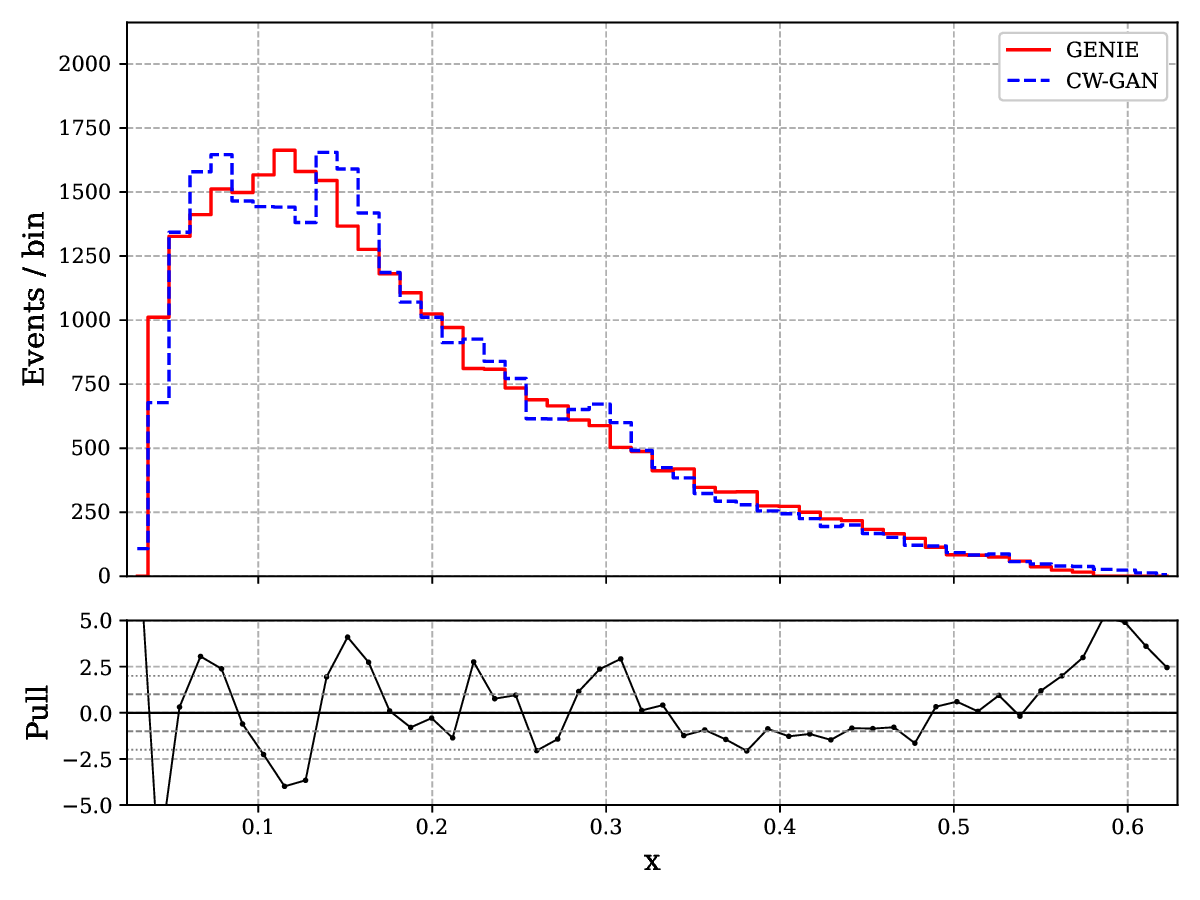} \caption{IBD-CC: x} \label{fig:ibd1d_x} \end{subfigure} \hfill

    \vspace{0.2cm}
    \begin{subfigure}{0.32\textwidth} \includegraphics[width=\linewidth]{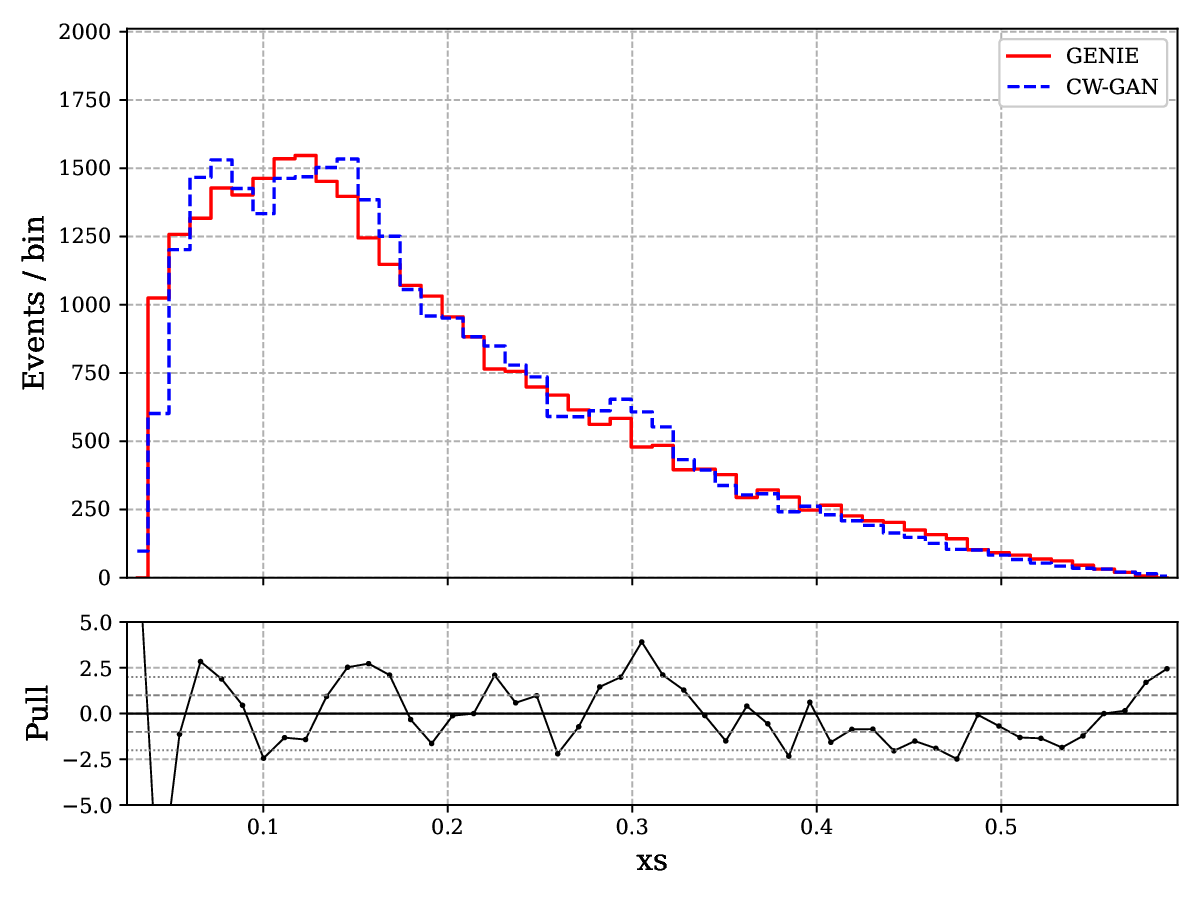} \caption{IBD-CC: xs} \label{fig:ibd1d_xs} \end{subfigure}
    \begin{subfigure}{0.32\textwidth} \includegraphics[width=\linewidth]{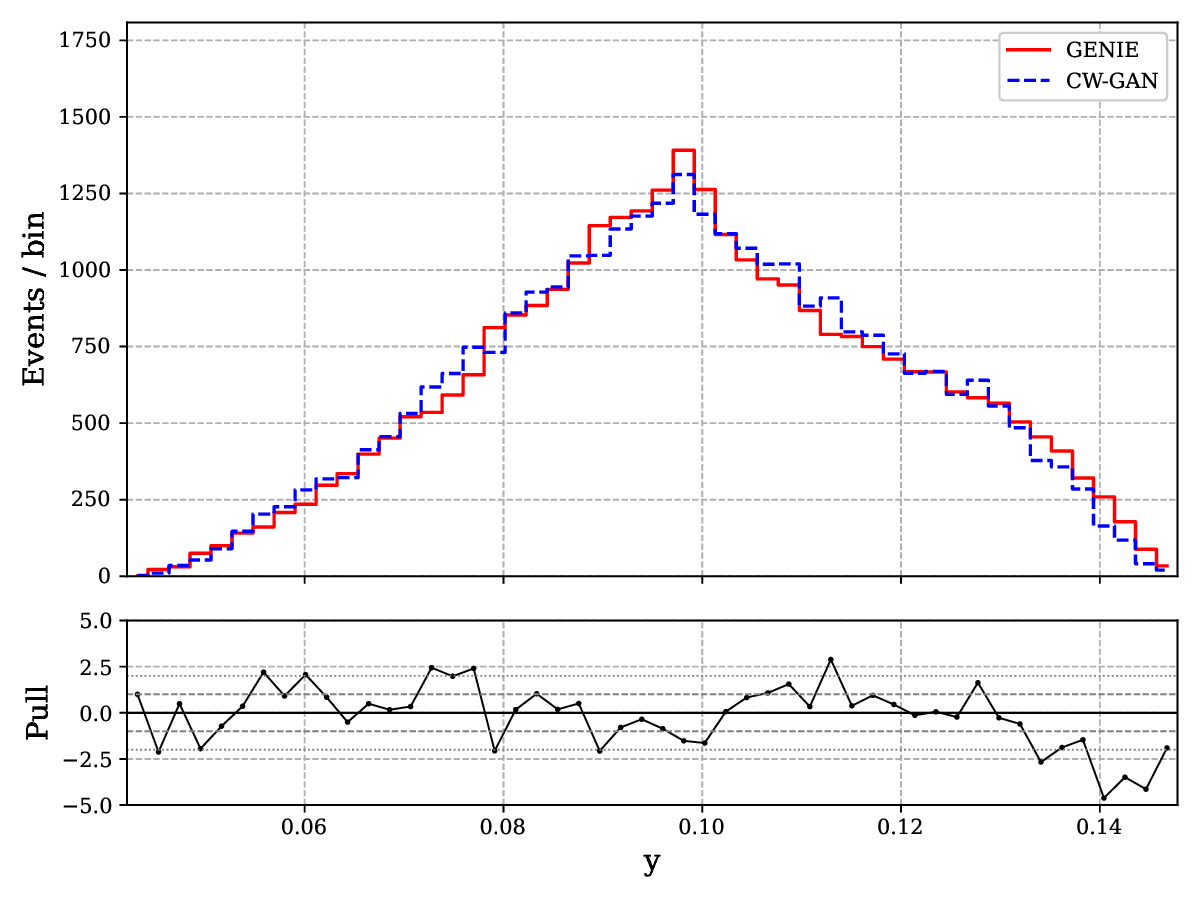} \caption{IBD-CC: y} \label{fig:ibd1d_y} \end{subfigure} \hfill
    \begin{subfigure}{0.32\textwidth} \includegraphics[width=\linewidth]{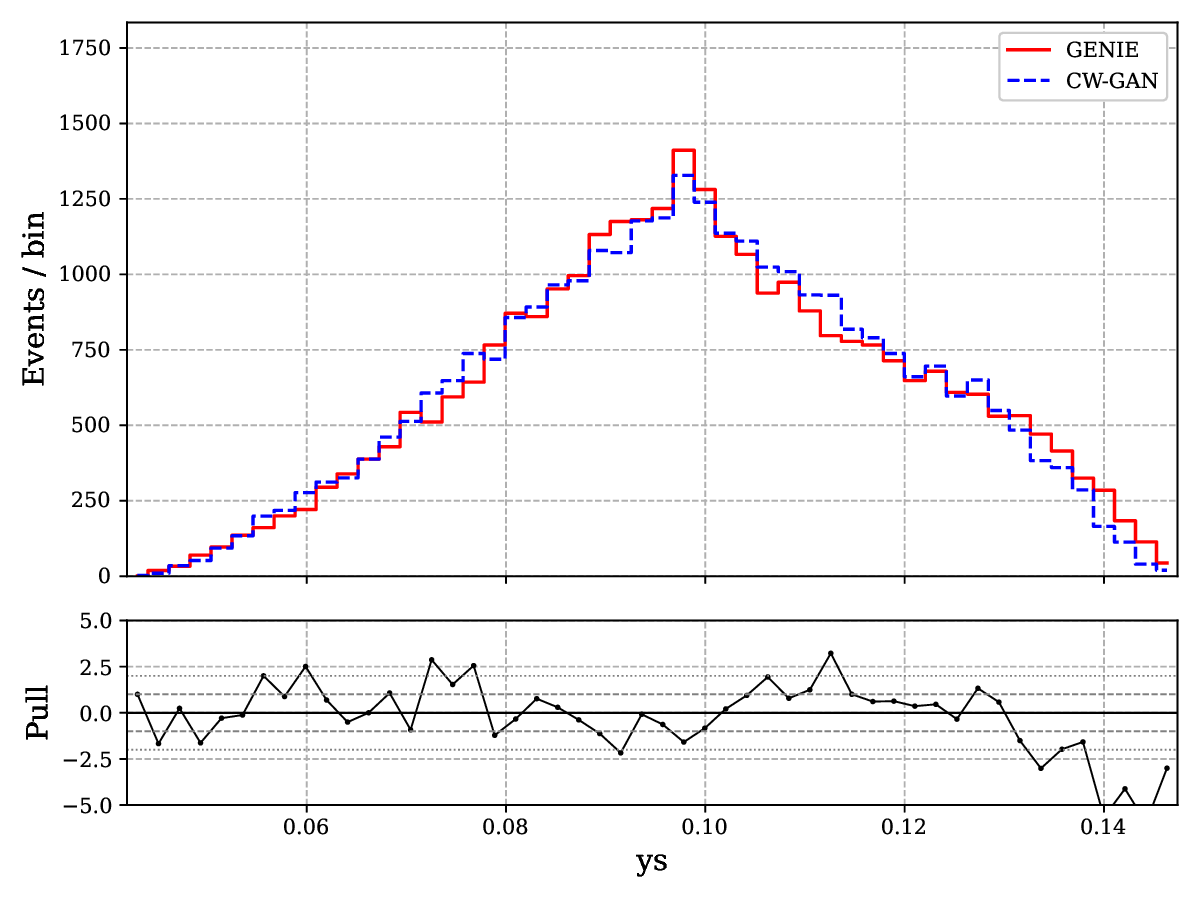} \caption{IBD-CC: ys} \label{fig:ibd1d_ys} \end{subfigure} \hfill
    
    \caption{(\subref{fig:ibd1d_Ef}) Inverse Beta Decay : 1D Marginal Distributions for the remaining 9 kinematic variables, including derived quantities ($Q^2, W, x, y$). The GAN captures the complex tail distributions and sharp features inherent to IBD interactions.}
    \label{fig:marginals_ibd_part2}
\end{figure*}

\begin{figure*}[p]
    \centering
    
    \begin{subfigure}{0.32\textwidth}
        \centering
        \includegraphics[width=\linewidth]{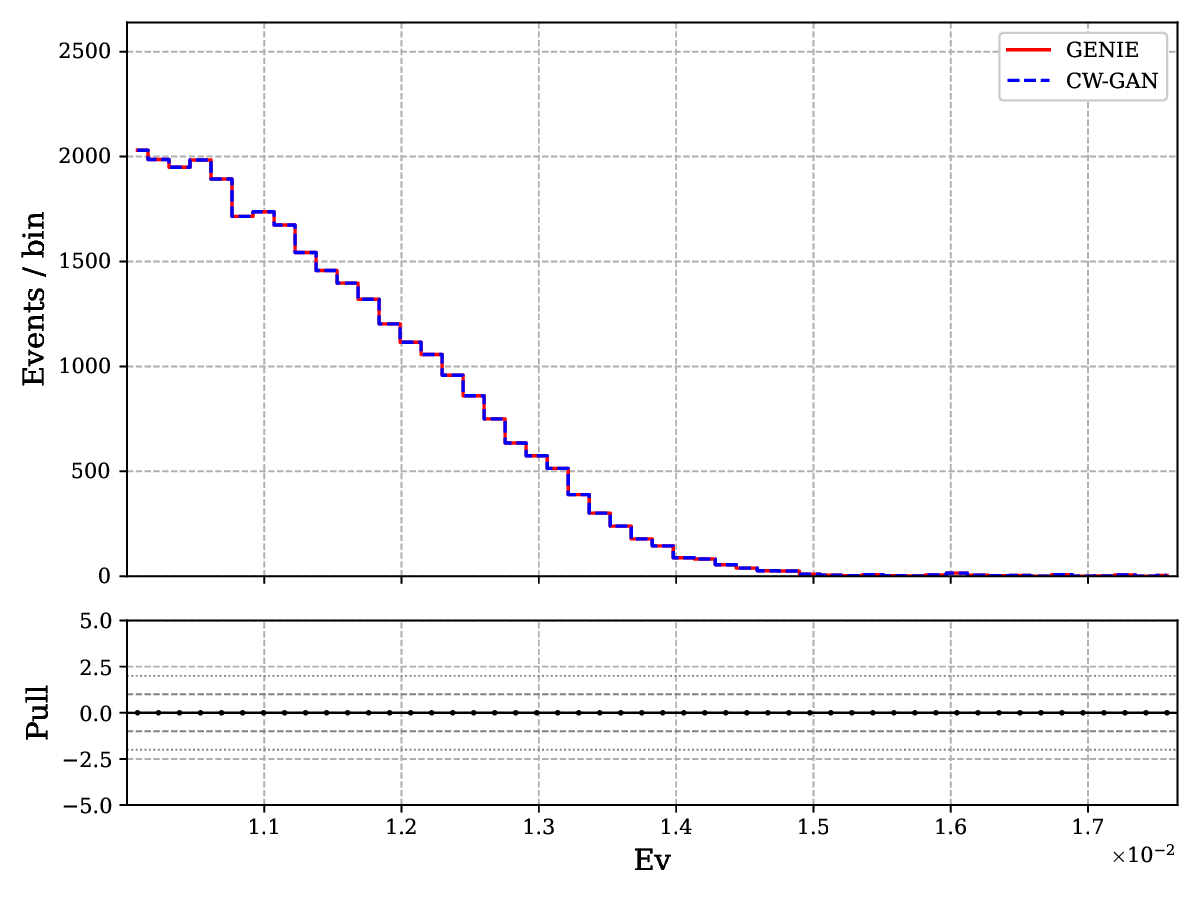}
        \caption{NC: $E_\nu$}
        \label{fig:nc-Ev}
    \end{subfigure}
    \hfill
    \begin{subfigure}{0.32\textwidth}
        \centering
        \includegraphics[width=\linewidth]{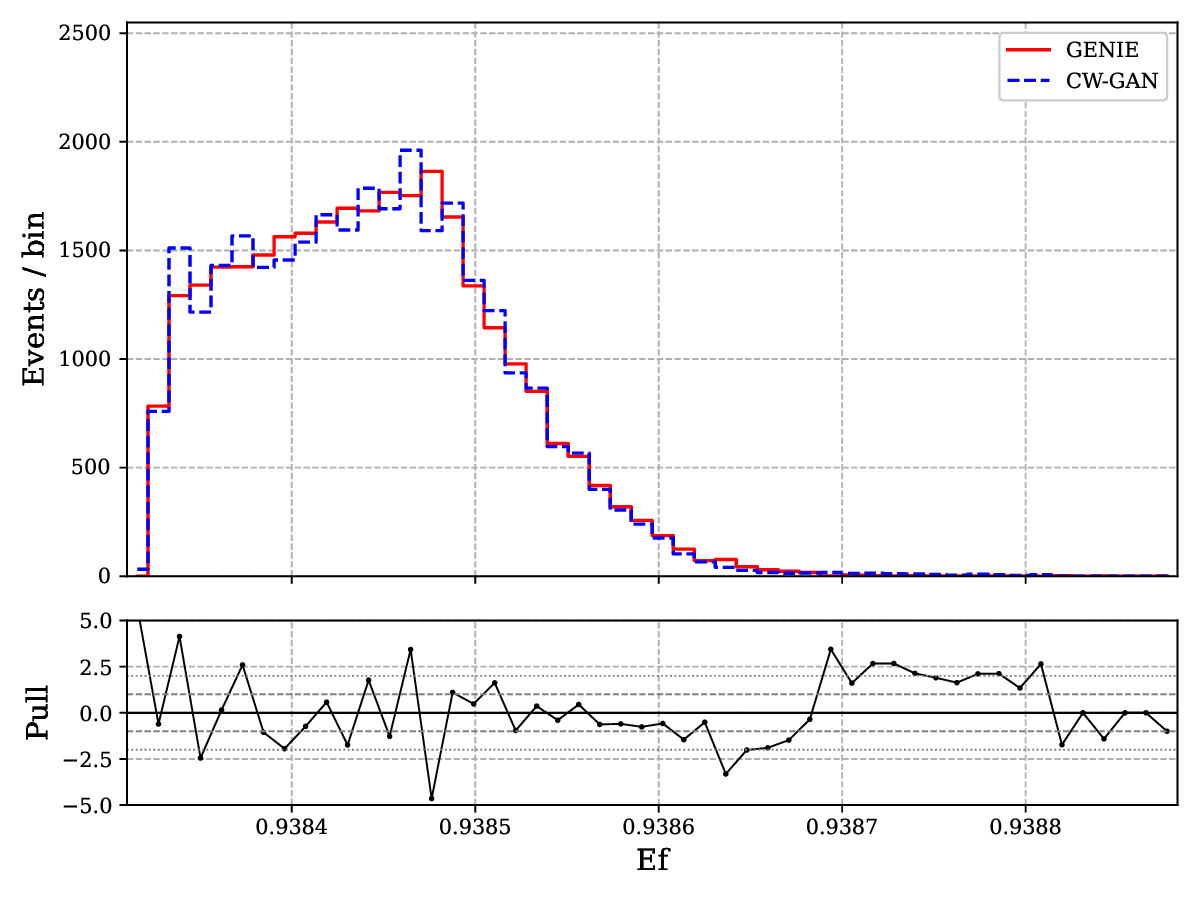}
        \caption{NC: Ef}
        \label{fig:nc-Ef}
    \end{subfigure}
    \hfill
    \begin{subfigure}{0.32\textwidth}
        \centering
        \includegraphics[width=\linewidth]{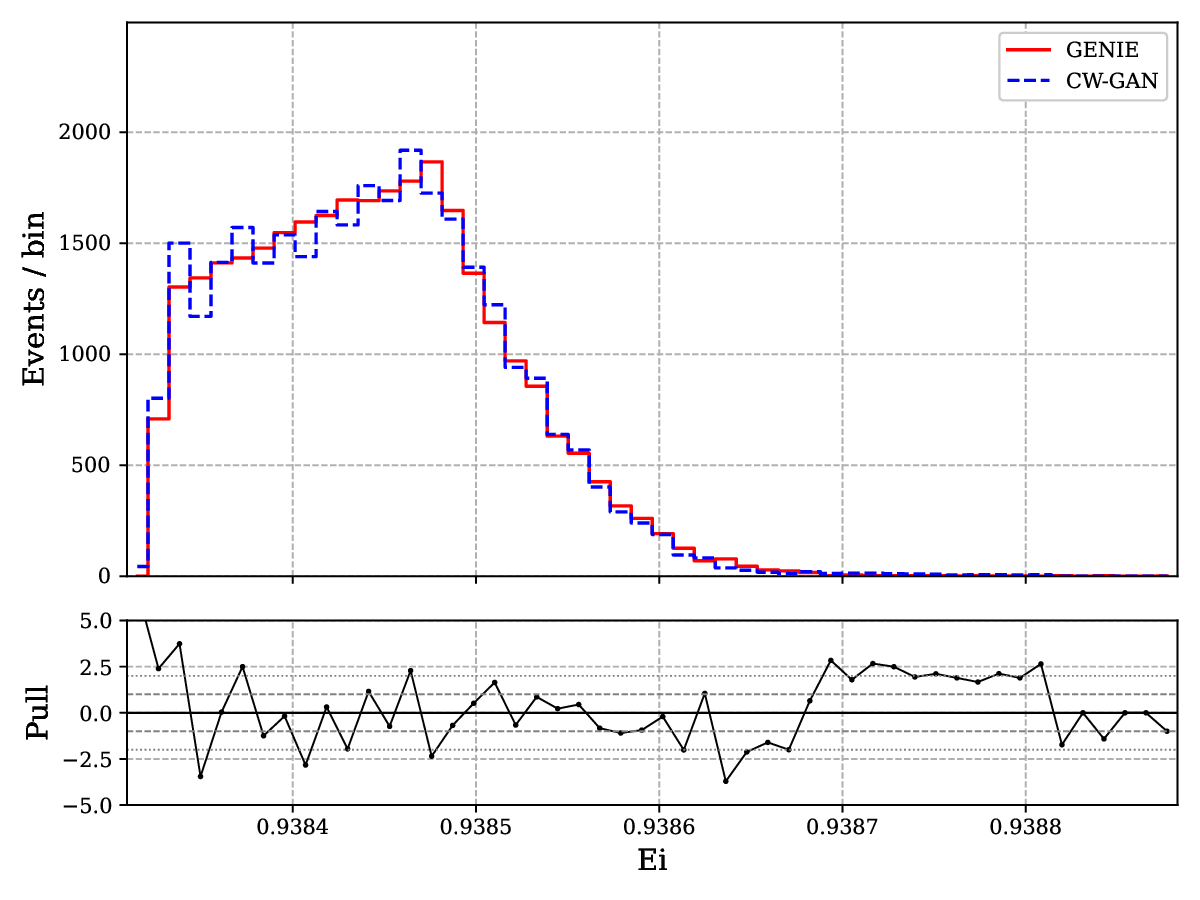}
        \caption{NC: Ei}
        \label{fig:nc-Ei}
    \end{subfigure}
    \vspace{0.3cm}
    \begin{subfigure}{0.32\textwidth}
        \centering
        \includegraphics[width=\linewidth]{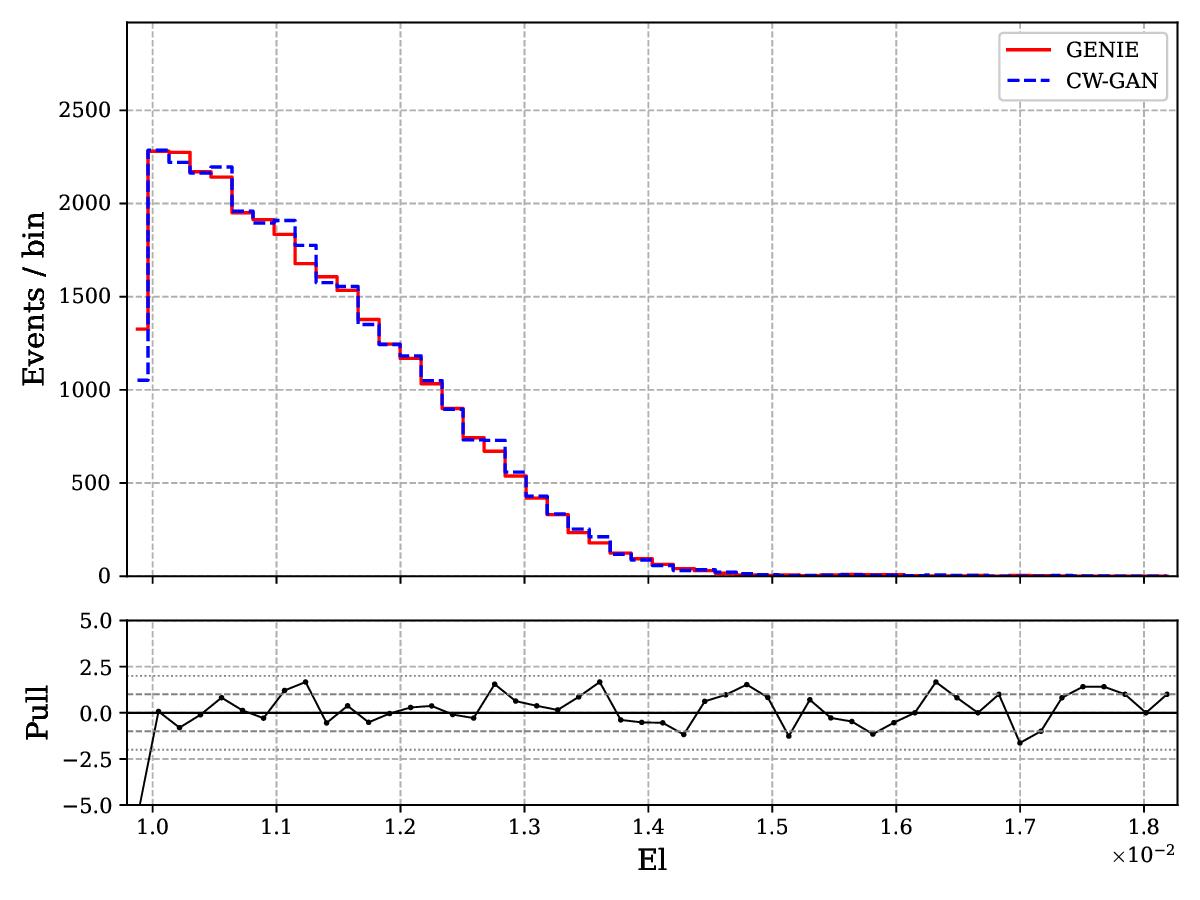}
        \caption{NC: El}
        \label{fig:nc-El}
    \end{subfigure}
    \hfill
    \begin{subfigure}{0.32\textwidth}
        \centering
        \includegraphics[width=\linewidth]{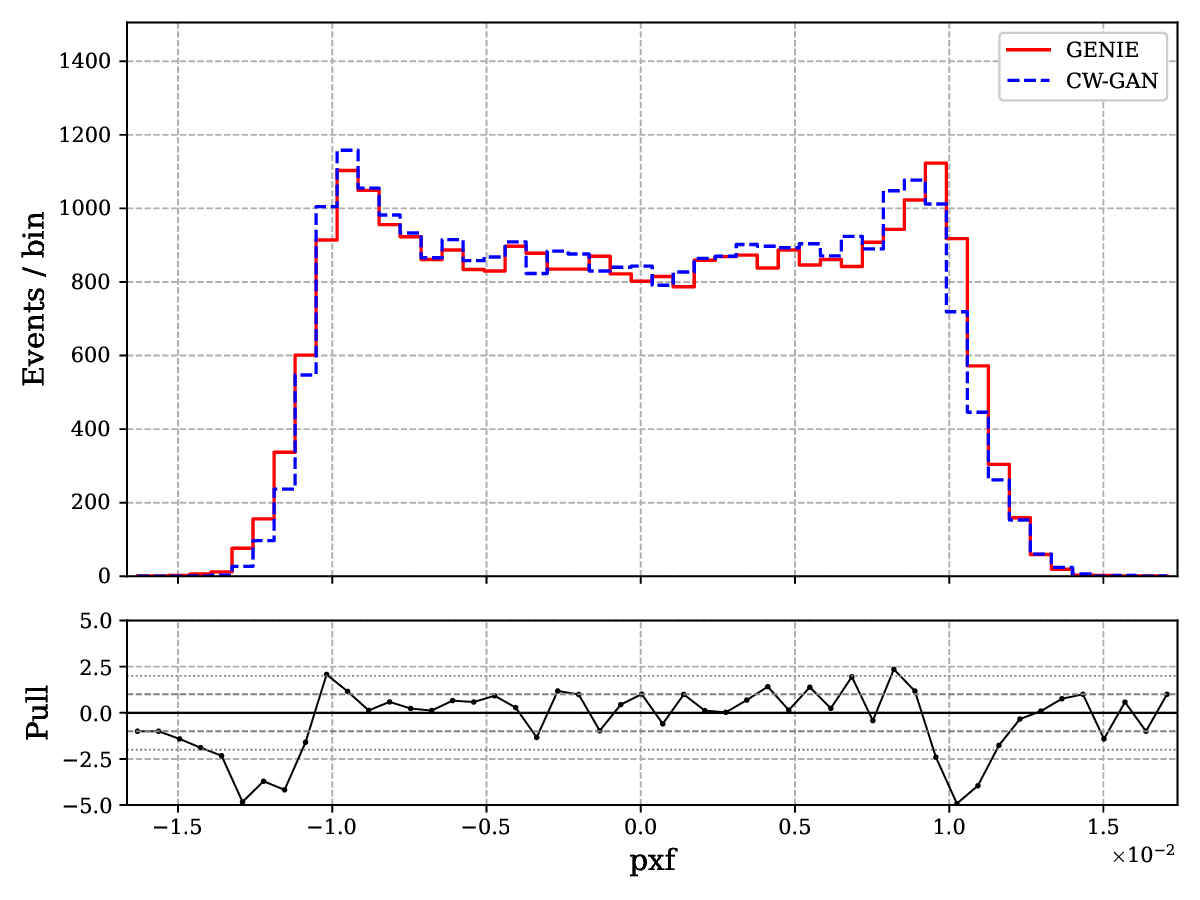}
        \caption{NC: pxf}
        \label{fig:nc-pxf}
    \end{subfigure}
    \hfill
    \begin{subfigure}{0.32\textwidth}
        \centering
        \includegraphics[width=\linewidth]{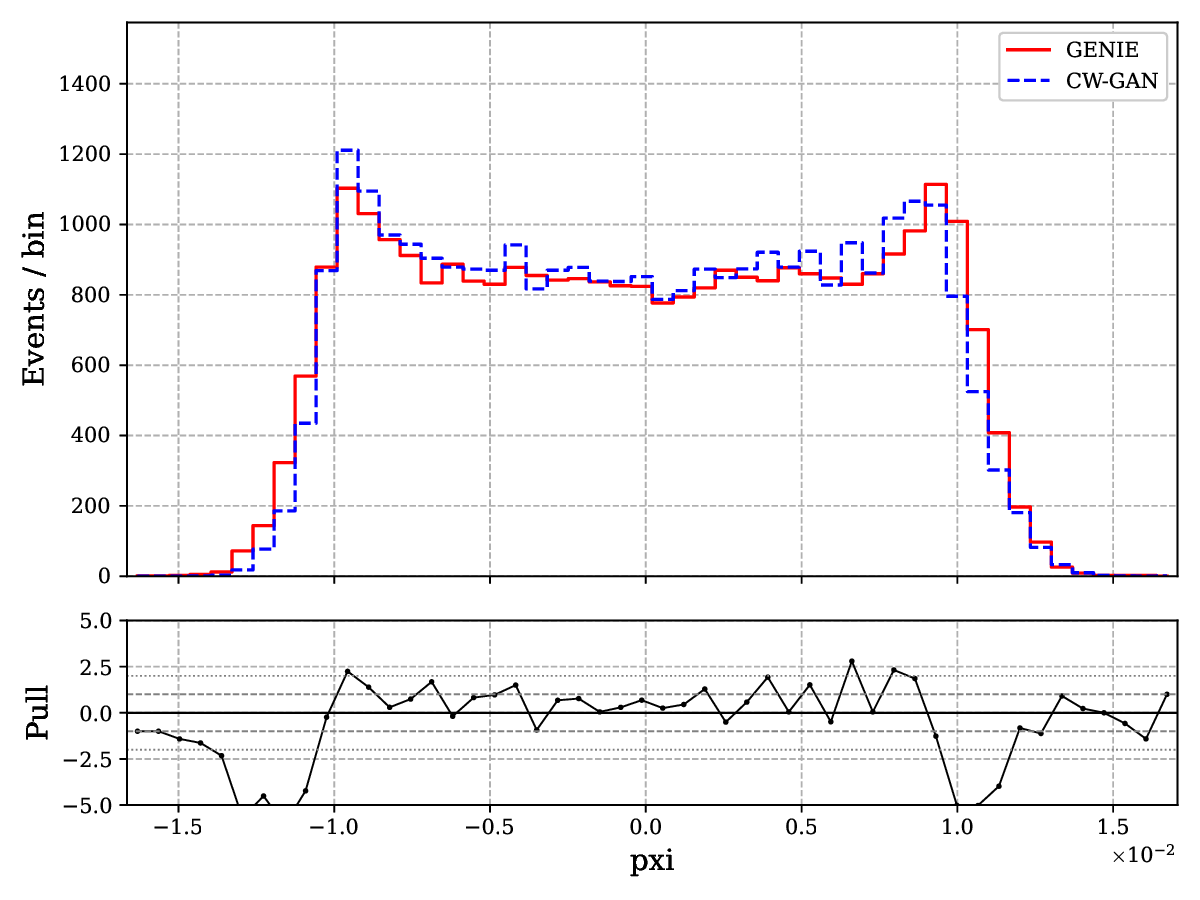}
        \caption{NC: pxi}
        \label{fig:nc-pxi}
    \end{subfigure}
    \vspace{0.3cm}
    \begin{subfigure}{0.32\textwidth}
        \centering
        \includegraphics[width=\linewidth]{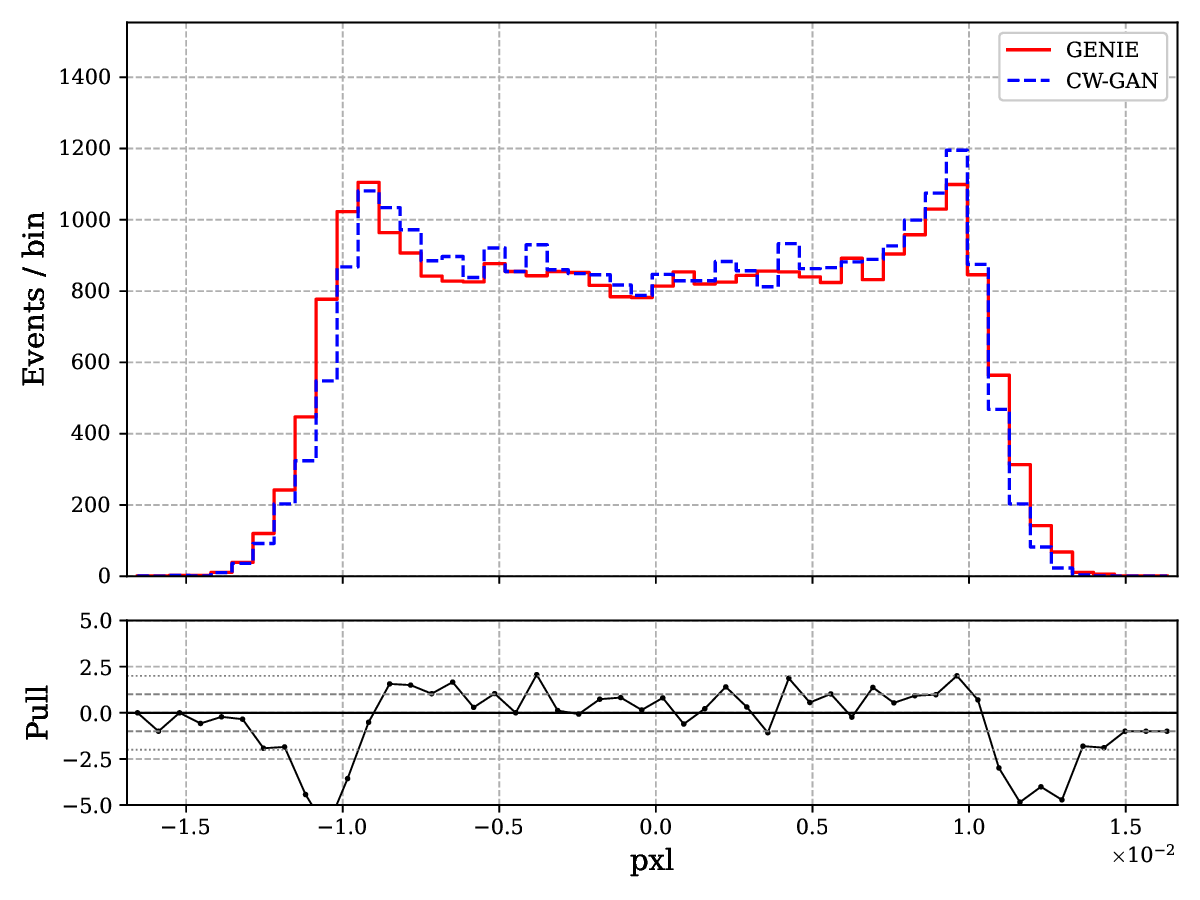}
        \caption{NC: pxl}
        \label{fig:nc-pxl}
    \end{subfigure}
    \hfill
    \begin{subfigure}{0.32\textwidth}
        \centering
        \includegraphics[width=\linewidth]{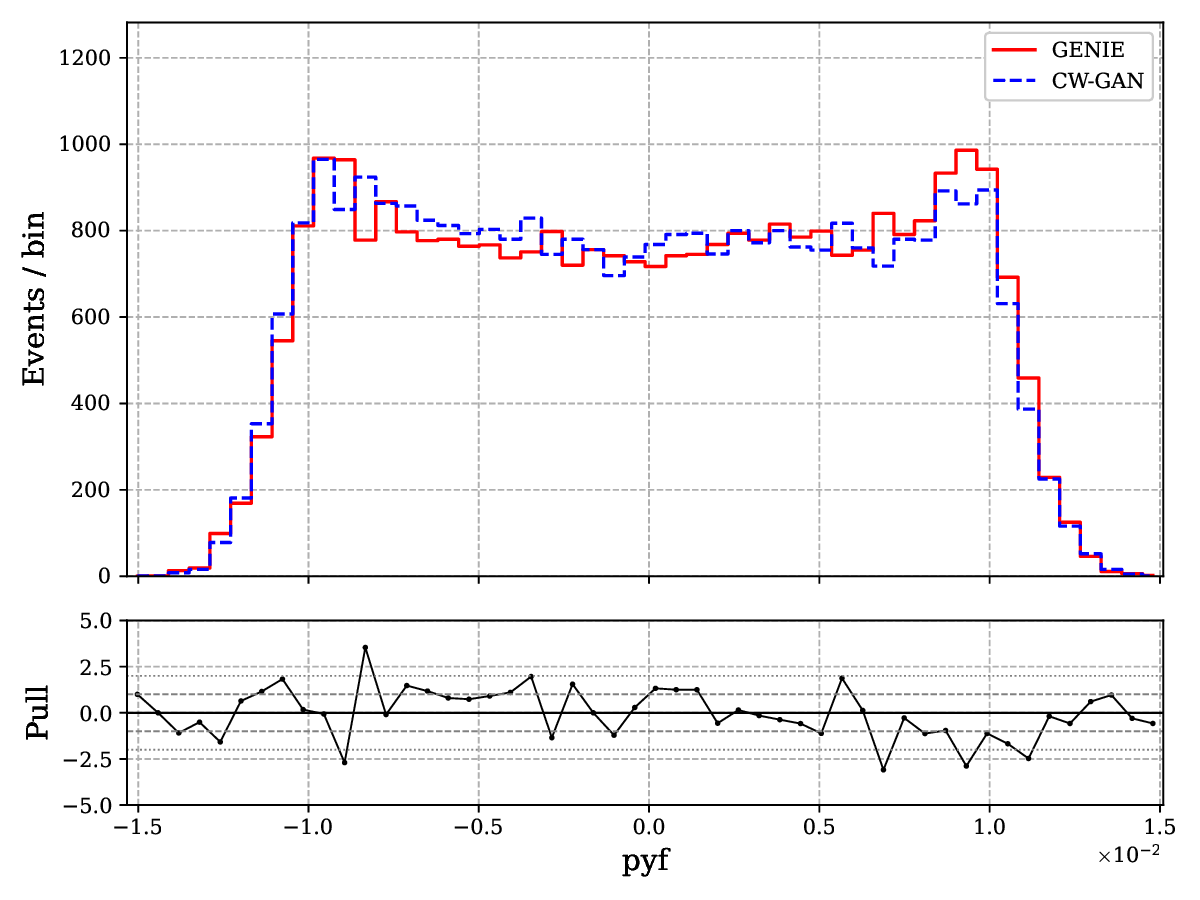}
        \caption{NC: pyf}
        \label{fig:nc-pyf}
    \end{subfigure}
    \hfill
    \begin{subfigure}{0.32\textwidth}
        \centering
        \includegraphics[width=\linewidth]{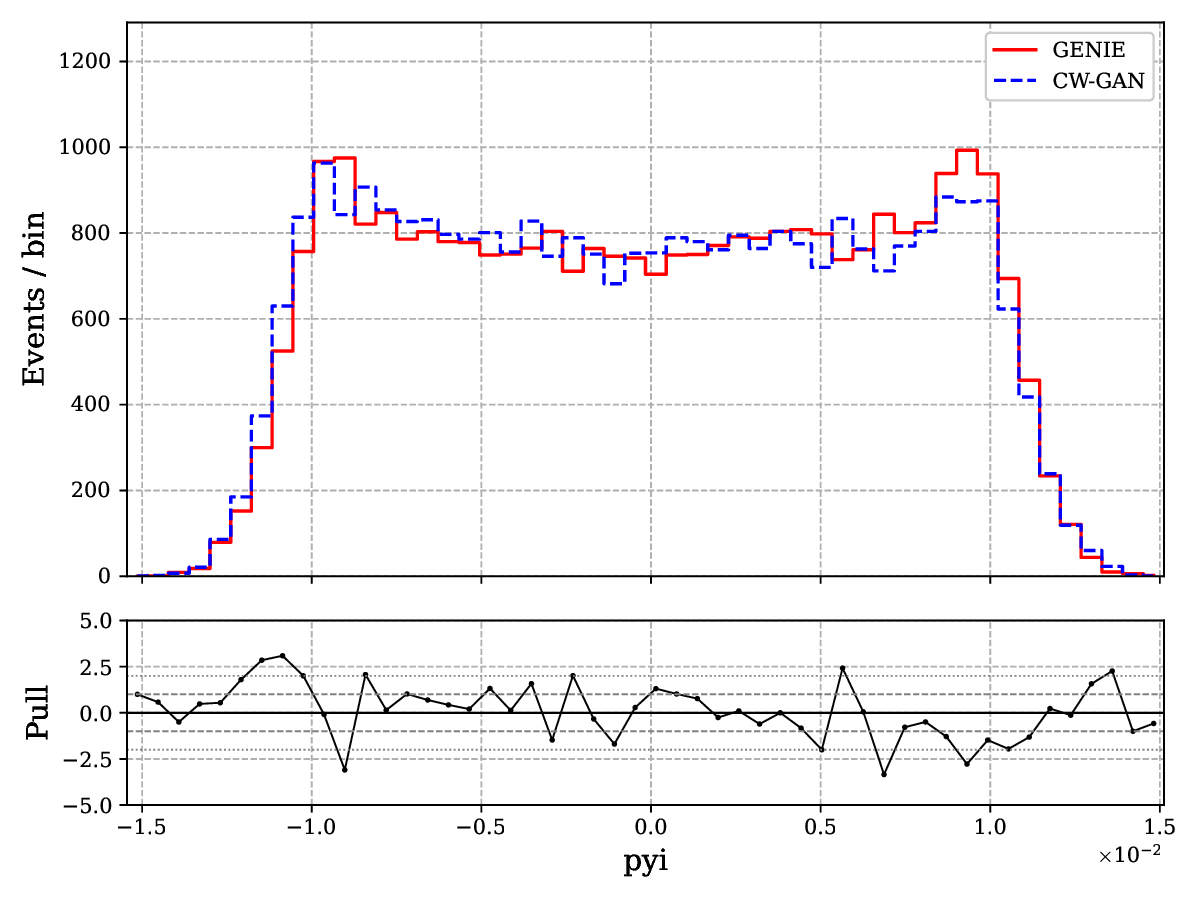}
        \caption{NC: pyi}
        \label{fig:nc-pyi}
        
    \end{subfigure}
    
    \caption{(\subref{fig:nc-Ev})Neutral Current: 1D Marginal Distributions for the first 9 kinematic variables. Agreement between generated (dashed) and truth (solid) data confirms the model correctly learns the momentum conservation laws governing NC interactions.}

    \label{fig:nc_1d_1}
\end{figure*}

\begin{figure*}[p]
    \ContinuedFloat
    \centering
    
    \begin{subfigure}{0.32\textwidth}
        \centering
        \includegraphics[width=\linewidth]{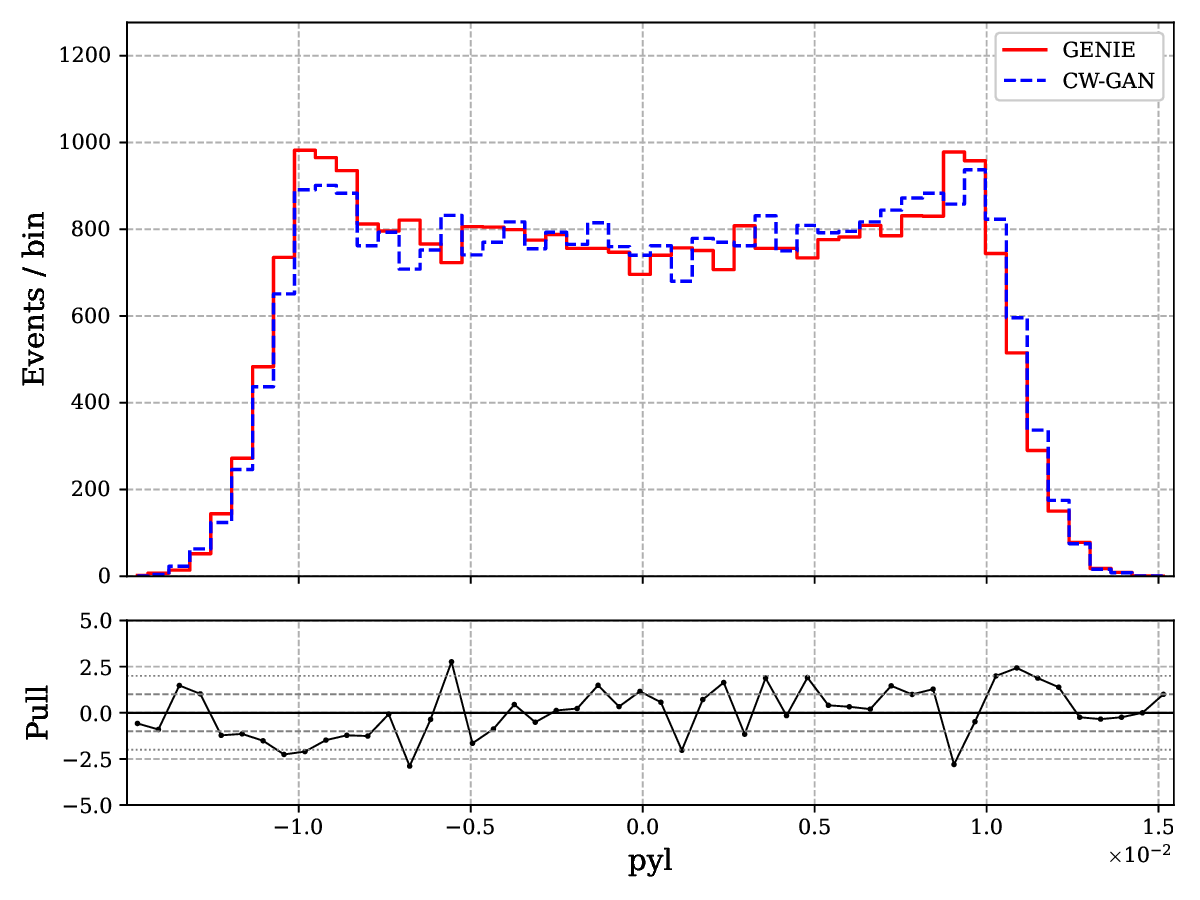}
        \caption{NC: pyl}
        \label{fig:nc-pyl}

    \end{subfigure}
    \hfill
    \begin{subfigure}{0.32\textwidth}
        \centering
        \includegraphics[width=\linewidth]{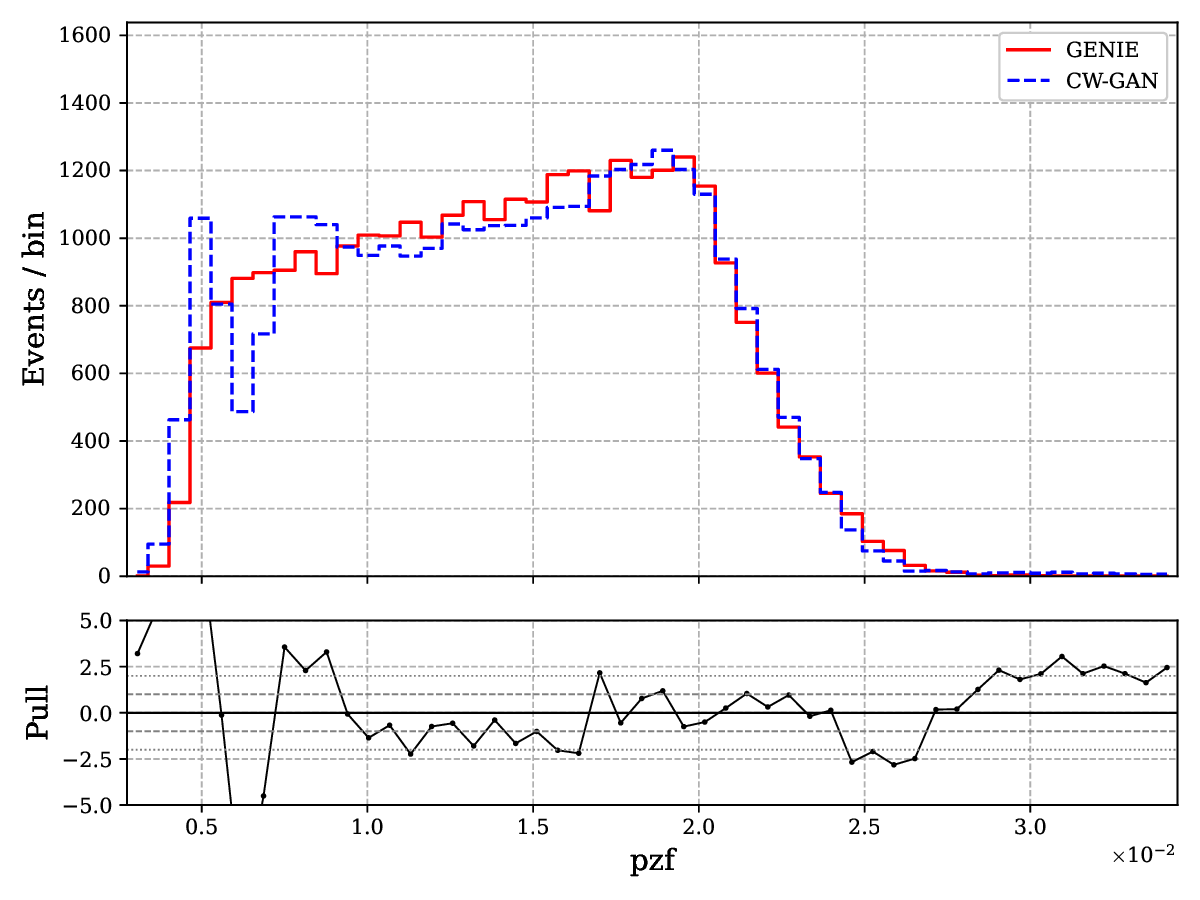}
        \caption{NC:pzf}
        \label{fig:nc-pzf}

    \end{subfigure}
    \hfill
    \begin{subfigure}{0.32\textwidth}
        \centering
        \includegraphics[width=\linewidth]{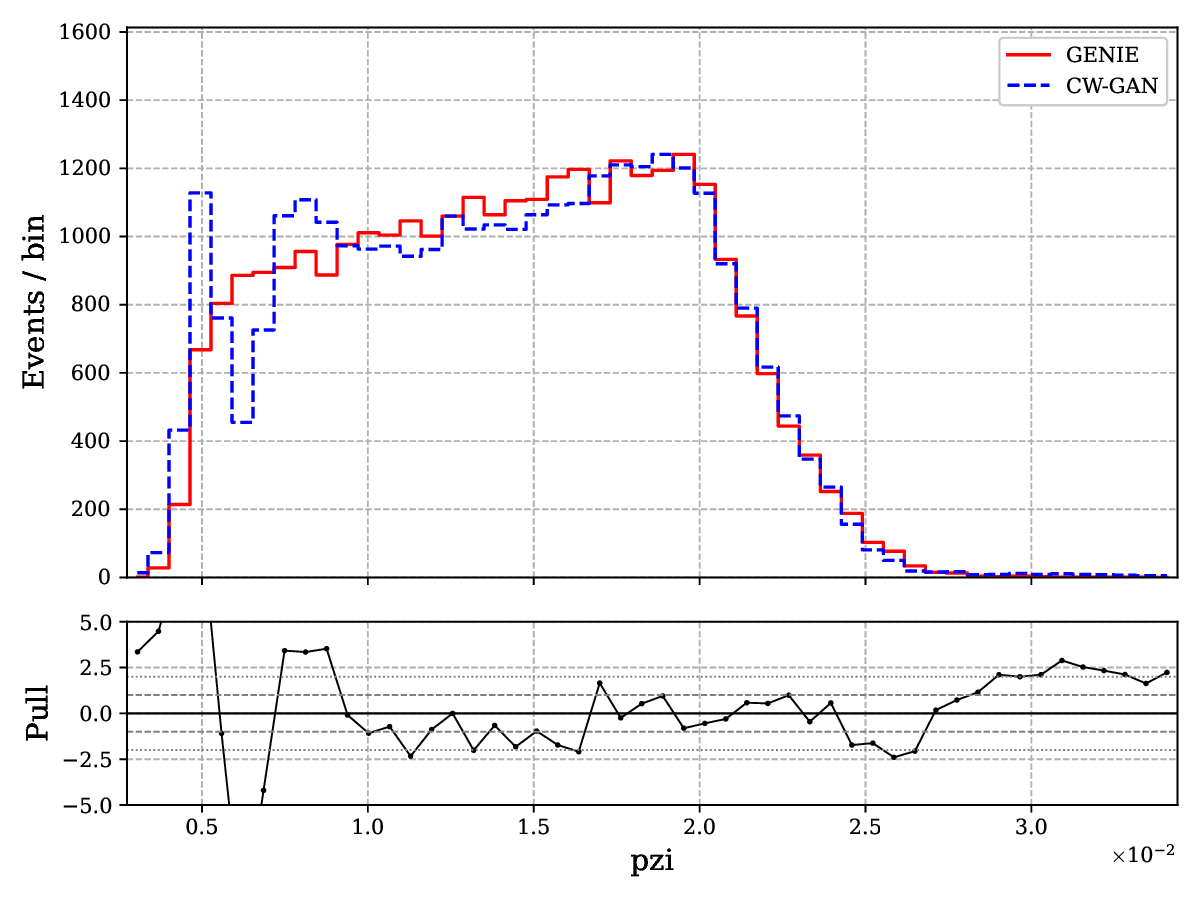}
        \caption{NC: pzi}
        \label{fig:nc-pzi}

    \end{subfigure}
    \vspace{0.3cm}
    \begin{subfigure}{0.32\textwidth}
        \centering
        \includegraphics[width=\linewidth]{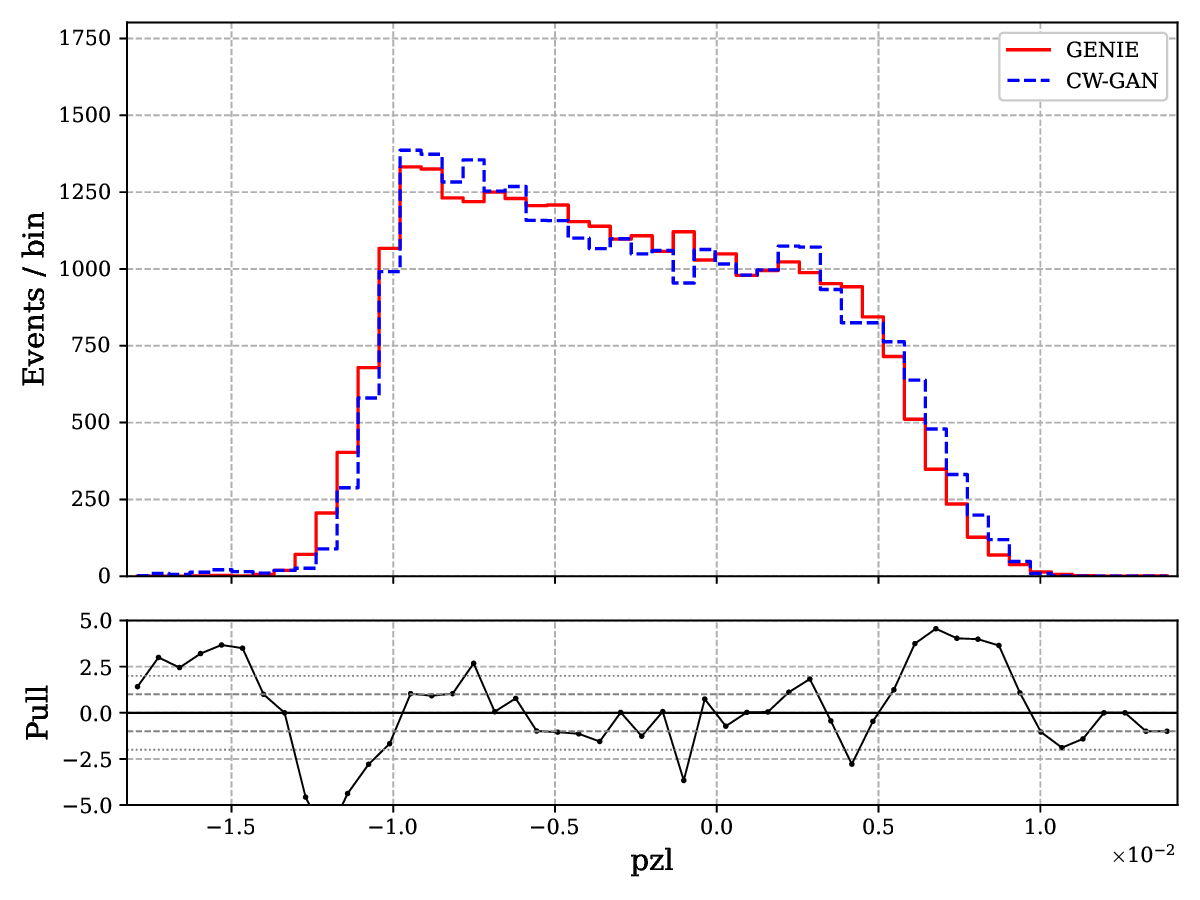}
        \caption{NC: pzl}
        \label{fig:nc-pzl}

    \end{subfigure}
    \hfill
    \begin{subfigure}{0.32\textwidth}
        \centering
        \includegraphics[width=\linewidth]{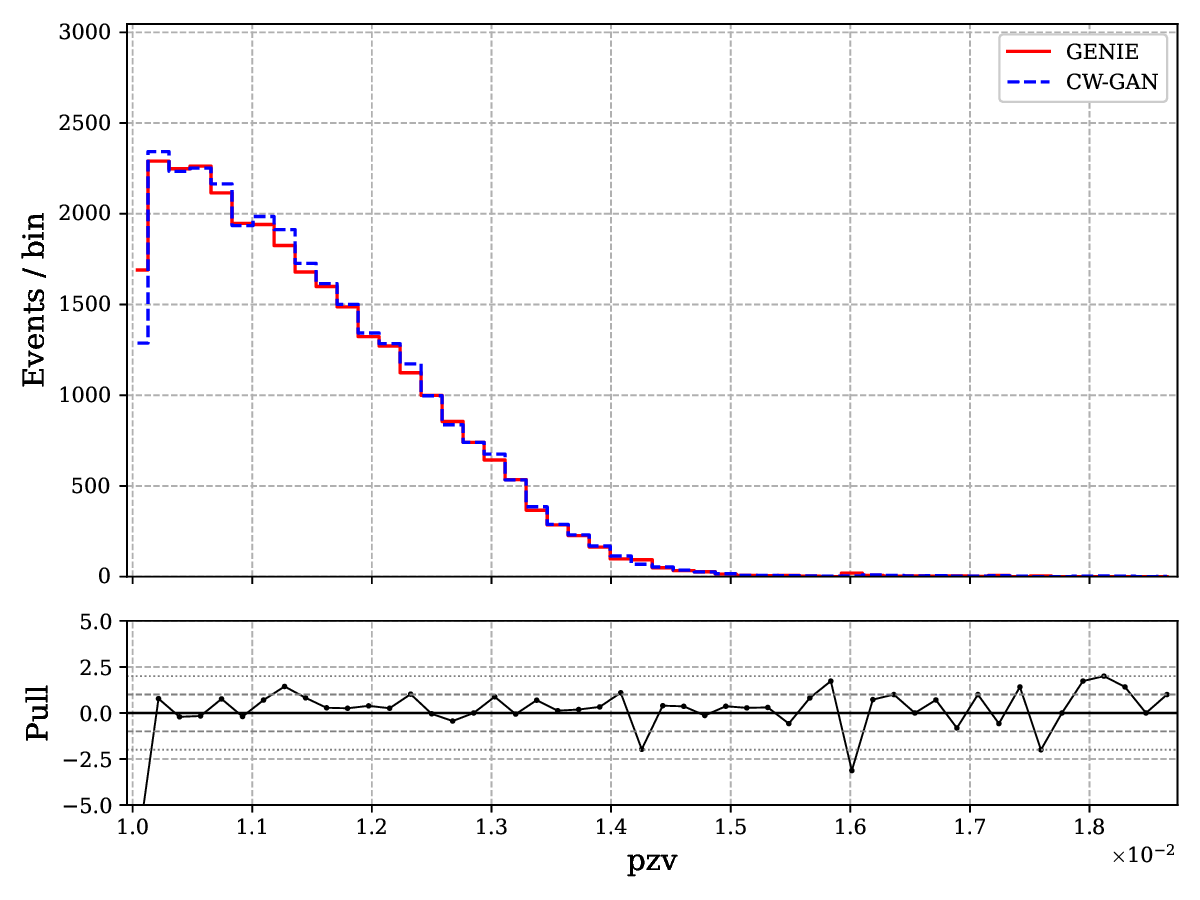}
        \caption{NC: pzv}
        \label{fig:nc-pzv}

    \end{subfigure}
    \hfill
    \begin{subfigure}{0.32\textwidth}
        \centering
        \includegraphics[width=\linewidth]{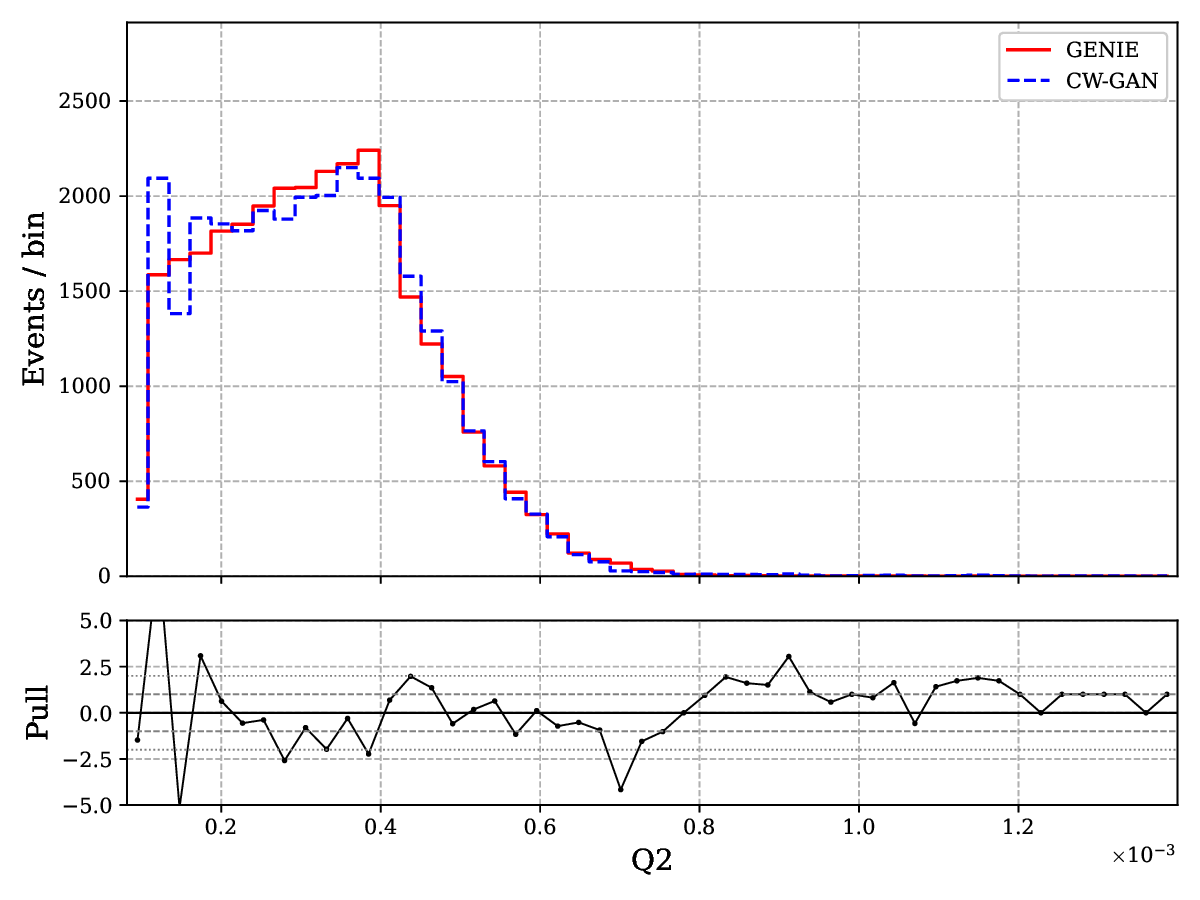}
        \caption{$Q^2$}
        \label{fig:nc-$Q^2$}

    \end{subfigure}
    \vspace{0.3cm}
    \begin{subfigure}{0.32\textwidth}
        \centering
        \includegraphics[width=\linewidth]{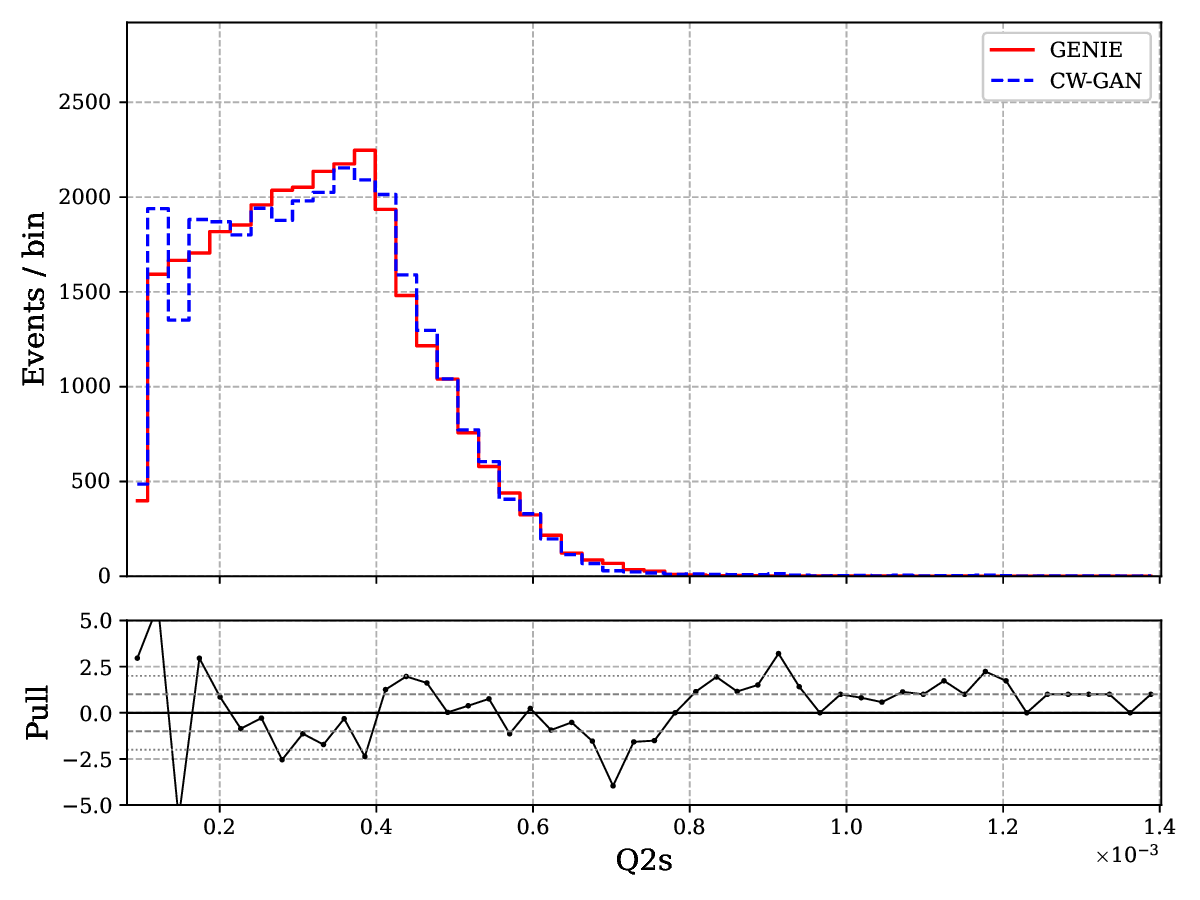}
        \caption{NC: $Q^2s$}
        \label{fig:nc-$Q^2s$}

    \end{subfigure}
    \hfill
    \begin{subfigure}{0.32\textwidth}
        \centering
        \includegraphics[width=\linewidth]{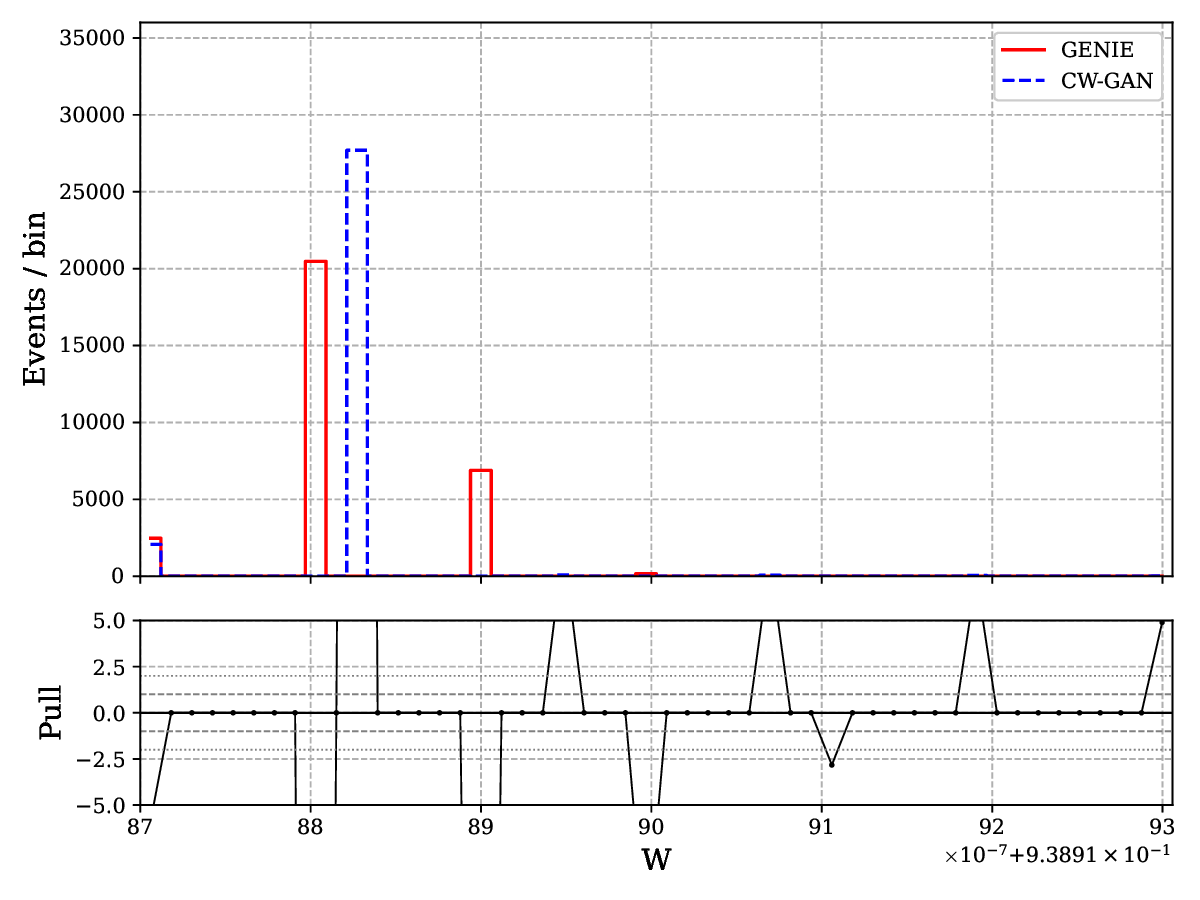}
        \caption{NC: W}
        \label{fig:nc-W}
    \end{subfigure}
    \hfill
    \begin{subfigure}{0.32\textwidth}
        \centering
        \includegraphics[width=\linewidth]{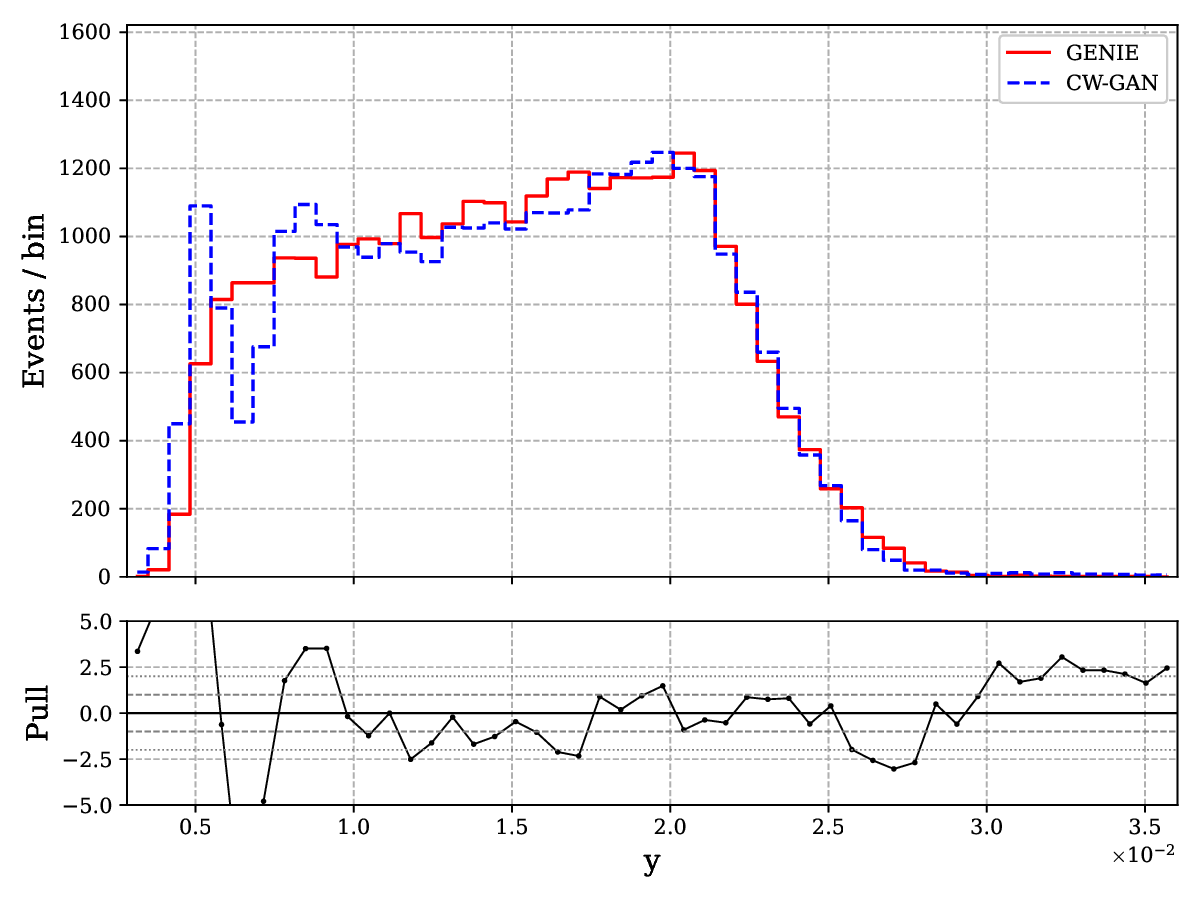}
        \caption{NC: y}
        \label{fig:nc-y}
    \end{subfigure}
    \hfill
    \begin{subfigure}{0.32\textwidth}
        \centering
        \includegraphics[width=\linewidth]{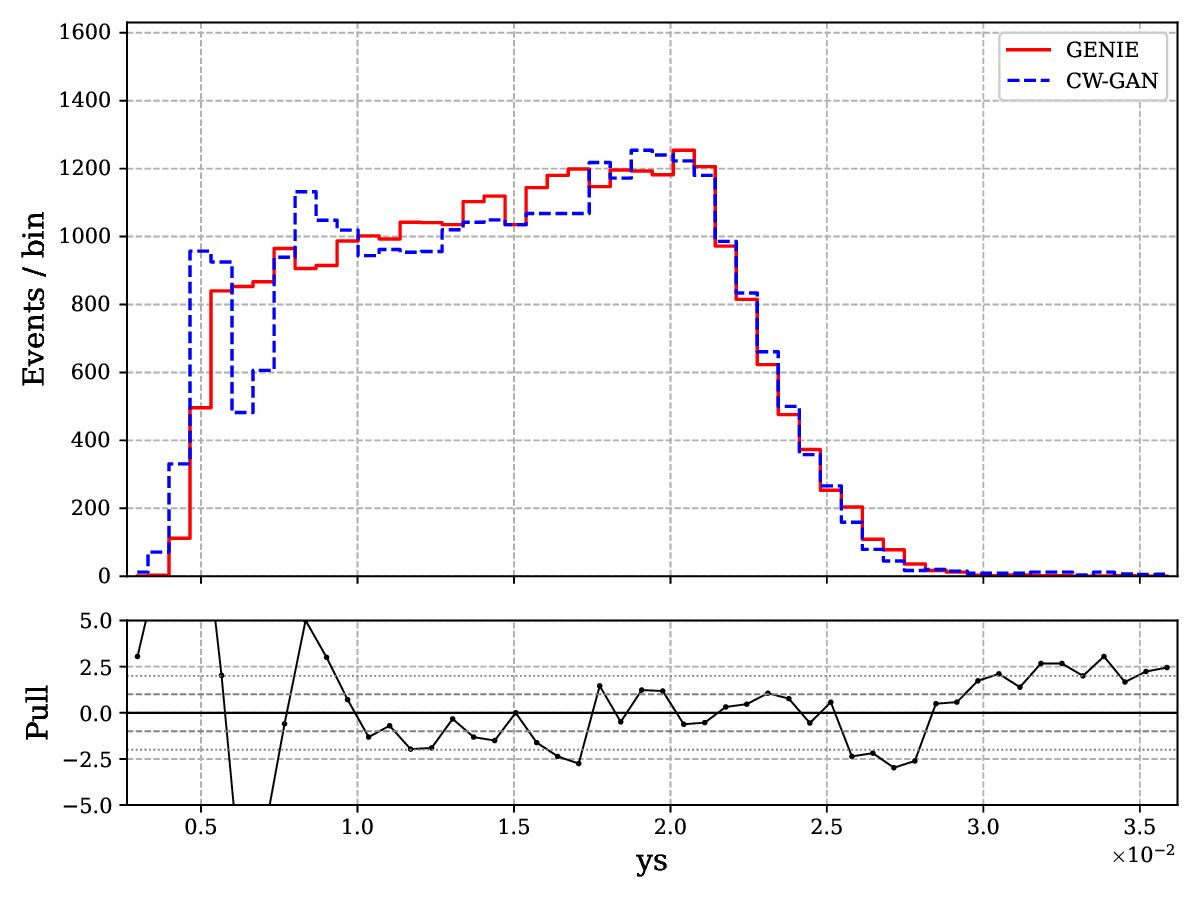}
        \caption{NC: ys}
        \label{fig:nc-ys}
        
    \end{subfigure}
    
    \caption{(\subref{fig:nc-Ef}) Neutral Current: 1D Marginal Distributions for the remaining 9 kinematic variables. The minimal residuals in $Q^2$, $W$, and $y$ indicate the generator has learned the underlying physics governing the neutral current cross-section.}
    \label{fig:nc_1d_2}
\end{figure*}

\endgroup  

\begin{figure*}[p]
\centering

\begin{subfigure}{\textwidth}
    \centering
    \includegraphics[width=0.7\textwidth]{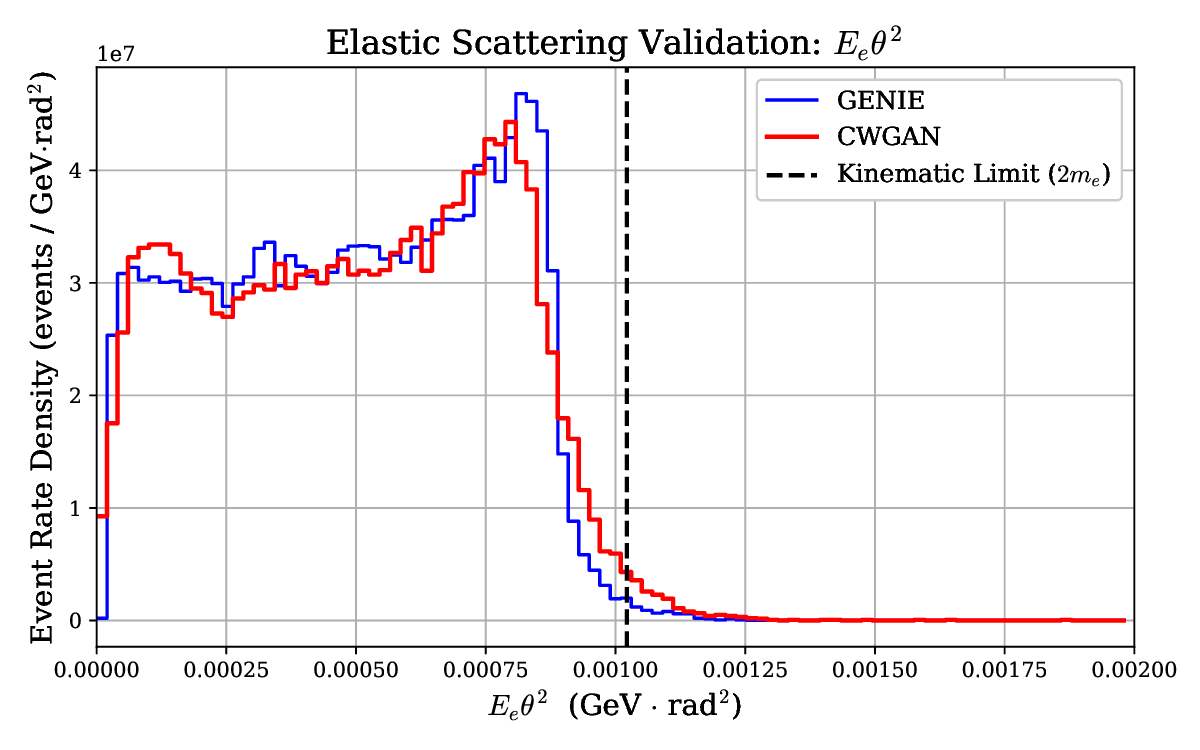}
    \caption{NuEElastic: $E_e \theta^2$ Limit}
    \label{fig:elastic}
\end{subfigure}

\begin{subfigure}{\textwidth}
    \centering
    \includegraphics[width=0.7\textwidth]{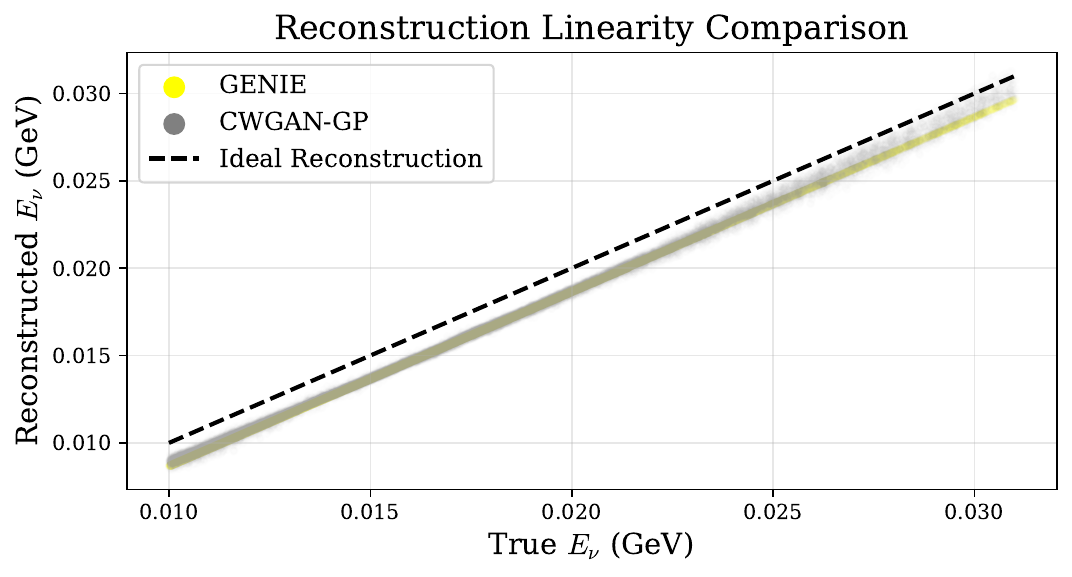}
    \caption{IBD: $E_\nu$ Reconstruction}
    \label{fig:ibd}
\end{subfigure}

\begin{subfigure}{\textwidth}
    \centering
    \includegraphics[width=0.7\textwidth]{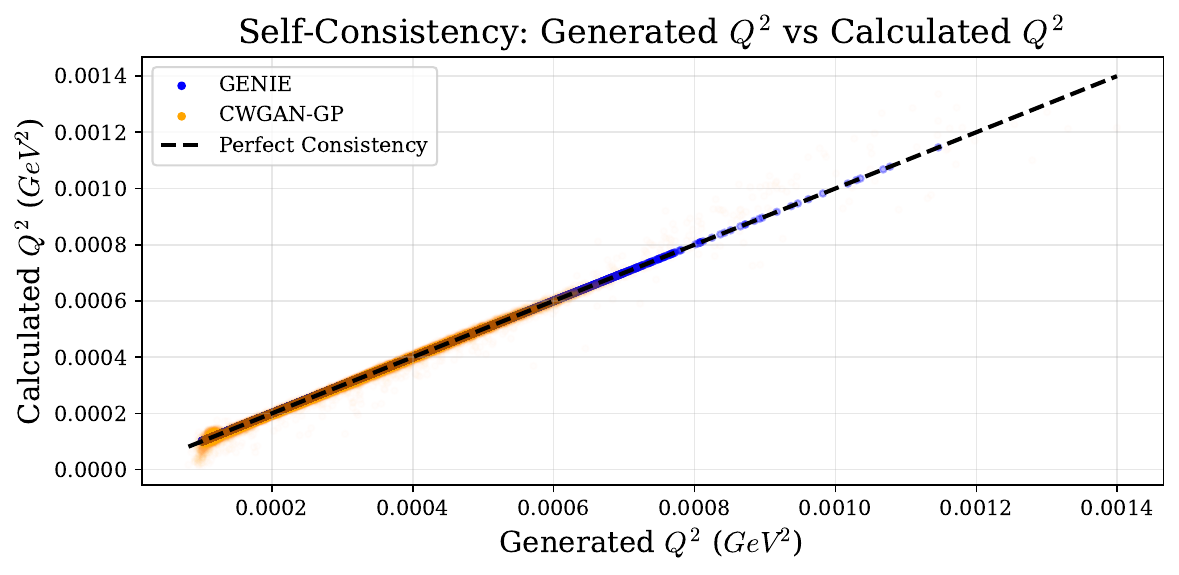}
    \caption{NC: $Q^2$ Consistency}
    \label{fig:nc}
\end{subfigure}

\caption{Physics Signature Tests. (\subref{fig:elastic}) Validation of the forward scattering limit ($E_e\theta^2$) in Elastic Scattering (NuEElastic). (\subref{fig:ibd}) Inverse Beta Decay (IBD-CC) linearity check showing the correlation between true and reconstructed neutrino energy ($E_\nu$). (\subref{fig:nc}) Verification of $Q^2$ conservation in Neutral Current events.}
\label{fig:physics_sig}
\end{figure*}

\subsection{ Physics Signatures }
To validate that the model is capturing the underlying physics rather than merely memorizing statistical patterns, we evaluate three interaction-specific signatures derived from fundamental conservation laws. 
\\ \\
In neutrino-electron elastic scattering, the kinematics are constrained by the conservation of four-momentum. For a neutrino of energy $E_\nu$ scattering off a stationary electron (mass $m_e$), the relationship between the electron's scattering angle $\theta$ and its energy $E_e$ is given by\cite{PhysRevD.100.092001}:
$$E_\nu = \frac{m_e E_e}{m_e - E_e (1 - \cos \theta)}$$In the high-energy limit ($E_\nu \gg m_e$) and small angle approximation, this imposes a strict upper bound on the quantity $E_e \theta^2$:

$$E_e \theta^2 \le 2m_e$$

where $m_e \approx 0.511$ MeV. This is a ``hard'' physical wall; no event can physically exist beyond this limit.
\\ \\
Figure \ref{fig:physics_sig}(\subref{fig:elastic}) shows the distribution of $E_e \theta^2$ for the generated events. The GENIE truth (blue) shows a sharp cutoff at $2m_e \approx 1.02 \times 10^{-3}$ GeV rad$^2$. The CW-GAN generated events (red) rigorously respect this boundary. The density of events drops to zero exactly at the physical limit. This is a profound result as the network was not explicitly programmed with either the electron mass or the scattering formula, yet it learned to define the boundary of the allowed phase space with absolute precision. This confirms the model's utility for background rejection studies, where the $E\theta^2$ cut is the primary discriminator between signal and background.
\\ \\
For Inverse Beta Decay, the incident neutrino energy is reconstructed from the visible positron energy and the scattering angle. The standard reconstruction formula is \cite{Munteanu_2020,Ricciardi_2022}:
$$E_\nu^{rec} \approx E_e + \frac{M_n^2 - M_p^2 - m_e^2}{2M_p}$$
This formula approximates the true energy, ignoring the small recoil kinetic energy of the neutron.

We compare the true conditioning energy $E_\nu$ (input to the CW-GAN) with the reconstructed energy $E_\nu^{rec}$ (calculated from the CW-GAN output). The result is a linear distribution along the diagonal(\ref{fig:physics_sig}(\subref{fig:ibd})). Crucially, the width of the distribution represents the intrinsic physical resolution of the IBD process (Fermi motion and recoil effects). The GAN accurately reproduces this width, meaning it correctly models the resolution limits of the IBD channel. This validates the model for use in energy spectrum unfolding analyses in reactor neutrino experiments.

In Neutral Current scattering, the momentum transfer squared $Q^2$ can be calculated in two independent ways: (1) The variable $Q^2$ generated directly by the network as a high-level feature. (2) Derived from the generated proton kinetic energy $T_p$ using the elastic scattering relation:
$$Q_{rec}^2 = 2 M_p T_p$$
where $M_p$ is the proton mass and $T_p$ is the proton energy.
\\ \\
Figure ~\ref{fig:physics_sig}(\subref{fig:nc}) compares the generated $Q^2$ against the value reconstructed solely from proton kinetic energy ($Q^2_{rec}$). In neutral current events ($\nu+p \to \nu+p$), the invisible outgoing neutrino dictates that momentum transfer must strictly equal the proton recoil ($Q^2 = 2M_p T_p$). The observed perfect diagonal confirms the CW-GAN has mastered this internal conservation law, accurately balancing the visible proton momentum against the unobservable neutrino.

\subsection{Joint Correlations (2D Maps)}
While 1D marginal distributions verify that the generator captures individual variable densities, they are insufficient for validating the physical coherence of the simulation. To confirm that the CW-GAN has learned the underlying conservation laws and kinematic constraints of the Standard Model, we analyze specific 2D projections of the phase space. 
\\
\newline
While 1D histograms ensure that individual variables are modeled correctly, they are insufficient for verifying high-dimensional correlations. To validate that the CW-GAN captures the multidimensional structure of the phase space, we analyze specific 2D projections chosen to probe the underlying kinematics of each interaction type.

\subsubsection{Elastic Scattering (NuEElastic) (6 Pairs)}
For $\nu_e - e^-$ elastic scattering, the correlations are strictly governed by the electron-neutrino scattering kinematics. We evaluated the 6 key pairs shown in Fig~\ref{fig: NuEElastic_corr1}(\subref{fig:NuEElastic2D_el_vs_ev}).

\textbf{Kinematic Boundaries:} The pairs $(E\nu, E_l)$, $(E_l, Q^2)$, and $(\nu, Q^2)$ verify the energy transfer logic. The recoil electron energy is constrained by the incident neutrino energy and the scattering angle $\theta$:
\begin{equation}
    E_l \theta^2 \le 2m_e
\end{equation}
This imposes a strict boundary in the $(E_l, Q^2)$ plane, as $Q^2 \approx 2m_e T_e$ \cite{Marciano_2003}. The CW-GAN ensures $E_l <E_\nu$ and respects these momentum transfer limits.

\textbf{Transverse Isotropy \& Longitudinal Boost:} The pair $(p_{x}^l, p_{y}^l)$ confirms the rotational symmetry of the outgoing electron in the transverse plane, while $(p_{z}^\nu, p_{z}^l)$ tests the forward-scattering nature of the interaction.

\textbf{Inelasticity:} The correlation $(Q^2, y)$ validates the relationship between momentum transfer and the fraction of energy transferred to the lepton, governed by $y = T_e / E_\nu$. 

The model maintains the physics based relations of the above pairs. 

\subsubsection{Inverse Beta Decay (7 Pairs)}
For IBD, validation requires checking the energy sharing between the positron and the neutron. We evaluated 7 pairs shown in Fig~\ref{fig: ibd-cc_corr1}. The resulting distributions align with ideal theoretical expectations, confirming that the model has successfully learned the underlying conservation laws and interaction dynamics of Inverse Beta Decay (IBD-CC).
\\
\newline
\textbf{Energy Conservation:} The pairs $(\nu, E_l)$ and $(E_l, E_f)$ probe energy partitioning. Neglecting recoil, energy conservation implies:
\begin{equation}
E_{\bar{\nu}} \approx E_{l} + (M_n - M_p)
\end{equation}
where $E_{\bar{\nu}}$ is the incoming antineutrino energy, $E_{l}$ is the outgoing lepton (positron) energy, $M_n$ is the neutron mass, and $M_p$ is the proton mass.
The $(\nu, E_l)$ plot shows this linear dependence, while $(E_l, E_f)$ demonstrates the necessary anti-correlation to conserve total energy \cite{PhysRevD.60.053003}.

\textbf{Interaction Regime:} The pairs $(E_l, Q^2)$ and $(Q^2, W)$ verify that the interaction remains within the quasi-elastic regime ($W \approx M_p$) and respects the dependence of lepton energy on momentum transfer.

\textbf{Symmetry \& Beam Correlation:} We verify $(pxl, pyl)$ and $(pxf, pyf)$ to ensure no artificial angular bias exists in the transverse plane for either particle. $(pzv, pzl)$ validates the longitudinal momentum transfer.

\subsubsection{Neutral Current (7 Pairs)}
Since the outgoing neutrino is invisible in NC events, correlations must be anchored to the recoil nucleon. We evaluate 7 pairs as shown in Fig~\ref{fig:nc_corr1}.

\textbf{Recoil Kinematics:} The pairs $(\nu, E_f)$, $(E_f, Q^2)$, and $(\nu, Q^2)$ are critical. Since $Q^2$ is experimentally derived from the nucleon kinetic energy $T_N$, the $(E_f, Q^2)$ plot must show a tight, deterministic correlation defined by \cite{LLEWELLYNSMITH1972261}:
\begin{equation}
    Q^2 = 2 M_N T_N = 2 M_N (E_f - M_N)
\end{equation}
where $Q^2$ is the four-momentum transfer squared, $M_N$ is the nucleon mass, $T_N$ is the kinetic energy of the recoiling nucleon, and $E_f$ is the total energy of the final-state nucleon.

\textbf{Hadronic Physics:} The pairs $(Q^2, W)$ and $(y, Q^2)$ constrain the invariant mass and inelasticity to physically allowed regions for elastic scattering \cite{Formaggio_2012}.

\textbf{Nucleon Geometry:} The pairs $(pxf, pyf)$ and $(pzv, pzf)$ validate that the recoil nucleon retains the correct angular distribution relative to the incident beam, confirming the GAN's handling of angular distributions.

Figures ~\ref{fig: NuEElastic_corr1}, ~\ref{fig: ibd-cc_corr1} and ~\ref{fig:nc_corr1} present side-by-side 2D histogram heatmaps of key kinematic variable pairs, comparing GENIE-simulated (left) and CW-GAN generated (right) distributions, where color intensity encodes the event count per bin. The close visual agreement in the overall shape, populated phase-space boundaries, and high-density regions between the two columns confirms that the model has successfully learned the joint physical correlations between the kinematic variables.
\begin{figure*}[htbp!]
    \centering
    \begin{subfigure}{\textwidth}
        \centering
        \includegraphics[width=0.9\textwidth]{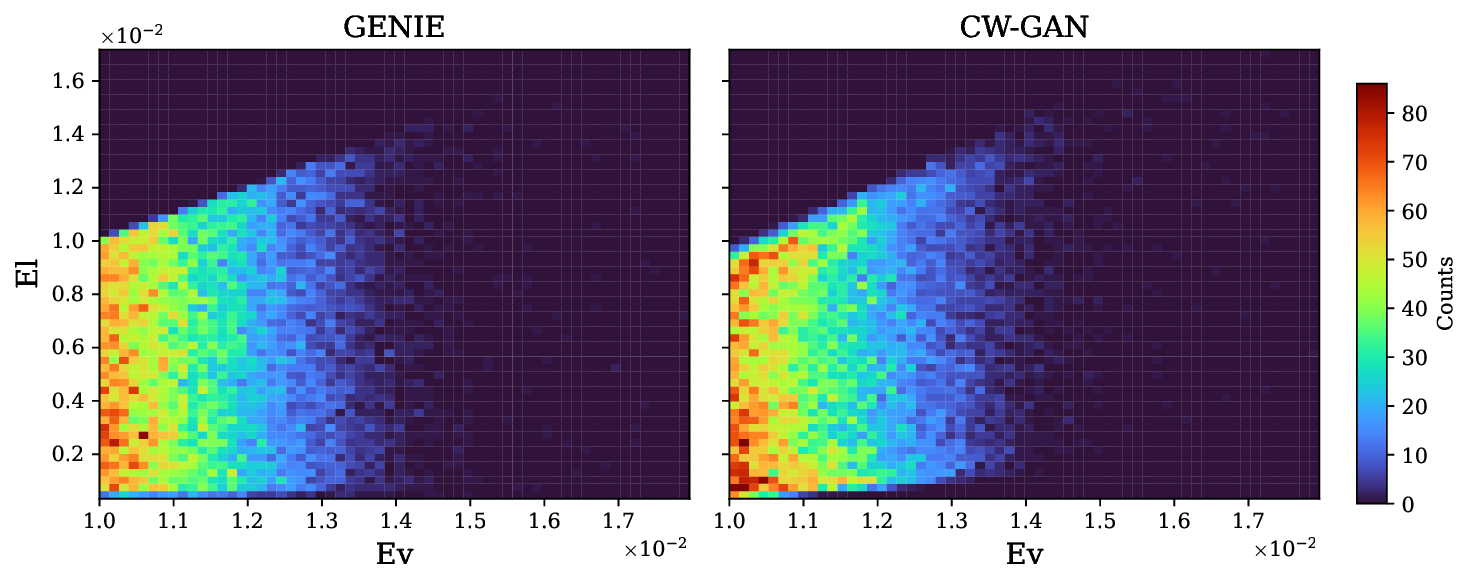}
        \caption{NuEElastic: $E_l$ vs $E_\nu$ Correlation}
        \label{fig:NuEElastic2D_el_vs_ev}
    \end{subfigure}
    
    \vspace{0.2cm}
    
    \begin{subfigure}{\textwidth}
        \centering
        \includegraphics[width=0.9\textwidth]{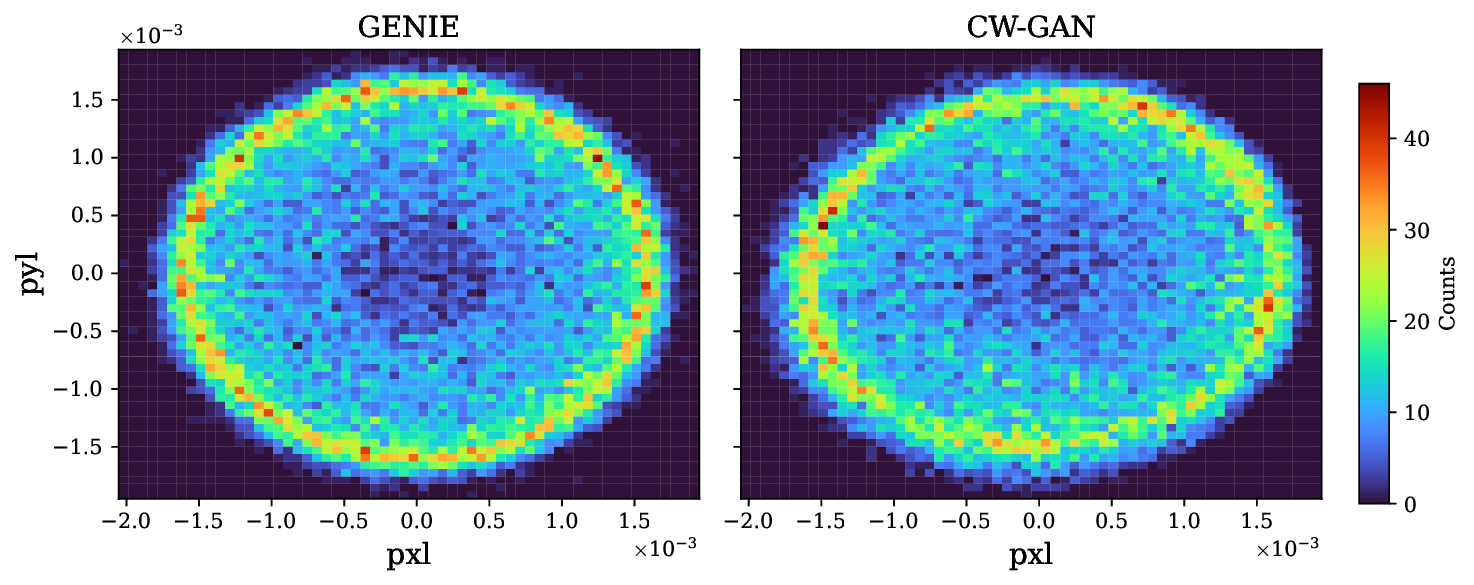}
        \caption{NuEElastic: $p_{yl}$ vs $p_{xl}$ Correlation}
        \label{fig:NuEElastic2D_pyl_vs_pxl}
    \end{subfigure}
    
    \vspace{0.2cm}
    
    \begin{subfigure}{\textwidth}
        \centering
        \includegraphics[width=0.9\textwidth]{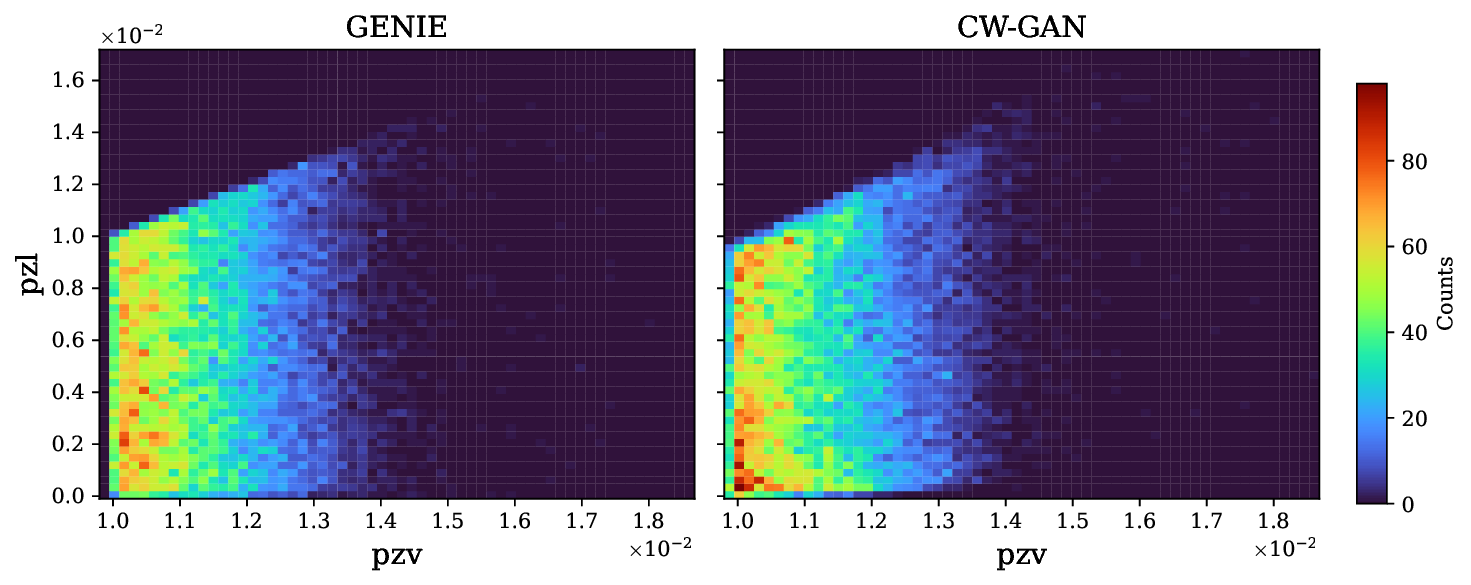}
        \caption{NuEElastic: $p_{zl}$ vs $p_{z\nu}$ Correlation}
        \label{fig:NuEElastic2D_pzl_vs_pzv}
    \end{subfigure}
    
    \caption{(\subref{fig:NuEElastic2D_el_vs_ev})Elastic Scattering : 2D Kinematic Correlations}
    \label{fig: NuEElastic_corr1}
\end{figure*}

\begin{figure*}[htbp!]
    \ContinuedFloat
    \centering
    \begin{subfigure}{\textwidth}
        \centering
        \includegraphics[width=0.9\textwidth]{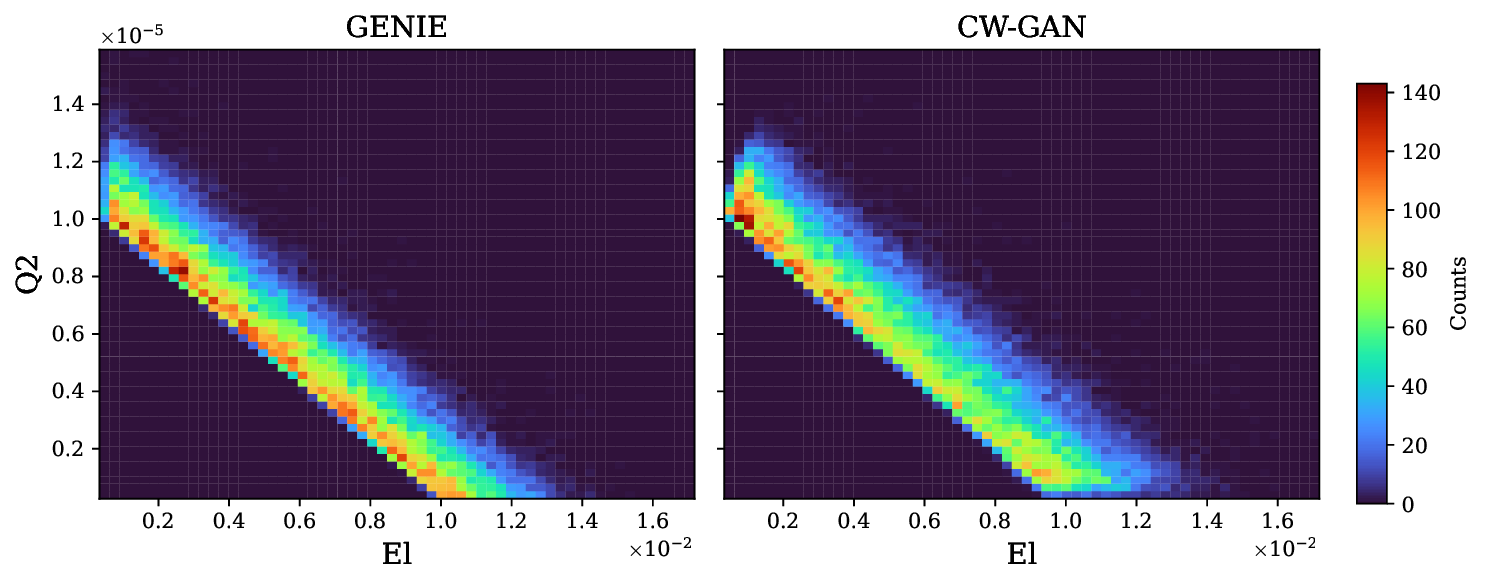}
        \caption{NuEElastic: $Q^2$ vs El Correlation}
        \label{fig:NuEElastic2D_Q2_vs_El}

    \end{subfigure}
    \vspace{0.2cm}
    \begin{subfigure}{\textwidth}
        \centering
        \includegraphics[width=0.9\textwidth]{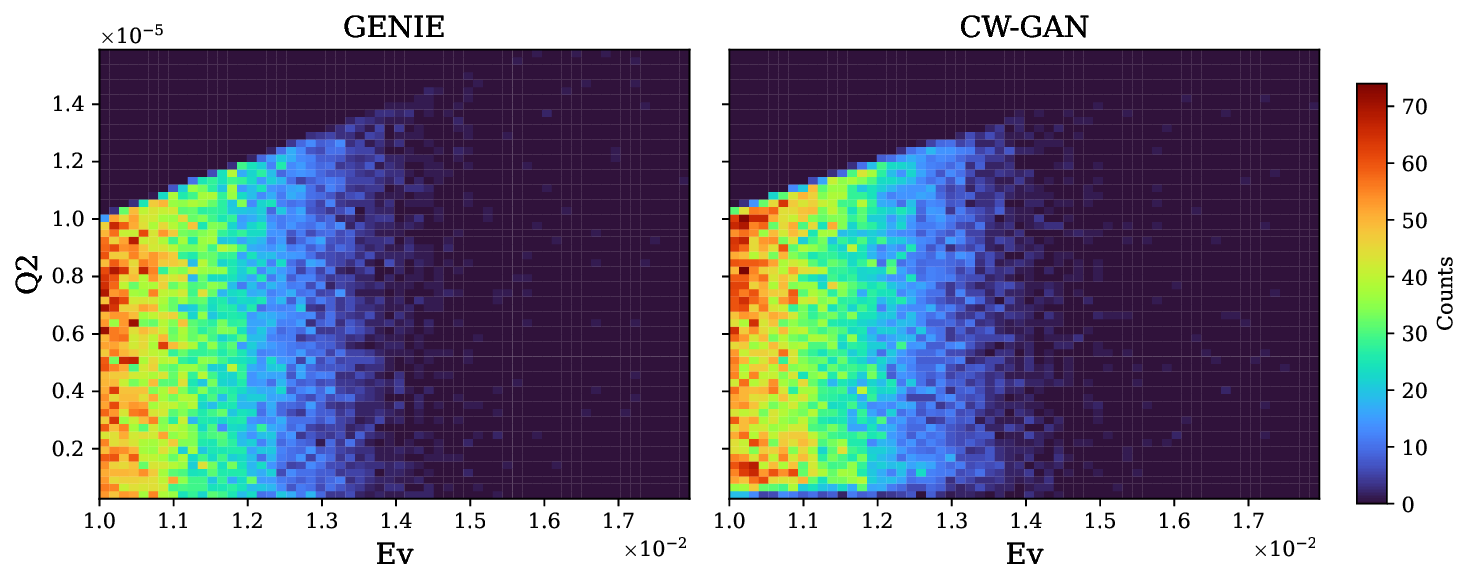}
        \caption{NuEElastic: $Q^2$ vs $Ev$ Correlation}
        \label{fig:NuEElastic2D_Q2_vs_Ev}

    \end{subfigure}
    \vspace{0.2cm}
    \begin{subfigure}{\textwidth}
        \centering
        \includegraphics[width=0.9\textwidth]{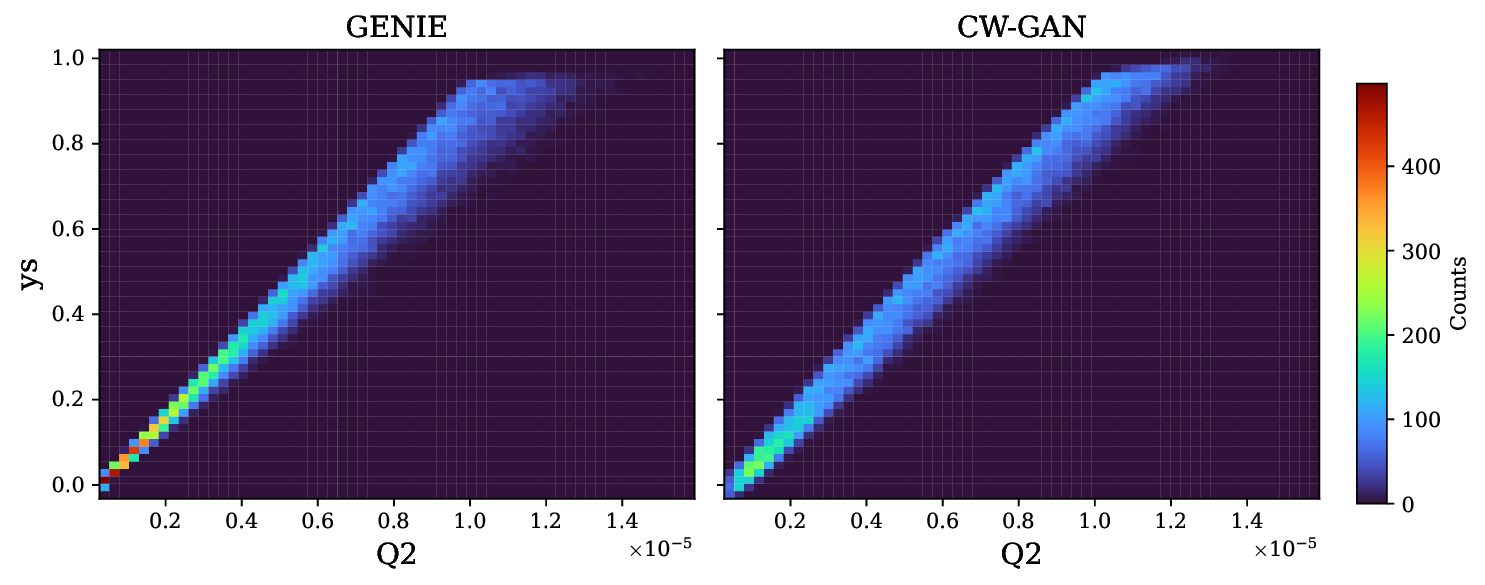}
        \caption{NuEElastic: ys vs $Q^2$ Correlation}
        \label{fig:NuEElastic2D_ys_vs_Q2}

    \end{subfigure}
    \caption{(\subref{fig:NuEElastic2D_pyl_vs_pxl}) Elastic Scattering :  2D Kinematic Correlations (Continued)}
    \label{fig:NuEElastic_corr2}
\end{figure*}

\begin{figure*}[htbp!]
    \centering
    \begin{subfigure}{\textwidth}
        \centering
        \includegraphics[width=0.9\textwidth]{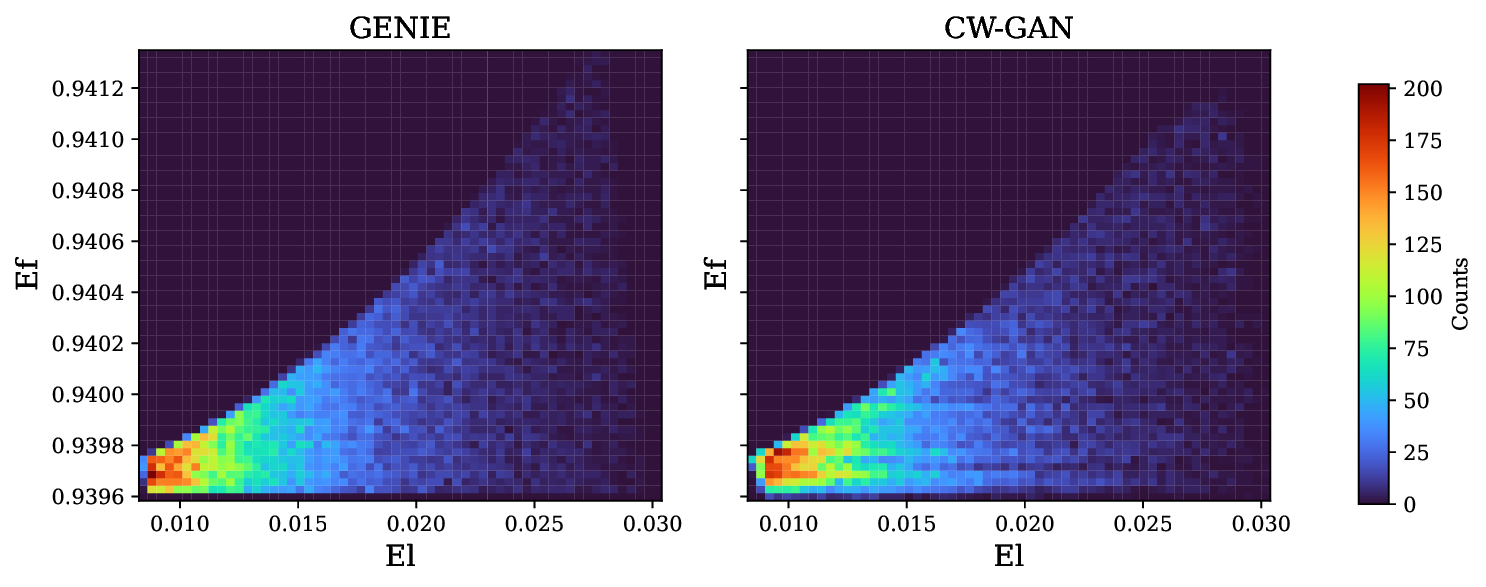}
        \caption{IBD-CC: Ef vs El}
        \label{fig:ibd_Ef_vs_El}
    \end{subfigure}
    
    \vspace{0.2cm}
    
    \begin{subfigure}{\textwidth}
        \centering
        \includegraphics[width=0.9\textwidth]{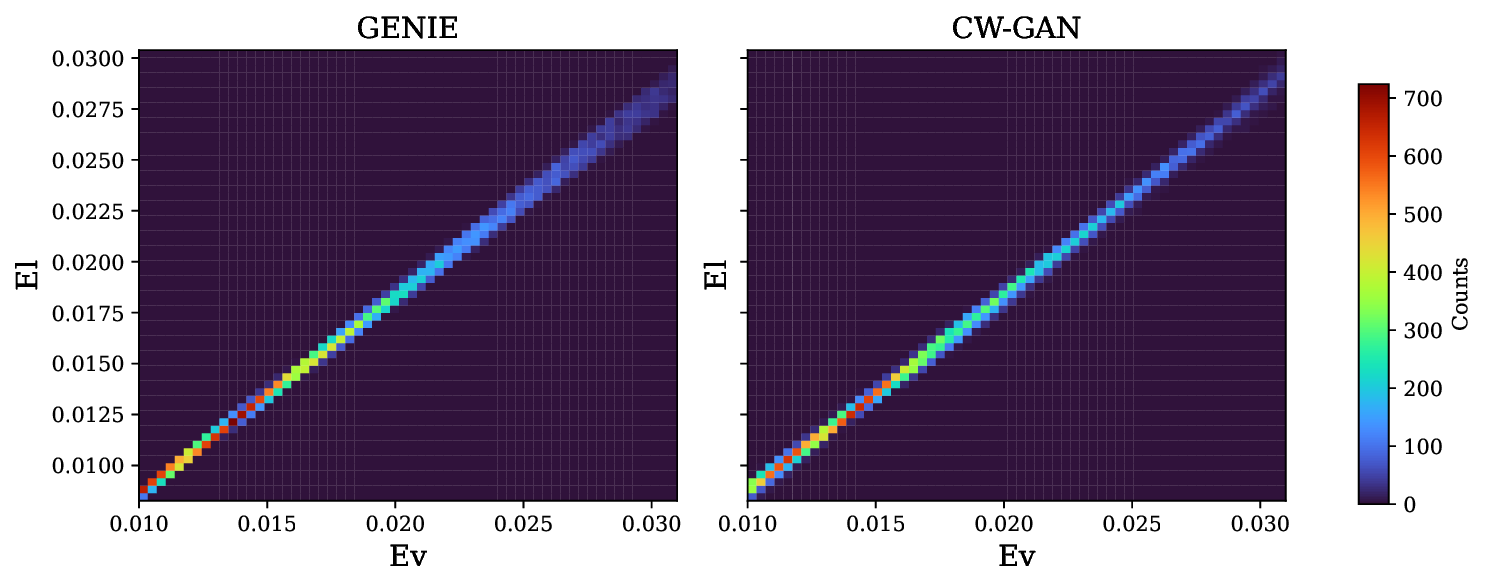}
        \caption{IBD-CC: El vs Ev}
        \label{fig:ibd_El_vs_Ev}
    \end{subfigure}
    
    \vspace{0.2cm}
    
    \begin{subfigure}{\textwidth}
        \centering
        \includegraphics[width=0.9\textwidth]{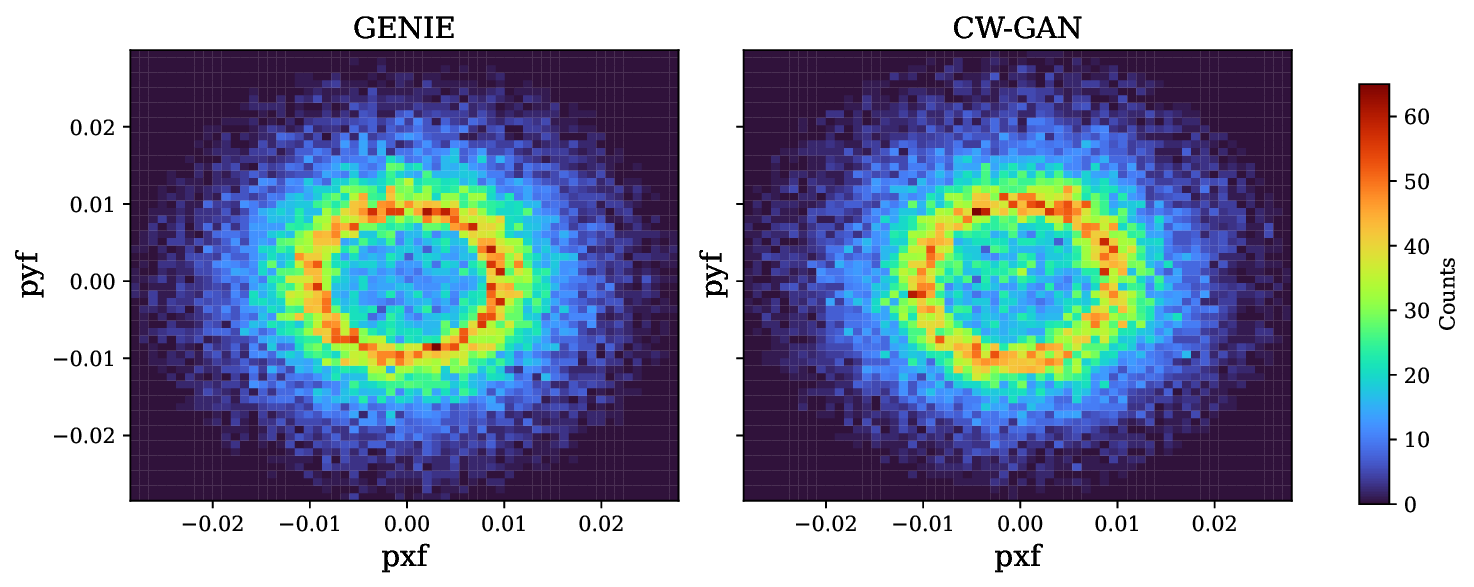}
        \caption{IBD-CC: pyf vs pxf}
        \label{fig:ibd_pyf_vs_pxf}
    \end{subfigure}
    \caption{(\subref{fig:ibd_Ef_vs_El}) IBD-CC : 2D Kinematic Correlations}
    \label{fig: ibd-cc_corr1}
\end{figure*}

\begin{figure*}[t!]
    \ContinuedFloat 
    \centering
    \begin{subfigure}{\textwidth}
        \centering
        \includegraphics[width=0.9\textwidth]{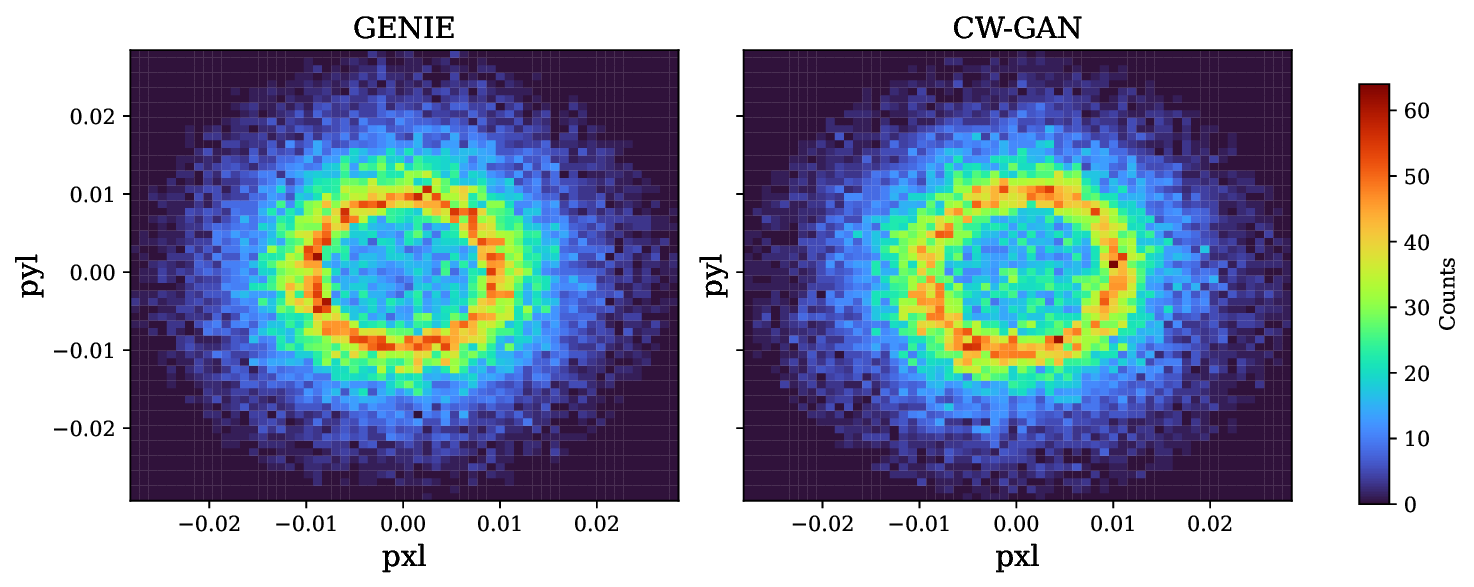}
        \caption{IBD-CC: pyl vs pxl}
        \label{fig:ibd_pyl_vs_pxl}
    \end{subfigure}
    
    \vspace{0.2cm}
    
    \begin{subfigure}{\textwidth}
        \centering
        \includegraphics[width=0.9\textwidth]{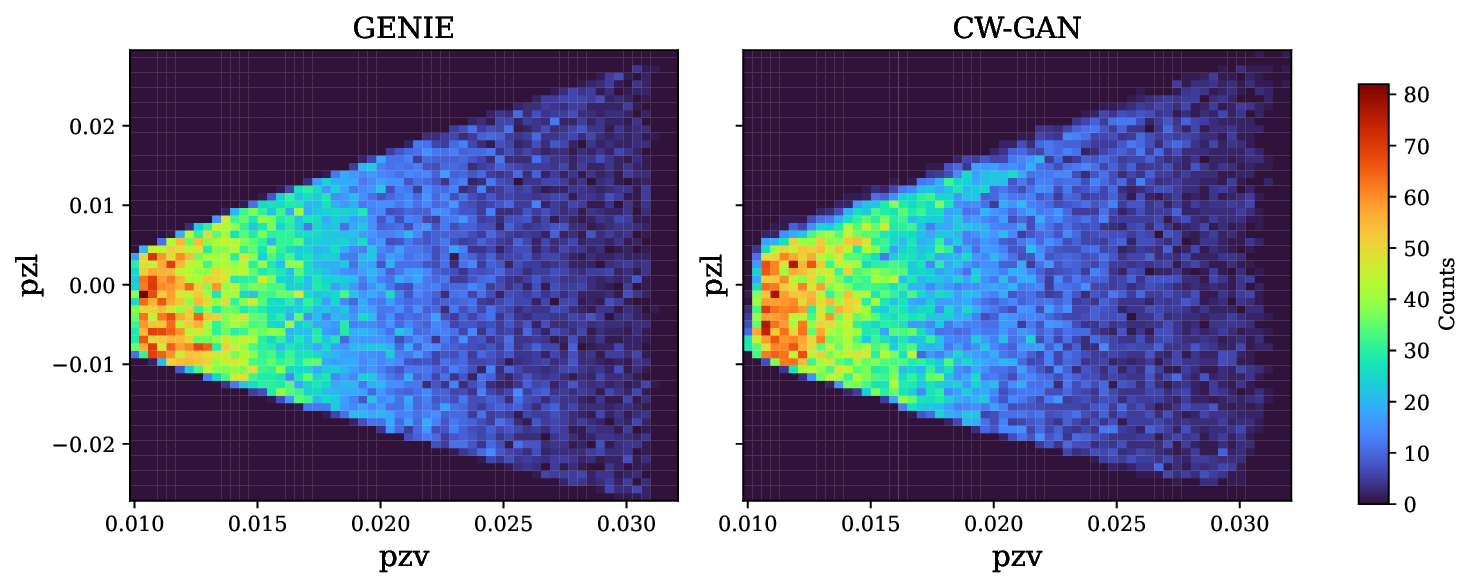}
        \caption{IBD-CC: pzl vs pzv}
        \label{fig:ibd_pzl_vs_pzv}
    \end{subfigure}
    
    \vspace{0.2cm}
    
    \begin{subfigure}{\textwidth}
        \centering
        \includegraphics[width=0.9\textwidth]{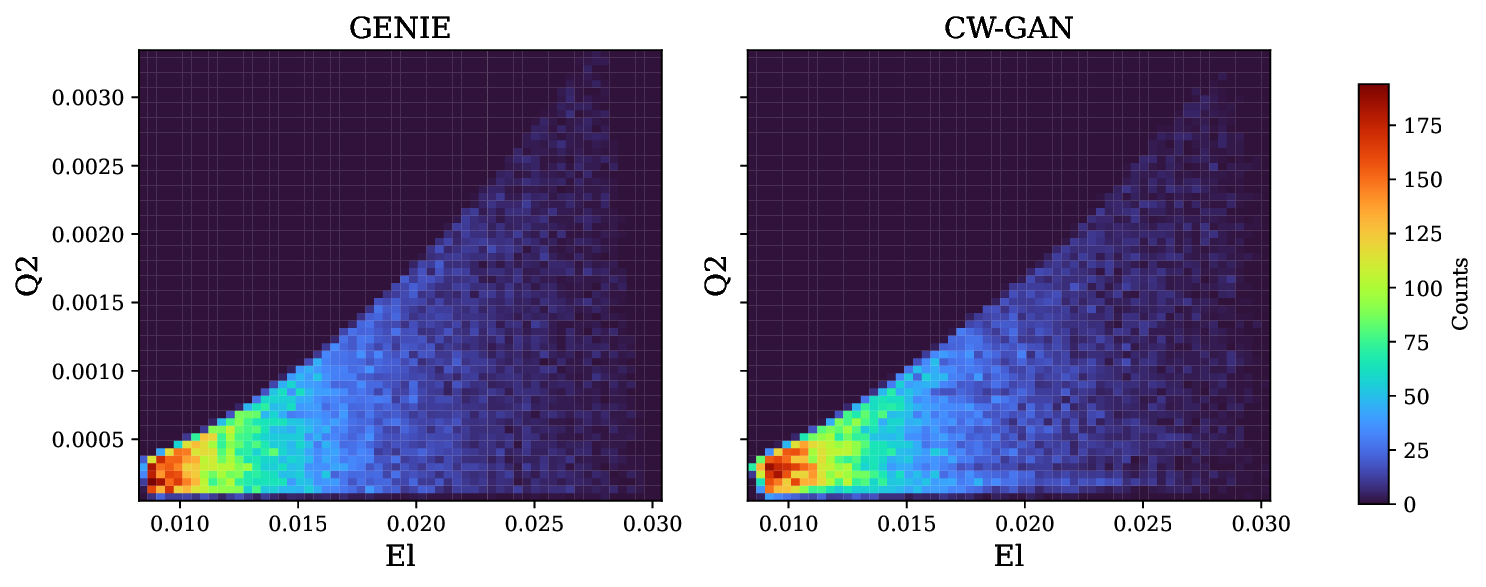}
        \caption{IBD-CC: Q2 vs El}
        \label{fig:ibd_Q2_vs_El}
    \end{subfigure}
    
    \caption{(\subref{fig:ibd_El_vs_Ev}) IBD-CC: 2D Kinematic Correlations (Continued)}
    \label{fig:ibd-cc_corr2}
\end{figure*}

\begin{figure*}[t!]
    \ContinuedFloat 
    \centering
    \begin{subfigure}{\textwidth}
        \centering
        \includegraphics[width=0.9\textwidth]{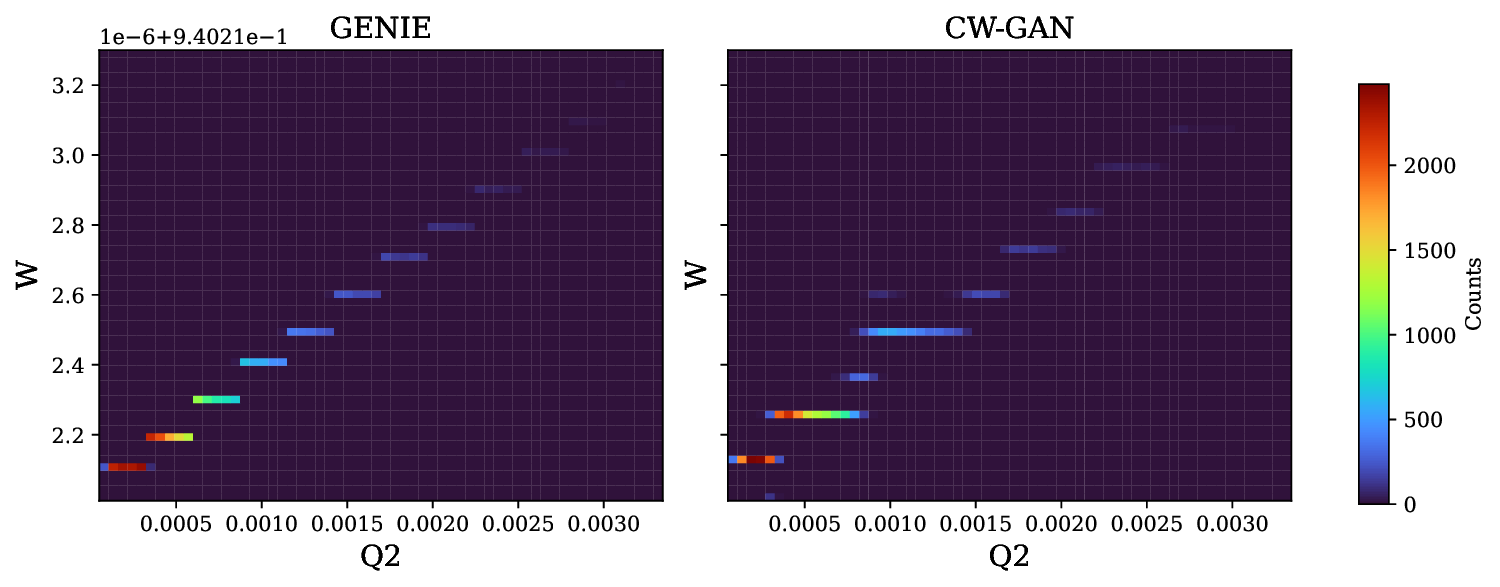}
        \caption{IBD-CC: W vs Q2}
        \label{fig:ibd_W_vs_Q2}
    \end{subfigure}
    \caption{(\subref{fig:ibd_pyf_vs_pxf}) IBD-CC: 2D Kinematic Correlations (Continued)}
    \label{fig:ibd-cc_corr3}
\end{figure*}

\vspace{-\floatsep}

\begin{figure*}[t!]
    \centering
    \begin{subfigure}{\textwidth}
        \centering
        \includegraphics[width=0.9\textwidth]{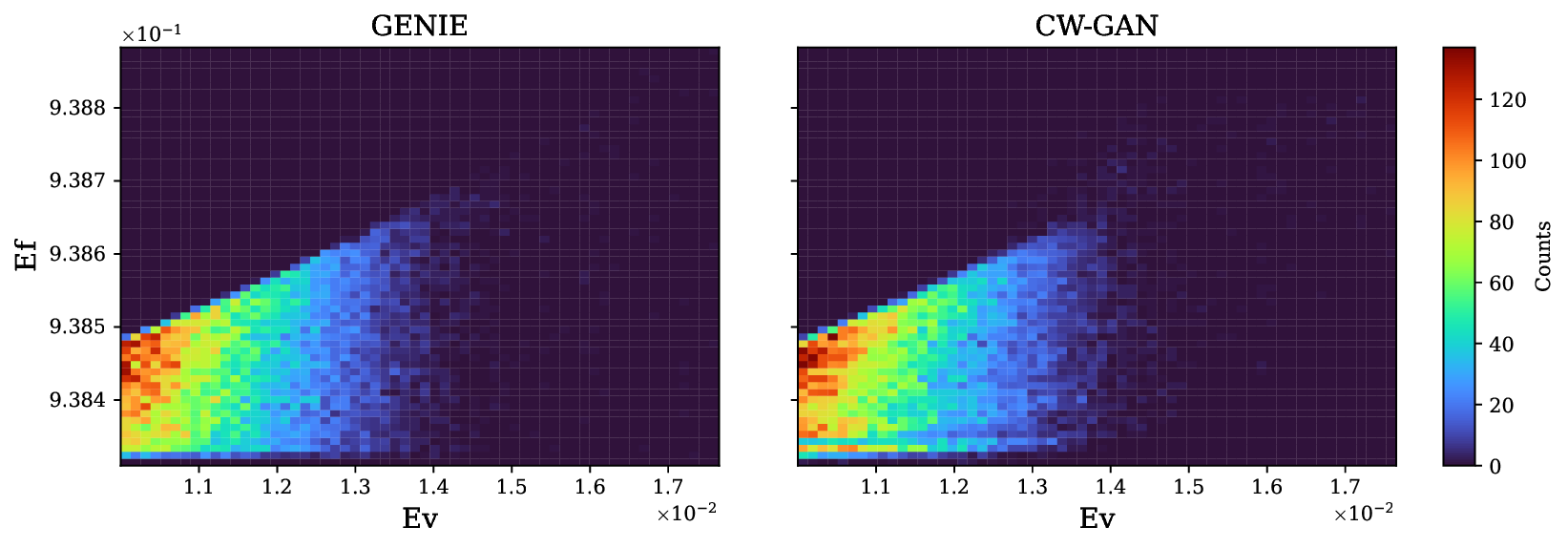}
        \caption{NC: Ef vs Ev}
        \label{fig:nc_Ef_vs_Ev}
    \end{subfigure}
    
    \vspace{0.2cm}
    
    \begin{subfigure}{\textwidth}
        \centering
        \includegraphics[width=0.9\textwidth]{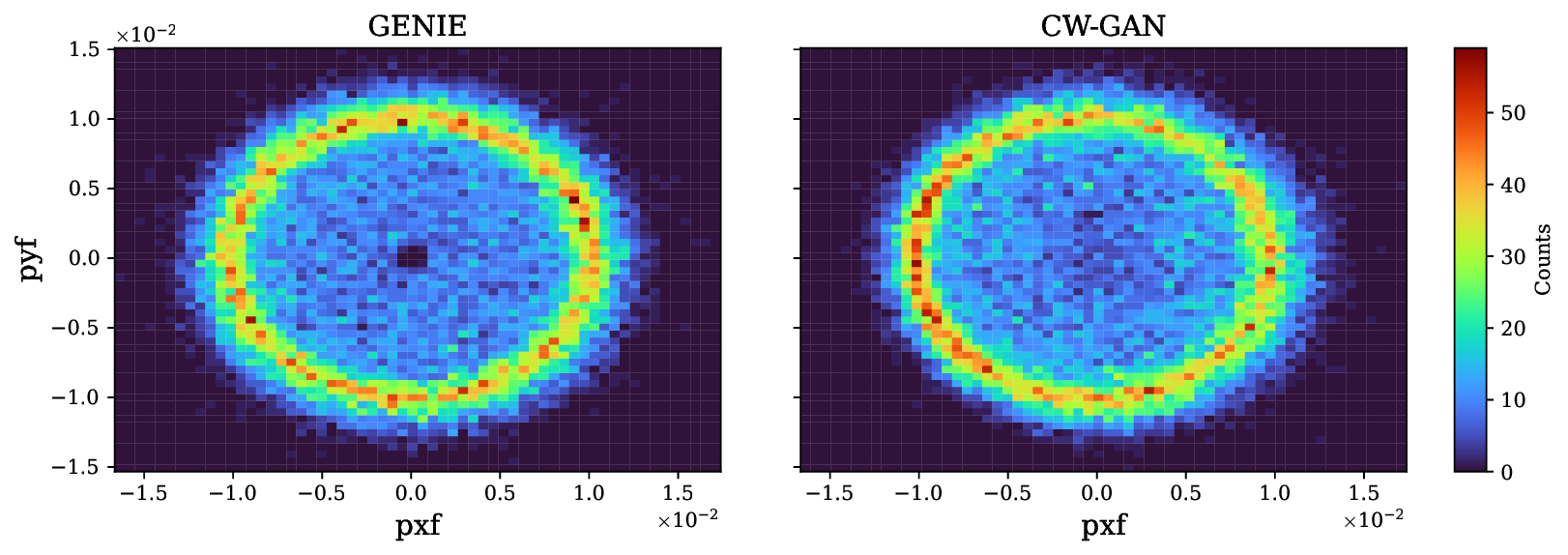}
        \caption{NC: pyf vs pxf}
        \label{fig:nc_pyf_vs_pxf}
    \end{subfigure}
    
    \caption{(\subref{fig:nc_Ef_vs_Ev}) NC: 2D Kinematic Correlations }
    \label{fig:nc_corr1}
\end{figure*}

\begin{figure*}[htbp!]
    \ContinuedFloat 
    \centering
    \begin{subfigure}{\textwidth}
        \centering
        \includegraphics[width=0.9\textwidth]{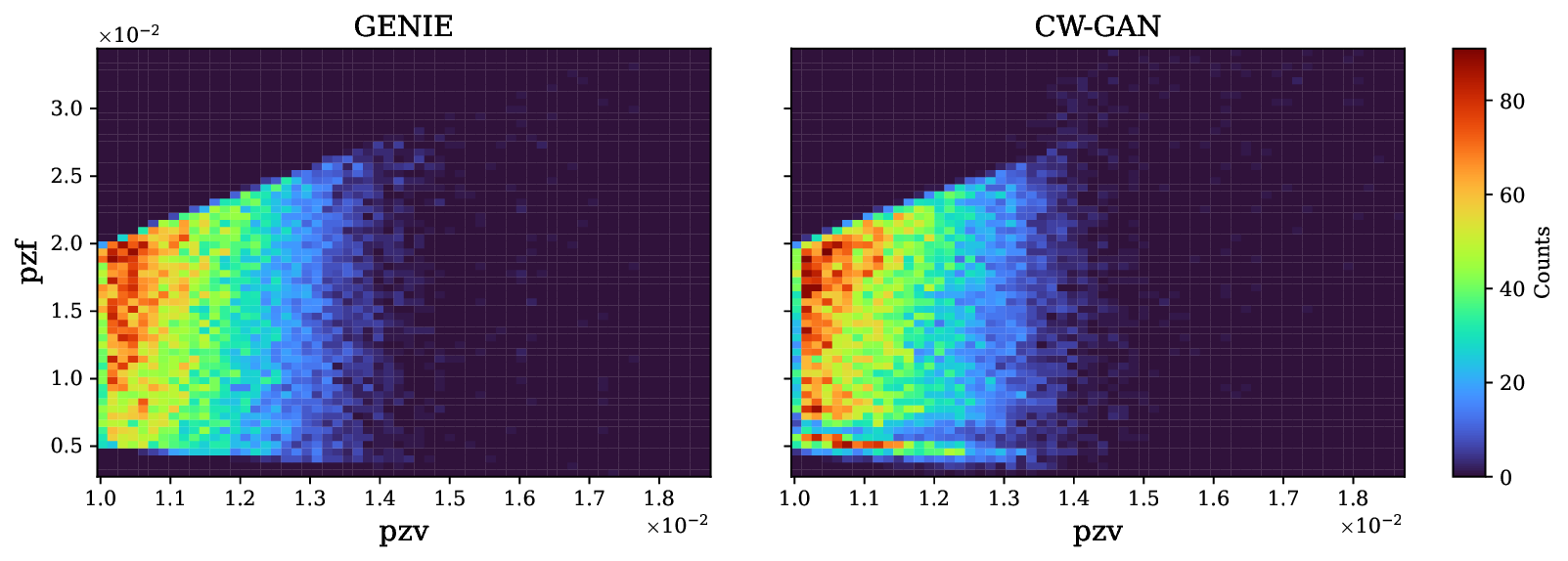}
        \caption{NC: pzf vs pzv}
        \label{fig:nc_pzf_vs_pzv}
    \end{subfigure}
    \vspace{0.2cm}
    \begin{subfigure}{\textwidth}
        \centering
        \includegraphics[width=0.9\textwidth]{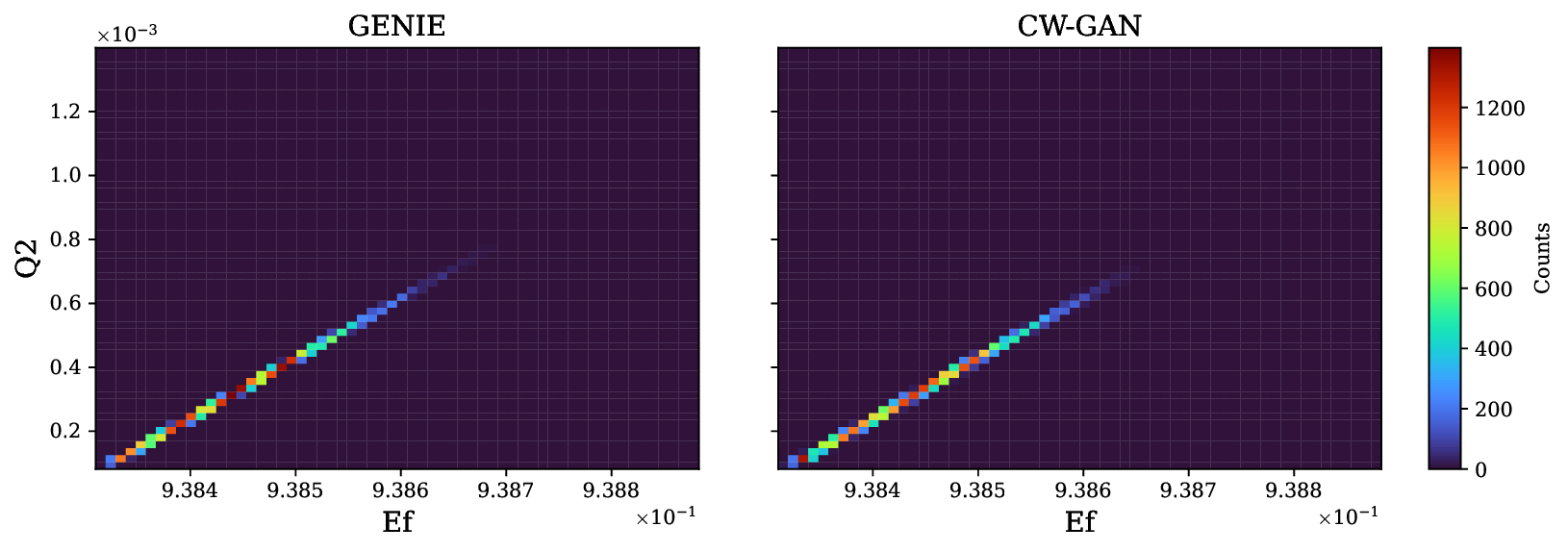}
        \caption{NC: $Q^2$ vs Ef}
        \label{fig:nc_Q2_vs_Ef}
    \end{subfigure}
    \vspace{0.2cm}
    \begin{subfigure}{\textwidth}
        \centering
        \includegraphics[width=0.9\textwidth]{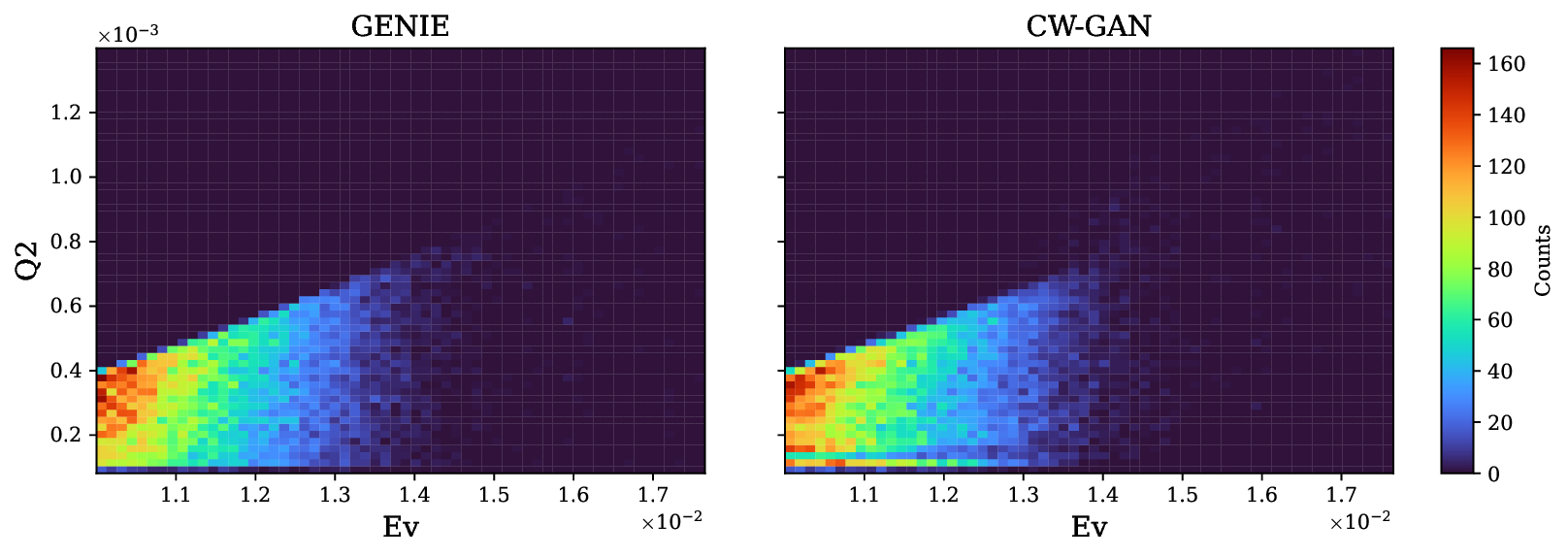}
        \caption{NC: $Q^2$ vs Ev}
        \label{fig:nc_Q2_vs_Ev}
    \end{subfigure}
    \caption{(\subref{fig:nc_pyf_vs_pxf}) NC: 2D Kinematic Correlations (Continued)}
    \label{fig:nc_corr2}
\end{figure*}

\begin{figure*}[htbp!]
    \ContinuedFloat 
    \centering
    \begin{subfigure}{\textwidth}
        \centering
        \includegraphics[width=0.9\textwidth]{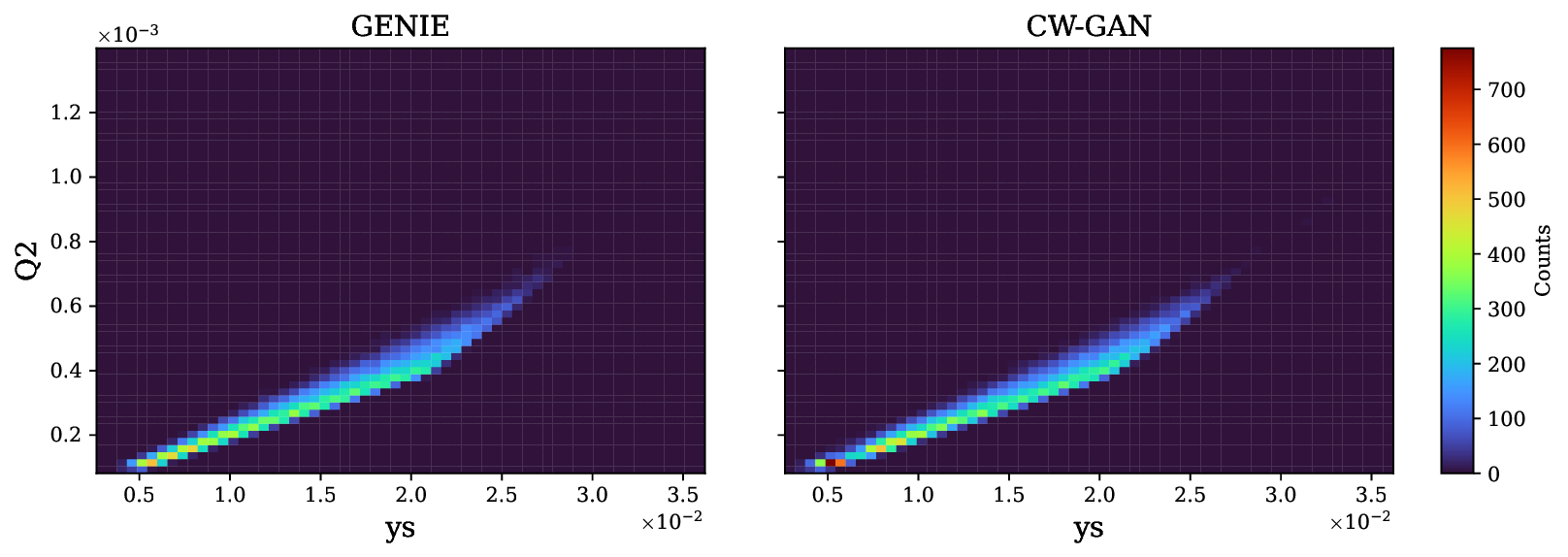}
        \caption{NC: $Q^2$ vs ys}
        \label{fig:nc_Q2_vs_ys}
    \end{subfigure}
    \vspace{0.2cm}
    \begin{subfigure}{\textwidth}
        \centering
        \includegraphics[width=0.9\textwidth]{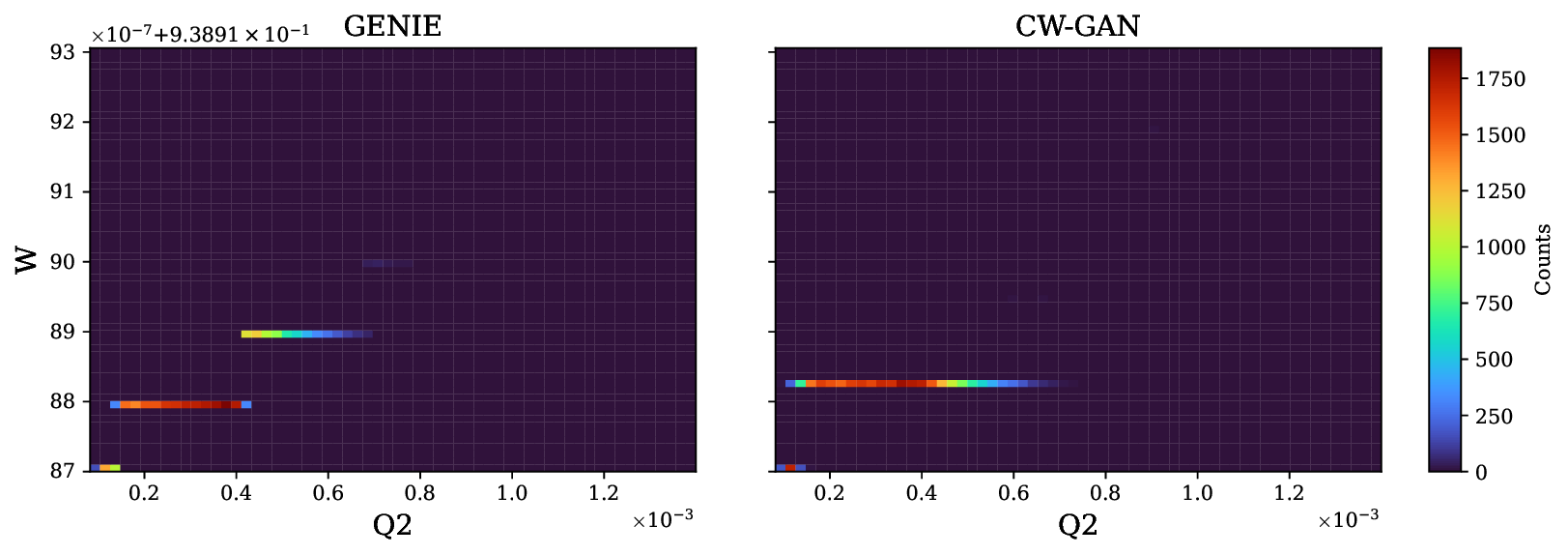}
        \caption{NC: W vs $Q^2$}
        \label{fig:nc_W_vs_Q2}
    \end{subfigure}
    \caption{(\subref{fig:nc_pzf_vs_pzv}) NC: 2D Kinematic Correlations (Continued)}
    \label{fig:nc_corr3}
\end{figure*}

\section{Conclusion}
\label{sec:conclusion}
We have introduced a fast, physics-informed Conditional Wasserstein GAN with Gradient Penalty (CW-GAN) for simulating neutrino-nucleus interactions in the 10-31 MeV energy range. By conditioning the generator on incident neutrino energy, the model successfully replicates the distinct kinematic signatures of Elastic Scattering (NuEElastic), Inverse Beta Decay (IBD-CC), and Neutral Current (NC) interactions defined by the GENIE generator.
\\
\newline
A primary advantage of this framework is its operational efficiency. The CW-GAN functions as a rapid inference engine, generating complete datasets in approximately \textbf{5 seconds} (using a 2-core CPU), a task that requires \textbf{10.13 minutes} using the standard GENIE generator. This acceleration significantly removes bottlenecks for large-scale sensitivity studies. 
Crucially, the model demonstrates high fidelity in both statistical precision and physical consistency. Our analysis of 2D joint distributions confirms that the network has implicitly learned fundamental conservation laws without explicit numerical integration.
In the NuEElastic channel, the generator respects the maximum allowable energy transfer ($E_l \le E_\nu$) imposed by center-of-mass limits.
The model autonomously reproduces the rotational invariance of scattering processes, evidenced by perfect symmetry in the transverse momentum planes ($p_x, p_y$) across all channels.
In the NC channel, the generator maintains a deterministic relationship between momentum transfer $Q^2$ and recoil proton energy, effectively balancing the visible proton against the invisible outgoing neutrino.
\\
\newline
Quantitative validation using Earth Mover's Distance (EMD) and Mean Absolute Pull (MAP) metrics indicates that the generated samples are statistically indistinguishable from the ground truth, with global pull distributions centered at $\mu \approx 0$ with $\sigma \approx 1$.
In summary, the CW-GAN serves as a surrogate for GENIE, reproducing its kinematic distributions within statistical precision. It does not reduce systematic uncertainties inherent in the GENIE physics model, but provides significant computational acceleration for applications where GENIE's physics assumptions are adequate.

Ongoing work is focused on transitioning this framework into a standardized, deployable software tool that allows users to input custom flux parameters directly. Future developments will also integrate detector response functions into the generator's loss landscape, enabling ``end-to-end'' simulation from incident flux to detector hits. This work establishes generative adversarial networks as a scalable, high-precision tool for the next generation of neutrino physics analyses.

\section{ACKNOWLEDGEMENTS}
This research is supported by the Department of Science and Technology (DST), India, under the Fund for Improvement of S\&T Infrastructure in Universities and Higher Educational Institutions (FIST) Program [Grant No.~SR/FST/ET-I/2022/1079], and a matching grant from VIT University. The authors are grateful to DST-FIST and VIT management for their financial support and the resources provided for this work.
\\
\newline
The authors also thank Dr. Suprabh Prakash for his mentorship during the course of this study and for his critical review of the physical framework presented herein.
\clearpage
\appendix

\onecolumngrid 

\section{GENIE Summary Tree (.gst) Parameter Definitions}
\label{app:gst_definitions}

The GENIE Summary Tree (\texttt{.gst}) format provides a comprehensive record of the generated event kinematics. Table \ref{tab:gst_full_defs} details the full list of 81 parameters available in the ntuple structure.


\begin{longtblr}[
  caption = {Complete definition of parameters in the GENIE \texttt{.gst} output format.},
  label = {tab:gst_full_defs},
  entry = {Complete definition of parameters in the GENIE .gst output format.},
]{
  colspec = {l p{0.15\textwidth} p{0.65\textwidth}},
  rowhead = 1, 
  hlines,
  row{1} = {font=\bfseries},
}
Parameter & Type & Definition \\

\texttt{iev} & \texttt{int} & Event number. \\
\texttt{neu} & \texttt{int} & Neutrino PDG code. \\
\texttt{tgt} & \texttt{int} & Nuclear target PDG code (10LZZZAAAI). \\
\texttt{Z} & \texttt{int} & Nuclear target Z. \\
\texttt{A} & \texttt{int} & Nuclear target A. \\
\texttt{hitnuc} & \texttt{int} & Hit nucleon PDG code. \\
\texttt{hitqrk} & \texttt{int} & Hit quark PDG code (DIS only). \\
\texttt{sea} & \texttt{bool} & Hit quark is from sea (DIS only). \\
\texttt{resid} & \texttt{bool} & Produced baryon resonance ID. \\

\texttt{qel} & \texttt{bool} & Quasi-elastic scattering event. \\
\texttt{res} & \texttt{bool} & Resonance neutrino-production event. \\
\texttt{dis} & \texttt{bool} & Deep-inelastic scattering event. \\
\texttt{coh} & \texttt{bool} & Coherent meson production event. \\
\texttt{dfr} & \texttt{bool} & Diffractive meson production event. \\
\texttt{imd} & \texttt{bool} & Inverse muon decay event. \\
\texttt{nuel} & \texttt{bool} & \(\nu e^-\) elastic event. \\
\texttt{cc} & \texttt{bool} & Charged Current (CC) event. \\
\texttt{nc} & \texttt{bool} & Neutral Current (NC) event. \\
\texttt{charm} & \texttt{bool} & Charm production flag. \\

\texttt{neut\_code} & \texttt{int} & Equivalent NEUT reaction code. \\
\texttt{nuance\_code} & \texttt{int} & Equivalent NUANCE reaction code. \\
\texttt{wght} & \texttt{double} & Event weight. \\

\texttt{xs} & \texttt{double} & Bjorken \(x\) (selected/off-shell). \\
\texttt{ys} & \texttt{double} & Inelasticity \(y\) (selected/off-shell). \\
\texttt{ts} & \texttt{double} & Energy transfer to nucleus (selected). \\
\texttt{Q2s} & \texttt{double} & Momentum transfer \(Q^2\) (selected) [GeV\(^2\)]. \\
\texttt{Ws} & \texttt{double} & Hadronic invariant mass \(W\) (selected). \\

\texttt{x} & \texttt{double} & Bjorken \(x\) (from event record). \\
\texttt{y} & \texttt{double} & Inelasticity \(y\) (from event record). \\
\texttt{t} & \texttt{double} & Energy transfer to nucleus (from event record). \\
\texttt{Q2} & \texttt{double} & Momentum transfer \(Q^2\) (from event record) [GeV\(^2\)]. \\
\texttt{W} & \texttt{double} & Hadronic invariant mass \(W\) (from event record). \\

\texttt{Ev} & \texttt{double} & Incoming neutrino energy [GeV]. \\
\texttt{pxv, pyv, pzv} & \texttt{double} & Incoming neutrino momentum components [GeV]. \\

\texttt{En} & \texttt{double} & Initial state hit nucleon energy [GeV]. \\
\texttt{pxn, pyn, pzn} & \texttt{double} & Initial state hit nucleon momentum components [GeV]. \\

\texttt{$E_l$} & \texttt{double} & Final state primary lepton energy [GeV]. \\
\texttt{pxl, pyl, pzl} & \texttt{double} & Final state primary lepton momentum components [GeV]. \\

\texttt{nfp, nfn} & \texttt{int} & Number of final state \(p, \bar{p}\) and \(n, \bar{n}\). \\
\texttt{nfpip, nfpim} & \texttt{int} & Number of final state \(\pi^+, \pi^-\). \\
\texttt{nfpi0} & \texttt{int} & Number of final state \(\pi^0\). \\
\texttt{nfkp, nfkm} & \texttt{int} & Number of final state \(K^+, K^-\). \\
\texttt{nfk0} & \texttt{int} & Number of final state \(K^0, \bar{K}^0\). \\
\texttt{nfem} & \texttt{int} & Number of final state \(\gamma, e^-, e^+\). \\
\texttt{nfother} & \texttt{int} & Number of heavier final state hadrons. \\

\texttt{nip, nin} & \texttt{int} & Number of primary \(p, \bar{p}\) and \(n, \bar{n}\). \\
\texttt{nipip, nipim} & \texttt{int} & Number of primary \(\pi^+, \pi^-\). \\
\texttt{nipi0} & \texttt{int} & Number of primary \(\pi^0\). \\
\texttt{nikp, nikm} & \texttt{int} & Number of primary \(K^+, K^-\). \\
\texttt{nik0} & \texttt{int} & Number of primary \(K^0, \bar{K}^0\). \\
\texttt{niem} & \texttt{int} & Number of primary \(\gamma, e^-, e^+\). \\
\texttt{niother} & \texttt{int} & Number of other primary hadrons. \\

\texttt{nf} & \texttt{int} & Count of particles in hadronic system. \\
\texttt{pdgf} & \texttt{int[]} & PDG codes of final state particles. \\
\texttt{Ef} & \texttt{double[]} & Energies of final state particles [GeV]. \\
\texttt{pxf, pyf, pzf} & \texttt{double[]} & Momenta of final state particles [GeV]. \\

\texttt{ni} & \texttt{int} & Count of particles in primary hadronic system. \\
\texttt{pdgi} & \texttt{int[]} & PDG codes of primary particles. \\
\texttt{Ei} & \texttt{double[]} & Energies of primary particles [GeV]. \\
\texttt{pxi, pyi, pzi} & \texttt{double[]} & Momenta of primary particles [GeV]. \\

\texttt{vtxx, vtxy, vtxz} & \texttt{double} & Vertex position \((x, y, z)\) in detector [m]. \\
\texttt{vtxt} & \texttt{double} & Vertex time \(t\) in detector [s]. \\
\texttt{calresp0} & \texttt{double} & Approximate calorimetric response. \\

\end{longtblr}

\twocolumngrid 

\bibliography{apssamp}

\end{document}